\documentclass[preprint,flushrt]{aastex}
\usepackage{natbib}
\def\aro{{$\alpha_{\rm ro}$}}
\def\aox{{$\alpha_{\rm ox}$}}
\def\arx{{$\alpha_{\rm rx}$}}

\def\axg{{$\alpha_{\rm x\gamma}$}}

\def\ax{{$\alpha_{\rm x}$~}}

\def\gr{{$\gamma$-ray}}

\def\nupS{$\nu_{\rm peak}^S$}
\def\nupIC{$\nu_{\rm peak}^{IC}$}

\newcommand{\lsim}{{\lower.5ex\hbox{$\; \buildrel < \over \sim \;$}}}
\newcommand{\gsim}{{\lower.5ex\hbox{$\; \buildrel > \over \sim \;$}}}
\newcommand\pdflux{\mbox{${\rm \, photons \,\, MeV^{-1} \, cm^{-2} \, s^{-1}}$}}
\newcommand{\etal}{et al.}
\newcommand{\fermi}{{\it Fermi} }
\newcommand{\frmi}{{\it Fermi}}
\newcommand{\swift}{{\it Swift} }
\newcommand{\sw}{{\it Swift}}

\slugcomment{Submitted to Astrophysical Journal}

\shorttitle{The SEDs of \fermi bright  blazars}
\shortauthors{Abdo et al.}

\title{The Spectral Energy Distribution of \fermi bright blazars\\}
\author{
A.~A.~Abdo\altaffilmark{2,3},
M.~Ackermann\altaffilmark{4},
M.~Ajello\altaffilmark{4},
M.~Axelsson\altaffilmark{5,6},
L.~Baldini\altaffilmark{7},
J.~Ballet\altaffilmark{8},
G.~Barbiellini\altaffilmark{9,10},
D.~Bastieri\altaffilmark{11,12},
B.~M.~Baughman\altaffilmark{13},
K.~Bechtol\altaffilmark{4}
% ...
}
% ...
\altaffiltext{2}{Space Science Division, Naval Research Laboratory, Washington, DC 20375, USA}
\altaffiltext{3}{National Research Council Research Associate, National Academy of Sciences, Washington, DC 20001, USA}
\altaffiltext{4}{W. W. Hansen Experimental Physics Laboratory, Kavli Institute for Particle Astrophysics and Cosmology, Department of Physics and SLAC National Accelerator Laboratory, Stanford University, Stanford, CA 94305, USA}
\altaffiltext{5}{Department of Astronomy, Stockholm University, SE-106 91 Stockholm, Sweden}
\altaffiltext{6}{The Oskar Klein Centre for Cosmoparticle Physics, AlbaNova, SE-106 91 Stockholm, Sweden}
\altaffiltext{7}{Istituto Nazionale di Fisica Nucleare, Sezione di Pisa, I-56127 Pisa, Italy}
\altaffiltext{8}{Laboratoire AIM, CEA-IRFU/CNRS/Universit\'e Paris Diderot, Service d'Astrophysique, CEA Saclay, 91191 Gif sur Yvette, France}
\altaffiltext{9}{Istituto Nazionale di Fisica Nucleare, Sezione di Trieste, I-34127 Trieste, Italy}
\altaffiltext{10}{Dipartimento di Fisica, Universit\`a di Trieste, I-34127 Trieste, Italy}
\altaffiltext{11}{Istituto Nazionale di Fisica Nucleare, Sezione di Padova, I-35131 Padova, Italy}
\altaffiltext{12}{Dipartimento di Fisica ``G. Galilei", Universit\`a di Padova, I-35131 Padova, Italy}
\altaffiltext{13}{Department of Physics, Center for Cosmology and Astro-Particle Physics, The Ohio State University, Columbus, OH 43210, USA}
% ...
\altaffiltext{a+2}{ASI Science Data Center, ASDC, via G. Galilei, I-00040 Frascati, Italy}
\altaffiltext{a+3}{Universit\'a degli studi di Milano-Bicocca, Dipartimento di Fisica, Piazza delle Scienze 3, I-20126 Milan, Italy}
\altaffiltext{a+4} {INAF Ð Osservatorio Astronomico di Brera, Via Bianchi 46, 23807 Merate, Italy}
\altaffiltext{a+5}{NASA,Goddard Space Flight Center, Greenbelt, MD 20771, USA}
\altaffiltext{a+6}{Department of Physics \& Astronomy, University of Leicester, University Road, Leicester, LE1 7RH, UK}
\altaffiltext{a+7}{Max-Planck-Institut f\"ur Radioastronomie, Auf dem H\"ugel 69, 53121 Bonn, Germany}
\altaffiltext{a+8}{Astro Space Center of Lebedev Physical Institute, Profsoyuznaya 84/32, 117997 Moscow, Russia}
\altaffiltext{a+9}{Special Astrophysical Observatory, Nizhnij Arkhyz, Karachai-Cherkessian Republic, 369167 Russia}
\altaffiltext{a+10}{INAF/Osservatorio Astronomico di Torino, via Osservatorio 20, 10025 Pino Torinese, Italy}
\altaffiltext{a+11}{Instituto de Astrof\'{\i}sica de Andaluc\'{\i}a, CSIC, Apartado 3004, 18080 Granada, Spain}
\altaffiltext{a+12}{Department of Astronomy, University of Michigan, Ann Arbor, MI 48109-1042, USA}
\altaffiltext{a+13}{Pulkovo Observatory, St.-Petersburg, Russia}
% ...
\begin{abstract}
We have conducted a detailed investigation of the broad-band spectral properties of the \gr ~selected blazars of the \fermi LAT Bright AGN Sample (LBAS).
By combining our accurately estimated \fermi \gr ~spectra with \sw, radio, infra-red, optical and other hard X-ray/\gr ~data, collected within three months of the LBAS data taking period,
we were able to assemble high-quality and quasi-simultaneous Spectral Energy Distributions (SED) for 48 LBAS blazars.
The SED of these \gr ~sources is similar to that of blazars discovered at other wavelengths, clearly showing, in the usual Log $\nu $ -- Log $\nu$ F$_\nu$ representation, the typical
broad-band spectral signatures normally attributed to a combination of low-energy synchrotron radiation followed by inverse Compton emission of one or more components. We have used these SEDs to characterize the peak intensity of both the low and the high-energy components.
The results have been used to derive empirical relationships that estimate the position of the two peaks from the broad-band colors (i.e. the radio to optical, \aro, and optical to X-ray, \aox,  spectral slopes) and from the \gr ~spectral index. Our data show that the synchrotron peak frequency (\nupS) is positioned between 10$^{12.5}$ and  10$^{14.5}$ Hz in broad-lined FSRQs and between $10^{13}$ and  $10^{17}$ Hz in featureless BL Lacertae objects.
We find that the \gr ~spectral slope is strongly correlated with the synchrotron peak energy and with the X-ray spectral index, as expected at first order in
synchrotron - inverse Compton scenarios. However, simple homogeneous, one-zone, Synchrotron Self Compton (SSC) models cannot explain most of our SEDs,
especially in the case of FSRQs and low energy peaked (LBL) BL Lacs. More complex models involving External Compton Radiation or multiple SSC components are required to reproduce the overall SEDs and the observed spectral variability. While more than 50\% of known radio bright high energy peaked (HBL) BL Lacs are detected in the LBAS sample,
only less than 13\% of known bright FSRQs and LBL BL Lacs are included. This suggests that the latter sources, as a class, may be much fainter \gr ~emitters than
LBAS blazars, and could in fact radiate close to the expectations of simple SSC models.
We categorized all our sources according to a new physical classification scheme based on the generally accepted paradigm for Active Galactic Nuclei (AGN) and on
the results of this SED study.  Since the LAT detector is more sensitive to flat spectrum \gr ~sources, the correlation between \nupS and \gr ~spectral index strongly favors the
detection of high energy peaked blazars thus explaining the \fermi overabundance of this type of sources compared to radio and EGRET samples.
This selection effect is similar to that experienced in the soft X-ray band where HBL BL Lacs are the dominant type of blazars.
\\

\end{abstract}


\begin{document}

\section{Introduction}

\label{sec:introduction}

The Large Area Telescope (LAT) an board of the \fermi {\it Gamma Ray Space Telescope}, launched on 11 June 2008, provides unprecedented sensitivity in the $\gamma$-ray band \citep[20 MeV to over 300 GeV,][]{Atwood2009} with a large increase over its predecessors EGRET \citep{Thompson93}, and AGILE, an Italian small  $\gamma$-ray astronomy mission launched in 2007 \citep{tavani08}. The first three months of operations in sky-survey mode led to the compilation of a list of 205 $\gamma$-ray sources with statistical significance larger than 10$\sigma$  \citep{AbdoLATpaper}.
As largely expected from the results of EGRET and AGILE most of the high Galactic latitude sources in this catalog are blazars \citep{AbdoAGNpaper}, a type of AGN well known to display extreme observational properties like large and rapid variability, apparent super-luminal motion, flat or inverted radio spectrum, large and variable polarization. According to a widely accepted scenario blazars are thought to be objects emitting non-thermal radiation across the entire electromagnetic spectrum from a relativistic jet that is viewed closely along the line of sight, thus causing strong relativistic amplification \citep[e.g.,][]{bla78,Urry95}.

Blazars are rare extragalactic objects as they are a subset of radio loud QSOs, which in turn are only $\approx $10\% of radio-quiet QSOs and Seyfert galaxies that are found in large numbers at optical and at X-ray frequencies. Despite that, the strong emission at all wavelengths that characterizes blazars, makes them the dominant type of extragalactic sources in those energy windows where the accretion onto a supermassive black hole, or other thermal mechanisms, do not produce significant radiation. For instance, in the microwave band, Giommi \& Colafrancesco (2004) showed that blazars are the largest population of extragalactic objects (see also Toffolatti \etal ~1998). The same is true in the \gr ~band  \citep{hartman,AbdoAGNpaper} and at TeV energies where BL Lac objects are the most frequent type of sources found in the high Galactic latitude sky \citep[e.g.,][]{costaghis02,colaf06}, see, e.g., the Web-based TeVCat \footnote{http://tevcat.uchicago.edu/} catalog for an up-to-date list of TeV sources and Weekes (2008) for a recent review.

Blazars have been known and studied in different energy windows for over 40 years, however, many questions still remain open about their physics and demographics.


One of the most effective ways of studying the physical properties of blazars is through the use of multi-frequency data. This approach has been followed by a number of authors  \citep[e.g.,][]{gioansmic,vonmontigny95, sambruna96, fossati98, giommisax,nieppola,pad06} who assembled the  Spectral Energy Distributions (SEDs) of many radio, X-ray, and \gr ~selected blazars.
In all cases, however, the effectiveness of the method was limited by the availability of only sparse, often non-simulteneous, flux measurements covering a limited portion of the electromagnetic spectrum. The need to build simultaneous and detailed SEDs is usually addressed through the organization of specific multifrequency observation campaigns. However, so far these large efforts have been carried out almost exclusively on the occasion of large flaring events of a few bright and well known blazars, e.g., 3C454.3 \cite{giommi454.3,Abdo454.3,vercellone09}, Mkn421, \cite{donnarumma09}, PKS2155-304 \cite{Aharonian09}.

With \frmi, \sw, and other high-energy astrophysics satellites simultaneously on orbit, complemented by other space and ground-based observatories, it is now possible to assemble high-quality data to build simultaneous and well sampled SEDs of large and unbiased samples of AGN.

In this paper we study the broad-band (radio to high-energy \gr) properties of the sample of  \fermi bright blazars recently presented by Abdo \etal (2009b)
and we derive the detailed SED of a subsample of 48 \fermi blazars using simultaneous or quasi-simultaneous data obtained from \swift and other ground and space based observatories. For the sake of brevity we will limit ourselves to presenting the data, to estimating some key parameters characterizing the SEDs, and to making some basic conclusions about the physics of blazars. Detailed fits, statistical studies, and more complete theoretical interpretations will be presented elsewhere.
Full analysis of the LAT $\gamma$-ray spectra and \gr ~variability of all the LBAS sources is presented in dedicated papers (Abdo et al. 2009d, Abdo et al. 2009e).

The paper is organized as follows: in Sect. \ref{sec:sample} we present the sample, in Sect. \ref{sec:observations} we describe the \fermi and \swift high energy data along with radio,
near-Infrared, optical and other multi-frequency data.  In Sect. \ref{sec:broadband} we build quasi-simultaneous SEDs for 48 LBAS AGN.
In Sect. \ref{sec:obsparameters} we use our SEDs to derive some key physical parameters such as the peak frequency of the Synchrotron and inverse Compton power (\nupS  and \nupIC) and the corresponding peak fluxes. We also describe an empirical  method that can be used for approximating the synchrotron bump parameters from simple observational quantities like \aox ~and \aro.
We then calculate physical parameters for all sources in the LBAS sample for which \aox ~and \aro ~are available from data in the literature.
In Sect. \ref{sec:classification} we derive a new physical classification of AGN based on our findings and we categorize all our blazars accordingly.
In Sect. \ref{sec:sedsparameters} we discuss some physical implications of our findings.
Finally in Sect. \ref{sec:discussion} we summarize and discuss our results.


\section{The sample}
\label{sec:sample}

The results of the first three months of operations of the \fermi \gr ~observatory, from 4 August to 31 October 2008, are described in Abdo \etal ~(2009a)
 who presented a list of 205 bright ($> 10\sigma$) \gr ~sources. In a companion publication Abdo \etal ~(2009b)
 studied the AGN content of this list associating with high confidence 106  sources at $|b| > 10\degr$ with AGN; 10 further sources were also associated with AGN but with a lower degree of confidence. This sample has been named the ``LAT Bright AGN Sample" or LBAS. The results of the
 Abdo \etal ~(2009b)  paper that are most relevant for this work are:

\begin{itemize}

\item About 90\% of the LBAS sources have been associated to AGN listed in radio catalogs (CRATES/CGRaBS, BZCat), thus implying that the
bright extragalactic $\gamma$-ray sky is confirmed to be dominated by radio-loud AGN (FSRQs, BL Lacs and radio galaxies).

\item The number of high-energy peaked (HBL) BL Lacs detected at GeV energies (even when not flaring) has risen to at least 10 (out of 42 BL Lacs) as compared
to only one (out of 14 BL Lacs) detected by EGRET. Seven LBAS BL Lacs are known TeV blazars.

\item Only about one third of the bright {\it Fermi} AGN were also detected by EGRET. This is a likely consequence of the strong variability and duty cycle of GeV blazars.

\item BL Lac objects make up almost half of the bright {\it Fermi} AGN sample, which consists of 58 FSRQs, 42 BL Lac objects, 2 radio galaxies, and 4 AGN of unknown type; the BL Lac fraction in the 3EG catalog was only $\sim$23\%. This is probably the result of a selection effect induced by the different response of the EGRET and LAT instruments.

\item HBL BL Lacs show significantly harder spectra than FSRQs and low energy peaked (LBL) BL Lacs.

\end{itemize}

Our purpose here is to study in detail the broad-band spectral properties of all blazars in the LBAS sample. The main properties of our sources are reported in
Table \ref{tab:sample}. Column 1 gives the \gr ~source name as it appears in Abdo \etal ~(2009a),
column 2 gives the name(s) of the blazar associated to the \gr ~source, columns 3 and 4 give the precise equatorial coordinates taken from the BZCat catalog \citep{Massaro09} or from NED,  column 5 gives the redshift (when known), columns 6 and 7 give the 5GHz radio flux density and the optical
apparent magnitude, Vmag, from the CRATES \cite{crates} and from the USNO-B1 \citep{usno} catalogs respectively; column 8  gives the 0.1-2.4 keV X-ray flux from the BZcat, or from recent \swift observations processed at the ASI Science Data Center (ASDC), as described in Section \ref{sec:swiftfillins}. All fluxes are as observed, that is not corrected for Galactic absorption. Finally columns 9, 10 and 11 give the broad-band (rest-frame) spectral slopes between 5~GHz and 5000~\AA ~(\aro), 5000~\AA  ~and 1 keV (\aox),   5~GHz and 1 keV (\arx), and 1 keV and 100 MeV (\axg) respectively, with $\alpha_{ab}$ defined as
\begin{equation}
{\alpha_{ab} = - {Log (f_a/f_b) \over Log(\nu_a/\nu_b)}}
\label{alpha_ab}
\end{equation}
where $f_a$ is the rest-frame flux at frequency $\nu_a$ properly de-reddened for Galaxy absorption.  The flux measurements and the redshifts used for the calculation of \aro, \aox, \arx ~and \axg ~are from Table  \ref{tab:sample} of this paper and from Table 3 of Abdo \etal ~(2009b).
For the case of BL Lac objects without known redshift we have assumed z=0.4.

\section{Multi-frequency observations}
\label{sec:observations}
In this Section we describe the multi-frequency observations of LBAS blazars carried out between August and October 2008 with \frmi, and between May 2008 and January 2009
with \swift and other space and ground-based facilities.

\subsection{\fermi LAT data analysis and $\gamma$-ray energy spectra}

The LAT $\gamma$-ray spectra of all the LBAS sources are studied in detail in a dedicated paper (Abdo et al. 2009f) based on 6 months of \fermi data.
Here we derive  the detailed $\gamma$-ray spectra of the 48 blazars for which we build the quasi simultaneous SEDs based on the three months data used to define the LBAS sample.

\label{sec:fermidata}

The Fermi-LAT data from 4 August to 31 October 2008 have been analyzed, selecting for each source only photons belonging to the
diffuse class (Pass6 V3 IRF) \cite{Atwood2009}. Events within a $15^\circ$ Region of Interest (RoI) centered around the source have been selected.
In order to discard photons from the Earth albedo, events with zenith angles larger than $105^\circ$ with respect to the Earth reference frame \cite{AbdoLATpaper} have been
excluded from the data samples.

A maximum likelihood analyis (\textit{gtlike})\footnote{http://fermi.gsfc.nasa.gov/ssc/data/analysis/documentation/Cicerone/Cicerone\_Likelihood} has been used to reconstruct the source energy spectrum.
A  model is assumed for the source spectrum as well as for the diffuse background components, depending on a set of free parameters. The Galactic diffuse emission is modeled using GALPROP package
while the extragalactic  one is described by a simple power law (Abdo et al. 2009a).  The method has been implemented to estimate the parameters in each individual energy bin (2 bins per decade, starting  from 100 MeV), and the parameters obtained from the fits are used to  evaluate the sources fluxes. For each energy bin the source under investigation and all nearby sources in the RoI are described  by one parameter representing the integral flux in that energy  bin. The diffuse background components are modeled with one single parameter each, describing the normalization. For each bin, only fit results with a significance larger than $3\sigma$ have been retained. Depending on the flux and energy spectrum, 4 to 7 bins had positive detections for each AGN in the sample.

As a cross check a deconvolution technique \citep[unfolding,][]{Unfolding} has been used to reconstruct the source energy spectra from the observed data, after background subtraction.
This method allows us to reconstruct the source spectrum from the data without assuming any spectral model, taking also into account  the finite energy dispersion of the detector.
The results of the two different methods are consistent as illustrated in the Appendix \ref{app}.

Once the differential flux in each energy bin $\phi(E)$ has been evaluated, the corresponding SED is then obtained by multiplying the differential flux by the square
of the central energy value of that bin, i.e., $\nu F(\nu) = E^2 \phi(E)$ where $E=h \nu$. The vertical error bars
represent only the statistical errors. The systematic uncertainties in the effective area for the Pass6 V3 DIFFUSE event selection have been estimated to be $10\%$ at 100 MeV, $5\%$ at 562 MeV and $20\%$ for energies
greater than 10 GeV \cite{lsi61}.

\subsection{\swift data}

\label{sec:swiftdata}

The \swift  Gamma-Ray-Burst (GRB) Explorer \citep{Gehrels04} is a multi-frequency, rapid response space observatory that was launched on 20 November 2004. To fulfill its purposes \swift carries three instruments on board: the Burst Alert Telescope \citep[BAT,][]{Barthelmy05} sensitive in the 15-150 keV band, the X-Ray Telescope \citep[XRT,][]{Burrows05} sensitive in the 0.3-10.0 keV band, and the UV and Optical Telescope (170-600 nm) \citep[UVOT,][]{Roming05}.
The very wide spectral range covered by these three instruments is of crucial importance for blazar issues as it covers where the transition between the
synchrotron and inverse Compton emission usually occurs.

The primary objective of the \swift scientific program is the discovery and rapid follow up of GRBs. However, as these elusive sources explode at random times and their
frequency of occurrence is subject to large statistical fluctuations, there are
periods when \swift  is not engaged with GRB observations and the observatory
can be used for different scientific purposes. The sources observed through this secondary science program are usually called \swift fill-in targets.
Since the beginning of its activities \swift has observed hundreds of blazars as part of the fill-in program \citep[e.g.,][] {GiommiWMAP07}.
With the launch of AGILE and  \frmi, the rate of \swift  blazar observations increased significantly, leading to the observation (and detection) of all but  6 blazars in the LBAS sample.

The \swift database currently includes 119 observations of 48 LBAS blazars that were carried out either simultaneously or within three months of the \fermi LBAS data taking period. We used the UVOT, XRT and BAT data of these observations to build our SEDs. Some blazars were observed several times in the period that we consider in this paper; in such cases we considered only the exposures where the
source was detected at minimum and maximum intensity by the XRT instrument.

\subsubsection{UVOT data analysis}

\swift observations are normally carried out so that UVOT produces a series of
images in each of the lenticular filters (V, B, U, UVW1, UVM2, and UVW2). The
photometry analysis of all our sources was performed using the standard UVOT
software distributed within the HEAsoft 6.3.2 package and the calibration
included in the latest release of the ``Calibration Database''. Counts were
extracted from an aperture of 5\arcsec\, radius for all filters and converted to
fluxes using the standard zero points \citep{poole08}. The fluxes were then de-reddened using the appropriate values of
$E(B-V)$ for each source taken from Schlegel \etal (1998) with $A_{\lambda}/E(B-V)$
ratios calculated for UVOT filters using the mean interstellar extinction curve
from Fitzpatrick (1999). No variability was detected within single
exposures in any filter.

The results of our analysis are summarized in Table \ref{tab:uvot_analysis} where Column 1 gives the source
name, Column 2 gives the observation date and the other Columns report the magnitudes in the five UVOT filters with the own errors.

\subsubsection{XRT data analysis}

The XRT is usually operated in Auto State mode which automatically adjusts the
readout mode of the CCD detector to the source brightness, in an attempt to
avoid  pile-up \citep[see][for details of the XRT observing modes]{Burrows05,Hill04}.
Given the low count rate of our blazars most of the data were
collected using the most sensitive Photon Counting (PC) mode while Windowed Timing
(WT) mode was used for bright sources with shorter exposures.

The XRT data were processed with the XRTDAS software package (v.~2.4.1) developed
at the ASI Science Data Center (ASDC) and distributed by the NASA High Energy
Astrophysics Archive Research Center (HEASARC) within the HEASoft package (v.~6.6.1).
Event files were calibrated and cleaned with standard filtering criteria with the
{\it xrtpipeline} task using the latest calibration files available in the \swift
CALDB. Events in the energy range 0.3--10 keV with grades 0--12
(PC mode) and 0--2 (WT mode) were used for the analysis.

Events for the spectral analysis were selected within a circle of 20 pixel
($\sim$47 arcsec) radius, which encloses about 90\% of the PSF at 1.5
keV \cite{Moretti05}, centered on the source position.
For PC mode data, when the source count rate is above $\sim$ 0.5 counts s$^{-1}$ data are
significantly affected by pile-up in the inner part of the Point Spread Function (PSF).
For such cases, after comparing the observed PSF profile with the analytical model derived by
Moretti \etal ~(2005), we removed pile-up effects by excluding events detected within up to 6 pixels from
the source position, and used an outer radius of 30 pixels. The value of the inner radius was evaluated individually
for each observation affected by pile-up, depending on the observed count-rate.

Ancillary response files were generated with the {\it xrtmkarf} task
applying corrections for the PSF losses and CCD defects. Source spectra were binned
to ensure a minimum of 20 counts per bin to utilize the $\chi^{2}$ minimization
fitting technique.

We fitted the spectra adopting an absorbed power law model with photon index $\Gamma_x$. When deviations
from a single power law model were found, we adopted a log-parabolic law of the form
$F(E)=K E^{(-a+b\cdot Log(E))}$ \cite{Massaro04} which has been shown to fit well the X-ray spectrum of
blazars \cite[e.g.,][]{giommi05,Tramacere09}. This spectral model is described by only two parameters: $a$, the photon index at 1 keV, and $b$,
the curvature of the parabola. For both models the amount of hydrogen-equivalent column density (N$_H$) was fixed
to the Galactic value along the line of sight \citep{Kalberla05}.

The results of the spectral fits are shown in Table \ref{tab:xrt_analysis} where Column 1 gives the source
name, Column 2 gives the observation date, Column 3 gives the net XRT exposure time, Column 4 gives the
2--10 keV X-ray flux, Column 5 gives the best fit photon index $\Gamma_x $ or the Log parabola parameter $a$ when a simple power law model was
not a good representation of the data, Column 6 gives the best fit curvature parameter $b$, Column 7 gives the number of degrees of freedom and Column 8 gives the value  of the reduced $\chi^{2}$.

\subsubsection{BAT hard X-ray data analysis}

We used survey data from the Burst Alert Telescope (BAT) on board \swift  to produce 15-200\,keV spectra of the blazars presented in
this analysis. In order to do so, we used three years of survey data  \citep[see][for details]{ajello09} and extracted the spectra of those blazars
that are significantly detected in the 15--55\,keV band. Because of the very long integration time these data are not simultaneous with our \fermi data.

Only 15 blazars, among those presented here, were detected by BAT at a significance $\geq 4$\,$\sigma$. The spectral extraction is performed as described in Ajello \etal ~(2008)
and the background-subtracted spectra represent the average emissions of the sources within the time spanned by the BAT survey.

\subsubsection{\swift observations of LBAS blazars carried out before May 2008 or after January 2009}
\label{sec:swiftfillins}
The \swift database includes a number of observations of LBAS blazars that were carried out outside the period that we consider useful to build our quasi-simultaneous SEDs.
These measurements are particularly important for the case of blazars that have never previously been observed by any X-ray astronomy satellite and were below the detection threshold of
the ROSAT all sky survey.  When these \swift observations have been analyzed and published by other authors we use the flux intensities reported in the literature, with
particular reference to the latest on-line version of the  BZcat catalog \footnote{http://www.asdc.asi.it/bzcat}.  For the cases where the \swift results have not yet appeared in the literature we list in column 8
of Table \ref{tab:sample} the X-ray fluxes estimated from the standard pipeline processing that is run at ASDC on all \swift XRT data shortly after they are added to the archive.
This ASDC processing makes use of the 'xrtpipeline'  task of the XRTDAS package that is run after applying very tight data screening criteria, e.g., a CCD temperature lower
than $-50 ^{\circ}C$ (instead of the standard limit of $-47^{\circ} C$) thus ensuring a very effective background reduction.
The calibrated and cleaned Photon Counting mode event files produced are then analyzed
with the XIMAGE package v.4.4.1 and the point sources present in each XRT field are searched using the XIMAGE detection algorithm.
For each source the net counts are corrected to account for CCD defects, effective exposure and vignetting using the exposure maps
and a PSF correction.
The count rates are finally converted into fluxes in the 0.1-2.4 keV band assuming a power law spectral model with energy slope of 0.9 and low-energy absorption due to
Galactic $N_H$.

\subsection{Other multi-frequency data}
\label{sec:multifrequency}

In order to improve the quality of our SEDs we complemented the \fermi and \swift quasi-simultaneous data with other multi-frequency flux measurements obtained
from a number of on-going programs from ground and space-based observatories. In the following Sections we describe each program and the corresponding data analysis.

\subsubsection{Effelsberg radio observations}

Quasi-simultaneous radio data for 25 sources of the first \fermi bright source
catalog were obtained within a \fermi-related monthly broad-band monitoring
program including the Effelsberg 100-m radio telescope of the MPIfR
\citep[F-GAMMA project][]{Fuhrmann07,Angelakis08}. From this program, radio
spectra covering the frequency range 2.6 to 42\,GHz were selected to be within the
time period 4 August 2008 to 31 October 31 2008, i.e., quasi simultaneous to the \fermi and \swift
observations presented in Sections \ref{sec:fermidata} and \ref{sec:swiftdata}.

The Effelsberg observations were conducted with cross-scans in
azimuth/elevation with the number of sub-scans matching the source brightness
at the given frequencies. The individual spectra were measured
quasi-simultaneously within $\le$\,40 minutes rapidly switching between the
various secondary focus receivers. The data reduction was done applying
standard procedures and post-observational corrections including (i) opacity
correction, (ii) pointing off-set correction, (iii) gain correction, and (iv)
sensitivity correction \citep[see][for details]{Fuhrmann08, Angelakis08}. The
sensitivity correction was done with reference to standard calibrators (e.g.,
3C\,286) and the measured antenna temperatures were linked to the absolute
flux-density scale \citep{Baars77,Ott94}. The precision ranges between $\lsim$1\% to a few percent.

The results are reported in Table \ref{tab:effelsberg} where Column 1 gives the source name, column 2 gives the observation date,
column 3 gives the frequency  and column 4 gives the flux density in units of Jansky.

\subsubsection{OVRO radio data}

Quasi-simultaneous 15~GHz observations of 24 \emph{Fermi} LBAS sources
were made using the Owens Valley Radio Observatory (OVRO) 40~m
telescope.  These observations were made as part of an ongoing
\fermi-LAT blazar monitoring program.  In this program, all 1158
CGRaBS blazars north of declination $-20\degr$ have been observed
approximately twice per week or more frequently since June~2007
\citep{Healey08}.

The OVRO flux densities are measured in a single 3~GHz wide band
centered at 15~GHz.  Observations were performed using azimuth double
switching as described in \cite{Readhead89}, which removes much
atmospheric and ground interference.  The relative uncertainties in
flux density result from a 5~mJy typical thermal uncertainty in
quadrature with a 1.6\% systematic uncertainty.  The absolute flux
density scale is calibrated to about 5\% via observations of the
steady calibrator 3C~286, using the  \cite{Baars77} model.

For each source, the maximum and minimum observed 15~GHz flux
densities during the 4~August to 31~October, 2008 period were included
in the quasi-simultaneous SEDs. The included OVRO 40~m observations
are summarized in Table~\ref{tab:OVRO}. Column 1 lists the 0FGL
source name. Columns 2 and 4 list the dates of the observed maximum
and minimum. Columns 3 and 5 list the measured maximum and flux
density in Jansky, including the 5\% absolute calibration uncertainty
in the quoted error.

\subsubsection{RATAN-600 1-22 GHz radio observations}

Among the 48 objects for which we present \swift and \fermi simultaneous SEDs 32 were observed
between September~10 and October 3, 2008 with the 600-meter ring radio telescope RATAN-600
\citep{RATANreview} of the Special Astrophysical Observatory, Russian
Academy of Sciences, located in Zelenchukskaya, Russia.
These observations, which produced  1--22~GHz instantaneous radio spectra, are part of a long-term program \citep[e.g.,][]{Kov02} to
monitor continuum spectra of active galactic nuclei with a strong
parsec-scale component of radio emission. The current list contains a
complete sample of more than 600 AGN with declination $\delta>-30^\circ$
and correlated VLBI flux density greater than 400~mJy selected from Kovalev \etal  (2007).

Broad-band radio continuum spectra were measured quasi-simultaneously
in a transit mode at six different bands with
the following central frequencies (and frequency bands):
0.95~GHz (0.03~GHz),
2.3~GHz  (0.25~GHz),
4.8~GHz  (0.6~GHz),
7.7~GHz  (1.0~GHz),
11.2~GHz (1.4~GHz),
21.7~GHz (2.5~GHz).
Each source was observed in the upper culmination with an unmoved antenna due to the Earth rotation collecting a multi-frequency source scan within several minutes.
Details on the method of observation, data processing, and amplitude calibration are described in Kovalev \etal (1999).
Presented data were
collected using the Southern ring sector with the Flat reflector of RATAN-600. The spectrum of every object was measured, typically, three
times during the observing set. Averaged flux density spectra used in our SEDs
are presented in Table \ref{tab:RATAN} where column 1 gives the source name, column 2 gives the frequency of observations and column 3 gives the radio flux density in units of Jansky.
During recent years, the radio frequency interference (RFI) became stronger at the two lowest frequency bands, 1 and 2.3 GHz. This results in higher measurement errors and
sometimes even loss of data, especially at the lowest frequency band. Bad weather conditions resulted in elevated errors at 22 GHz in a few cases.

\subsubsection{Radio, mm, NIR and optical data from the GASP-WEBT collaboration}

The GLAST-AGILE Support Program (GASP) originated from the Whole Earth Blazar Telescope (WEBT) \footnote{http://www/oato.inaf.it/blazars/webt/}   \citep[see e.g.,][]{villata07,raiteri08a}) and started its operation in September 2007 \citep[see][]{villata08}, with the aim of performing long-term optical-to-radio monitoring of 28 $\gamma$-loud blazars, to compare the low-energy flux behaviour with the behaviour observed at $\gamma$-ray energies.

In the period considered in this work, the GASP carried out $\sim 3000$ optical ($R$ band) observations of 19 LBAS blazars, while $\sim 700$ near-IR ($JHK$, Campo Imperatore), and $\sim 600$ microwave (230 and 345 GHz, SMA) and radio data (5 to 43 GHz, Medicina, Noto, UMRAO) observations were taken on the same sources.

The optical and near-IR magnitudes were de-reddened by assuming the Galactic extinction in the $B$ band from
Schlegel et al. (1998) and deriving the extinction in the other bands according to \cite{cardelli89}. The conversion to fluxes was performed adopting the zero-mag fluxes by Bessel (1998).

In the SED plots we report the average, maximum and minimum values at each observed frequency in the period 4 August -- 31 October, 2008\footnote{Average flux densities were calculated on the 1-day binned data-sets, to avoid giving too much weight to the days with denser sampling.}.

Table \ref{tab:webt} reports the plotted values directly as Log($\nu F_\nu$): the average, maximum, and minimum values are shown in Columns 3, 4, and 5, respectively. Column 6 displays the number of data available in the period. When the number of data available is reported as "0" this indicates that the data given in the Table and shown in the SED plots are not strictly inside the period (this happens for ON 231 and 3C 279 in the optical, and for 3C 273 in both the optical and near-IR, because of solar conjunction), but come from immediately outside and, due to the smoothness of the light curve, they can represent the state in between.
Note that Columns 4 and 5 report the error bar extremes instead of maximum and minimum values.

The optical data of Mkn 421 have been cleaned for the contribution of the host galaxy, according to \cite{Nilsson07}. As for PKS 0235+164, we corrected the fluxes for both the photometric contribution from the southern AGN and the additional extinction due to the intervening DLA system, according to \cite{raiteri05}; see also \citep{raiteri08b}.

\subsubsection{Mid-infrared VISIR observations}

The MIR observations were carried out from 2006 to 2008 using VISIR
\citep{2004Lagage}, the ESO/VLT mid-infrared imager and spectrograph,
composed of an imager and a long-slit spectrometer covering several
filters in N and Q bands and mounted on Unit 3 of the VLT
(Melipal). The standard ''chopping and nodding" MIR observational
technique was used to suppress the background dominating at these
wavelengths. Secondary mirror-chopping was performed in the
north-south direction with an amplitude of 16$\arcsec$ at a frequency
of 0.25\,Hz. Nodding technique, needed to compensate for chopping
residuals, was chosen as parallel to the chopping and applied using
telescope offsets of 16$\arcsec$. Because of the high thermal MIR
background for ground-based observations, the detector integration
time was set to 16\,ms.

We performed broad-band photometry in 3 filters, PAH1
($\lambda$=8.59$\pm$0.42 $\mu$m), PAH2 ($\lambda$=11.25$\pm$0.59
$\mu$m), and Q2 ($\lambda$=18.72$\pm$0.88 $\mu$m) using the small
field in all bands (19\farcs2x19\farcs2 and 0\farcs075 plate
scale). All the observations were bracketed with standard star
observations for flux calibration and PSF determination. The weather
conditions were good and stable during the observations.

Raw data were reduced using the IDL reduction package (E.
Pantin 2009, in prep). The elementary images were co-added in
real-time to obtain chopping-corrected data, then the different
nodding positions were combined to form the final image. The VISIR
detector is affected by stripes randomly triggered by some abnormal
high-gain pixels.  A dedicated destriping method was developed to
suppress them. The MIR fluxes and observation dates of all observed sources including the
1$\sigma$ errors are listed in Table \ref{table:MIR}.

\subsubsection{Non-simultaneous Spitzer Space Telescope observations}

The Spitzer Space Telescope is a 0.85-meter class telescope launched on 25 August 2003.
Spitzer obtains images and spectra in the spectral range between 3 and 180 micron through three instruments on board: the InfraRed Array Camera (IRAC), which provides images at 3.6, 4.5, 5.8 and 8.0 microns, the Multiband Imaging Photometer for Spitzer (MIPS), which performes imaging photometry at 24, 70 and 160 micron, and the InfraRed Spectrograph (IRS) which provides spectra over 5-38 microns in low (R $\sim$ 60-127) and high (R $\sim $ 600) spectral resolution mode.
The Spitzer Science Archive include MIPS observations of 8 sources belonging to the LBAS sample, all of them performed earlier than three months from the start of LBAS
data taking period. From the Spitzer Archive we retrieved the post-BCD (post Basic Calibrated Data), that is products generated after calibration of the individual BCD exposures.
The DAOPHOT package was used for the photometric analysis, which was carried out on the post-BCD using the method of aperture photometry and subtracting the background emission.
The results are reported in Table \ref{tab:spitzer} where Column 1 gives the source name, Column 2 gives the observation date, Column 3 gives the log of frequency Log($\nu$) and
Column 4 gives the log of $\nu$F$_{\nu}$.

\subsubsection{AGILE $\gamma$-ray data}

The AGILE satellite, launched in April 2007, is an Italian Space
Agency (ASI) mission devoted to $\gamma$-ray astrophysics in the 30 MeV--50 GeV
energy range, with simultaneous X-ray imaging in the 18--60 keV band.
The AGILE Instrument (Tavani et al. 2008, 2009) consists of
the Silicon Tracker (ST), the X-ray detector SuperAGILE (SA), the
CsI(Tl) Mini-Calorimeter (MCAL) and an anti-coincidence system (ACS).
The combination of ST, MCAL and ACS forms the Gamma-Ray Imaging
Detector (GRID).

The $\gamma$-ray data collected by the GRID for energies greater than 100 MeV used in this paper
(blue star symbols in the SED figures) are extracted from the First AGILE Catalog of high-confidence \gr ~sources detected by
AGILE during the first 12 months of operations,
from 9 July 2007 to 30 June 2008 \citep{Pittori09}.
The First AGILE Catalog includes only high-significance sources
characterized by a prominent mean \gr ~flux above 100 MeV when
integrated over the total one-year exposure period.

Flare detections and determination of source peak fluxes through dedicated
investigation over shorter timescales are not included in the First AGILE Catalog.
However, it should be noted that for some blazars, such as Mkn 421 (1AGL J1104+3754)
ON 231/W Comae (1AGL J1222+2851),  PKS 1510-089 (1AGL J1511-0908),
and 3C279 (1AGL J1256-0549), the effective AGILE exposure over the entire time period was quite low, only
a few effective days, but it  included Target of Opportunity or previously planned observations during a flaring state of the source.
In such cases the AGILE observed mean $\gamma$-ray flux may be close
to the source peak flux values.

The differential AGILE flux values appearing in the SED figures at fixed energy point (E=300 MeV) have been rescaled from the mean \gr ~flux
above 100 MeV, obtained with a simple power law source model with fixed spectral index $-2.1$.

Table \ref{tab:agilecomp} reports the results where Column 1 gives the source name and Column 2 gives the AGILE observed flux and Column 3 gives the mean exposure.

\section{Quasi-simultaneous radio to $\gamma$-ray SEDs of 48 LBAS blazars}
\label{sec:broadband}

In this Section we use the multi-frequency data described above to build quasi-simultaneous SEDs of 48 objects, in the usual Log $\nu $--Log $\nu$ F$_\nu$ representation. These
48 sources are a sizable subset ($\approx 45\%$) of LBAS that is representative of the entire sample since they were chosen only on the basis of  the availability of \swift observations carried out between May 2008 and January 2009 (which have been scheduled largely independently of \fermi results) and not on brightness level or on any other condition that could influence the shape of the SED. We checked this by verifying that the distributions of redshift, optical, X-ray and \gr ~fluxes are all consistent with being the same in the two subsamples.

We stress that there is one important difference between \gr ~and other multi-frequency data:  our  \fermi data were collected over a period of three months while all other data were collected over much shorter periods (typically less than a few hours) and are not necessarily simultaneous among themselves. This is clearly a limitation as flux and spectral variability in blazars often takes place on short timescales. Such a behavior is clearly visible, in fact in our multi-frequency data when more than one \swift observation is available (see e.g., Figs. \ref{fig:sed_56}, \ref{fig:sed_78}, \ref{fig:sed_1112}, etc.). Since our \gr ~data have been accumulated over the relatively long period of three months, they likely represent the
average of different intensity states.

\begin{figure}
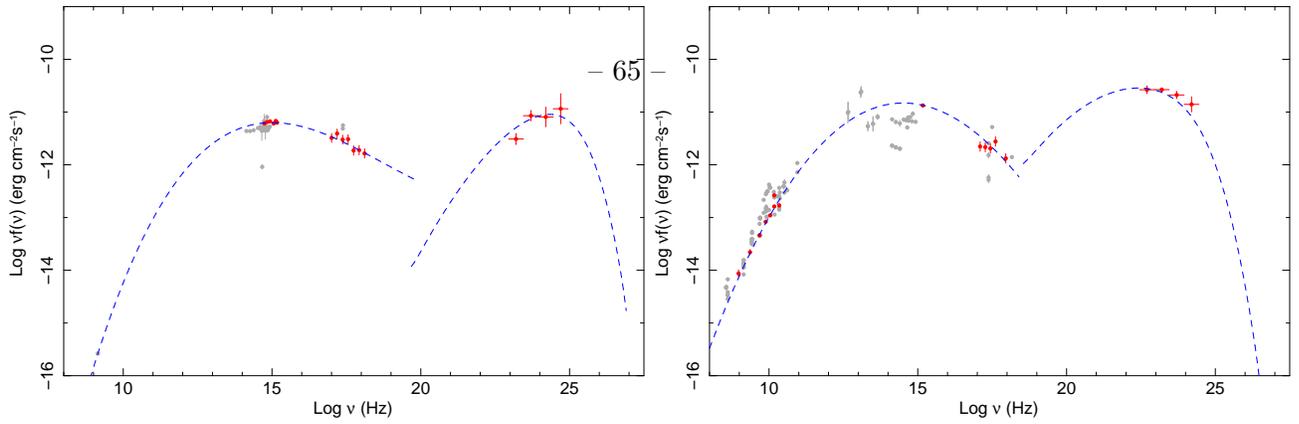

\epsscale{.80}
\vspace{-2cm}
\includegraphics[height=8.5cm,angle=-90]{BZBJ0033-1921_ASO0010.ps}
\includegraphics[height=8.5cm,angle=-90]{PKS0048-09_ASO0018.ps}
\caption{The SED of 0FGL J0033.6-1921 =  1RXS J003334.6-192130 = SHBL J003334.2-192133 (left) and
of 0FGL J0050.5-0928 =  PKS0048-09 (right). The quasi-simultaneous data appear as large filled red symbols, while non-simultaneous archival measurements
are shown as small open grey points. The dashed lines represent the best fits to the Synchrotron and Inverse Compton part of the quasi-simultaneous SEDs (see text for detail). }
\label{fig:sed_first}
\end{figure}

\begin{figure}
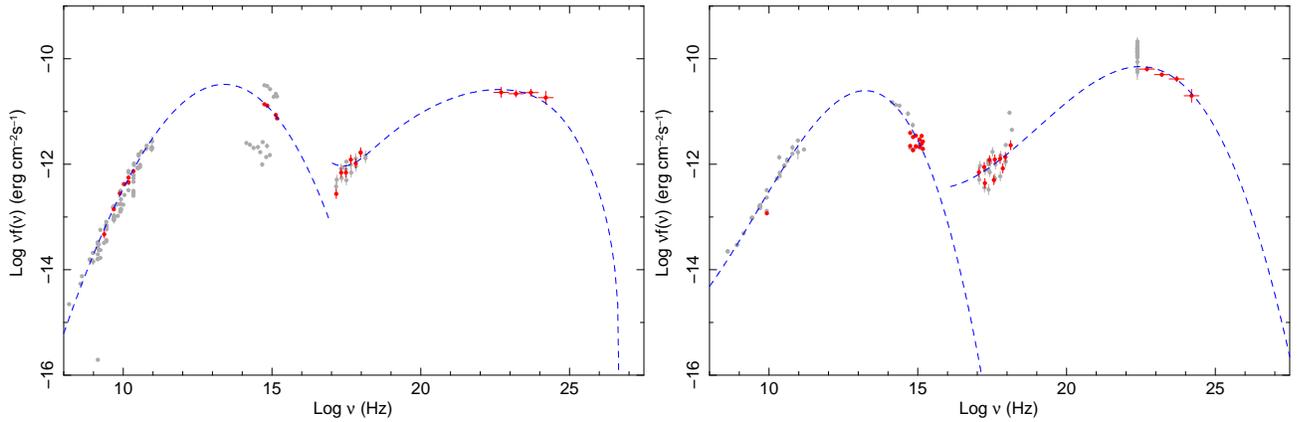

\epsscale{.80}
\includegraphics[height=8.5cm,angle=-90]{S40133+47_ASO0035.ps}
\includegraphics[height=8.5cm,angle=-90]{PKS0208-512_ASO0043.ps}
\caption{The SED of 0FGL J0137.1+4751 = S40133+47 (left) and
of 0FGL J0210.8-5100 = PKS0208-512 (right)}
 \label{fig:sed_34}
\end{figure}

\begin{figure}
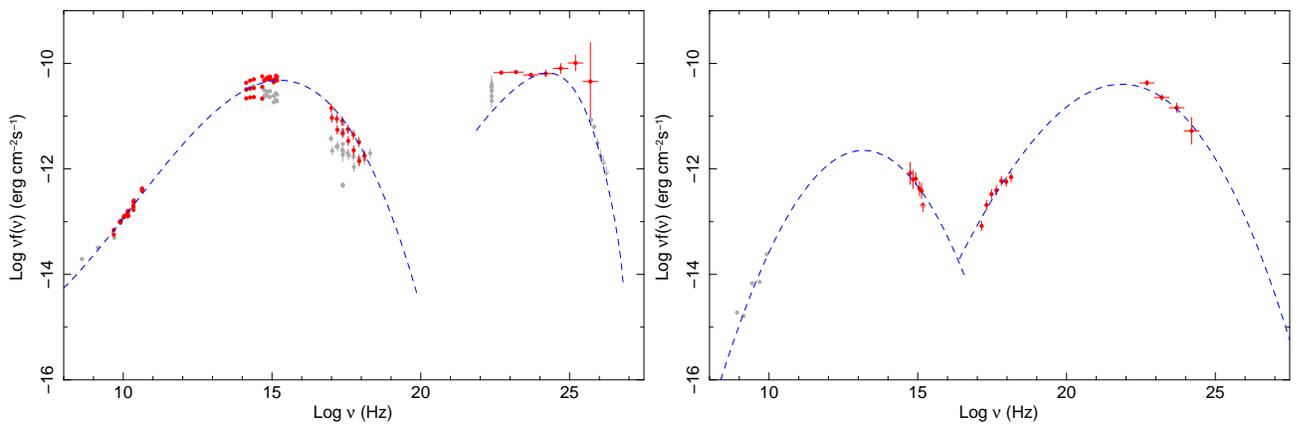

\epsscale{.80}
\includegraphics[height=8.5cm,angle=-90]{3C66A_ASO0048.ps}
\includegraphics[height=8.5cm,angle=-90]{PKS0227-369_ASO0051.ps}
\caption{The SED of 0FGL J0222.6+4302 = 3C 66A (left)
and of 0FGL J0229.5-3640 =  PKS0227-369 (right)}
 \label{fig:sed_56}
\end{figure}

\begin{figure}
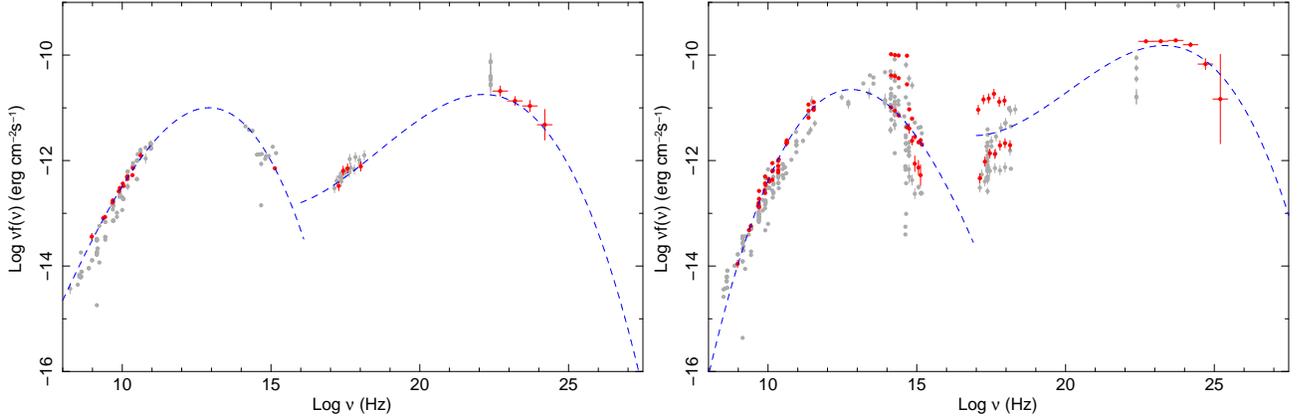

\epsscale{.80}
\includegraphics[height=8.5cm,angle=-90]{4C28.07_ASO0052.ps}
\includegraphics[height=8.5cm,angle=-90]{PKS0235+164_ASO0053.ps}
\caption{The SED of 0FGL J0238.4+2855 =  4C28.07  (left) and
of 0FGL J0238.6+1636 = PKS0235+164 (right).}
 \label{fig:sed_78}
\end{figure}

\begin{figure}
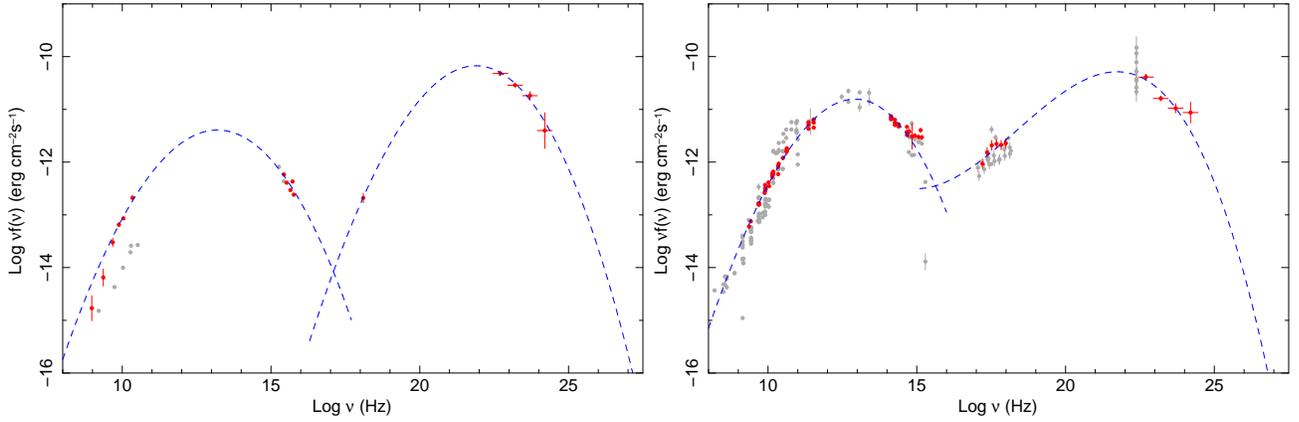

\epsscale{.80}
\includegraphics[height=8.5cm,angle=-90]{PKS0347-211_ASO0077.ps}
\includegraphics[height=8.5cm,angle=-90]{PKS0420-01_ASO0083.ps}
\caption{The SED of 0FGL J0349.8-2102 = PKS 0347-211 (left) and
of 0FGL J0423.1-0112 = PKS0420-01 (right)}
 \label{fig:sed_910}
\end{figure}

\begin{figure}
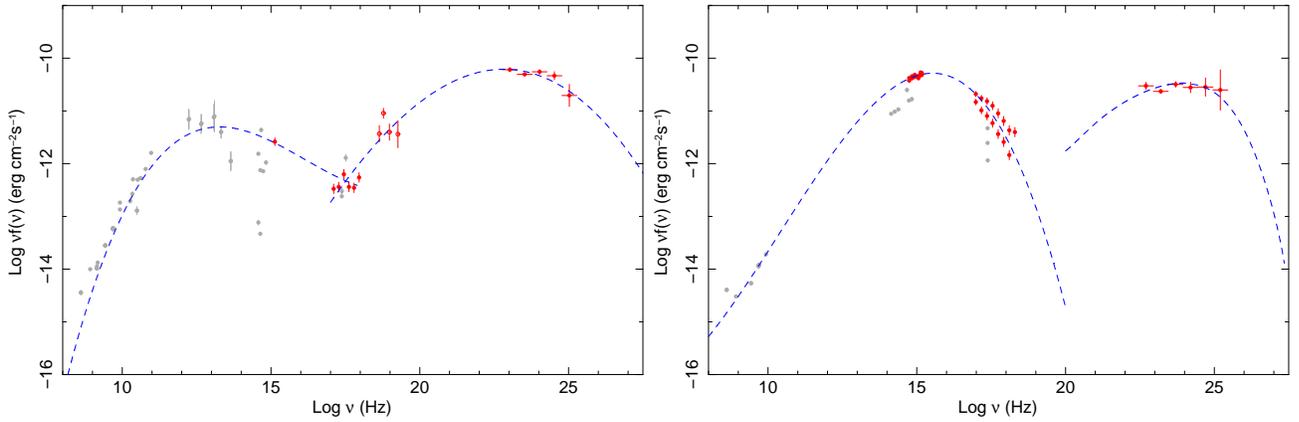

\epsscale{.80}
\includegraphics[height=8.5cm,angle=-90]{PKS0426-380_ASO0089.ps}
\includegraphics[height=8.5cm,angle=-90]{PKS0447-439_ASO0100.ps}
\caption{The SED of 0FGL J0428.7-3755 = PKS0426-380 (left) and
 of 0FGL J0449.7-4348 =  PKS0447-439 (right)}
\label{fig:sed_1112}
\end{figure}

\begin{figure}
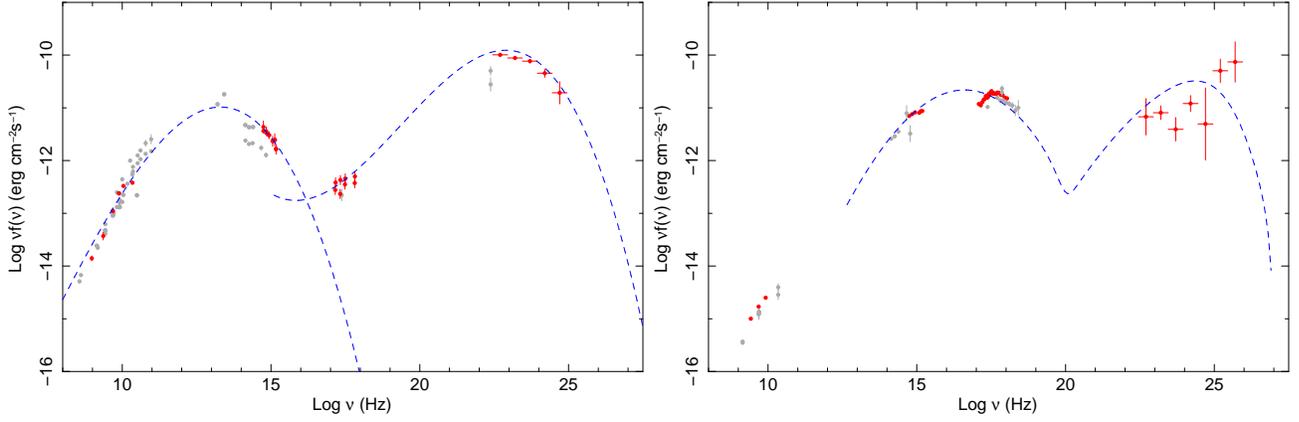

\epsscale{.80}
\includegraphics[height=8.5cm,angle=-90]{PKS0454-234_ASO0103.ps}
\includegraphics[height=8.5cm,angle=-90]{1ES0502+675_ASO0108.ps}
\caption{The SED of 0FGL J0457.1-2325 = PKS0454-234 (left) and
of 0FGL J0507.9+6739 = 1ES 0502+675 (right)}
\label{fig:sed_aso0103}
\end{figure}

\begin{figure}
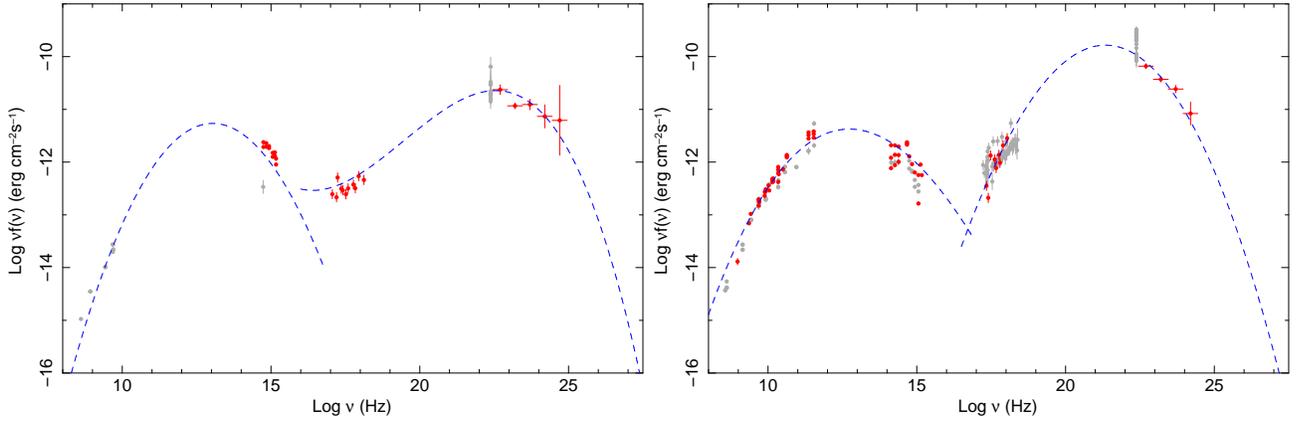

\epsscale{.80}
\includegraphics[height=8.5cm,angle=-90]{MC4_0516-6207_ASO0112.ps}
\includegraphics[height=8.5cm,angle=-90]{PKS0528+134_ASO0119.ps}
\caption{The SED of 0FGL J0516.2-6200 = MC4 0516-621 (left) and
of 0FGL J0531.0+1331 = PKS 0528+134 (right)}
\label{fig:sed_aso0112}
\end{figure}

\begin{figure}
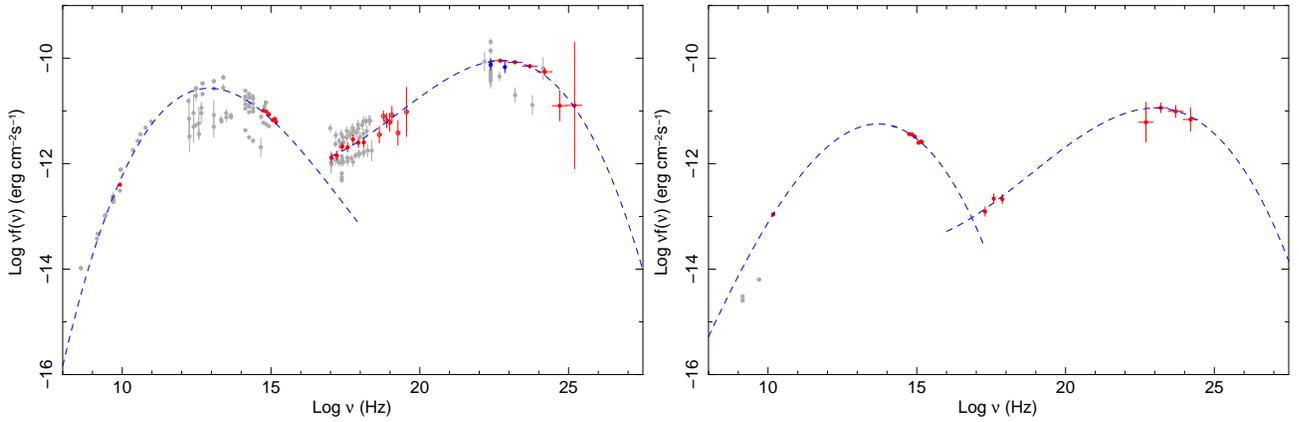

\epsscale{.80}
\includegraphics[height=8.5cm,angle=-90]{PKS0537-441_ASO0126.ps}
\includegraphics[height=8.5cm,angle=-90]{GB6J0712+5033_ASO0159.ps}
\caption{The SED of  0FGL J0538.8-4403 = PKS0537-441. (left) and
of 0FGL J0712.9+5034 = GB6 J0712+5033 (right)}
\label{fig:sed_aso0126}
\end{figure}

\begin{figure}
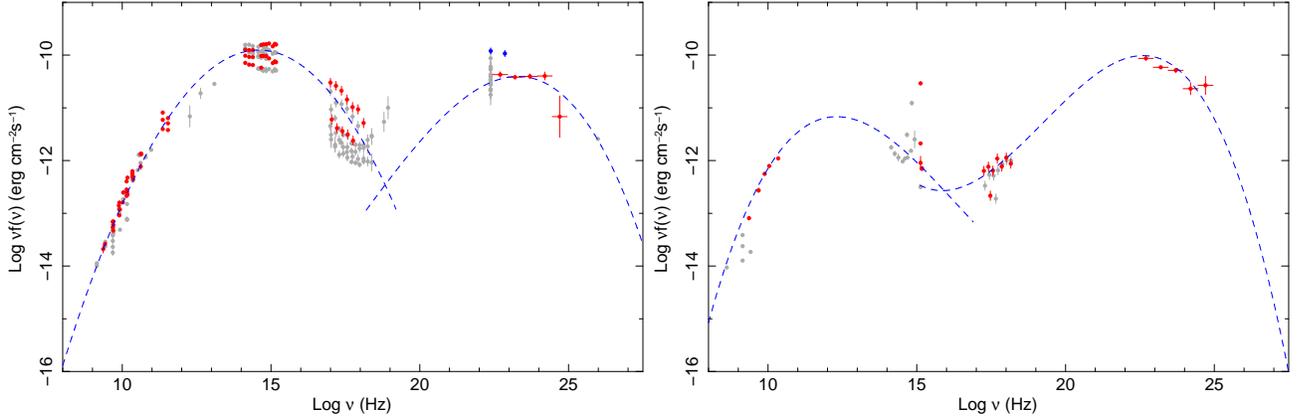

\epsscale{.80}
\includegraphics[height=8.5cm,angle=-90]{S50716+714_ASO0163.ps}
\includegraphics[height=8.5cm,angle=-90]{PKS0727-11_ASO0164.ps}
\caption{The SED of 0FGL J0722.0+7120 =  S50716+714 (left) and
of 0FGL J0730.4-1142 = PKS0727-11 (right)}
\label{fig:sed_aso0163}
\end{figure}

\begin{figure}
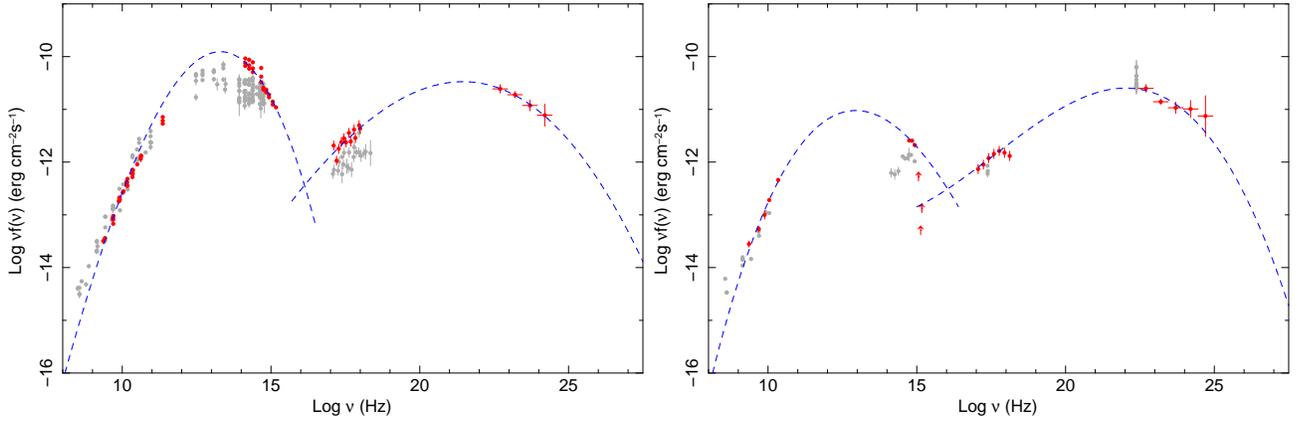

\epsscale{.80}
\includegraphics[height=8.5cm,angle=-90]{PKS0851+202_ASO0191.ps}
\includegraphics[height=8.5cm,angle=-90]{S40917+44_ASO0201.ps}
\caption{The SED of     0FGL J0855.4+2009 = PKS0851+202 (left) and
of 0FGL J0921.2+4437 =  S40917+44 (right)}
\label{fig:sed_aso0191}
\end{figure}

\begin{figure}
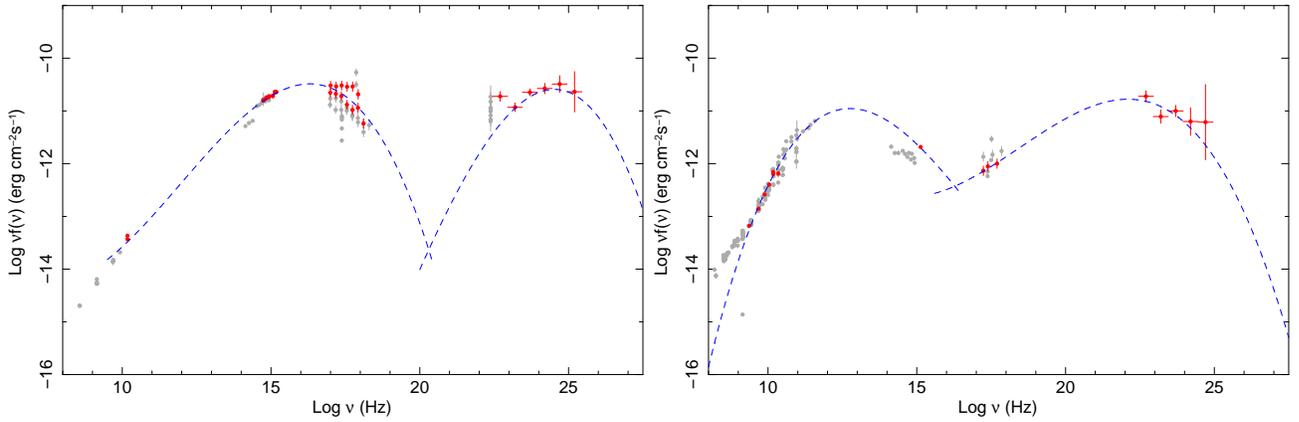

\epsscale{.80}
\includegraphics[height=8.5cm,angle=-90]{1H1013+498_ASO0213.ps}
\includegraphics[height=8.5cm,angle=-90]{4C01.28_ASO0230.ps}
\caption{The SED of 0FGL J1015.2+4927 = 1H 1013+498 (left) and
of 0FGL J1057.8+0138 = 4C01.28 (right)}
\label{fig:sed_aso0213}
\end{figure}

\begin{figure}
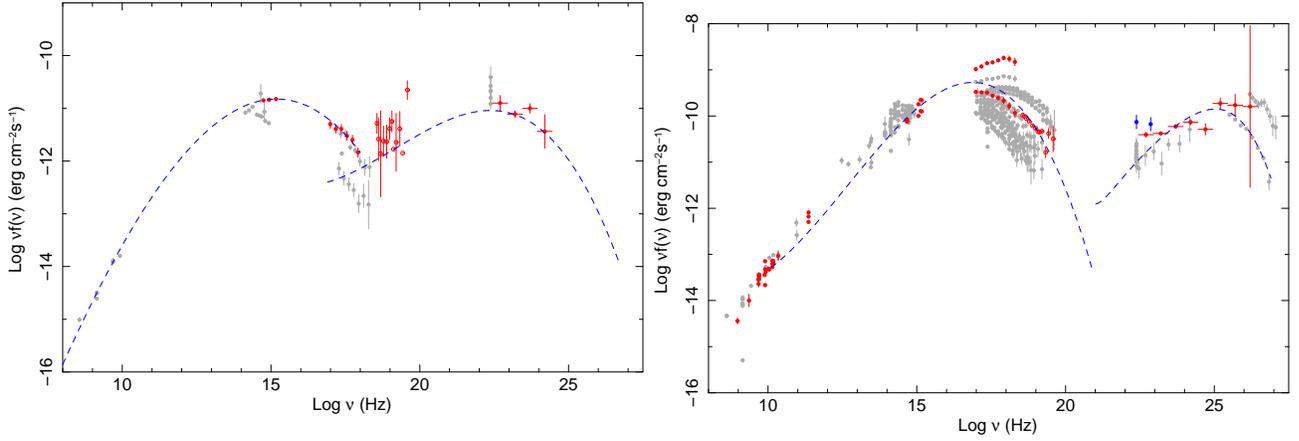

\epsscale{.80}
\includegraphics[height=8.5cm,angle=-90]{GB6J1058+5628_ASO0232.ps}
\includegraphics[height=8.5cm,angle=-90]{Mkn421_ASO0236.ps}
\caption{The SED of 0FGL J1058.9+5629 = GB6 J1058+5628 (left)
and of 0FGL J1104.5+3811 = Mkn 421 (right)}
\label{fig:sed_aso0232}
\end{figure}

\begin{figure}
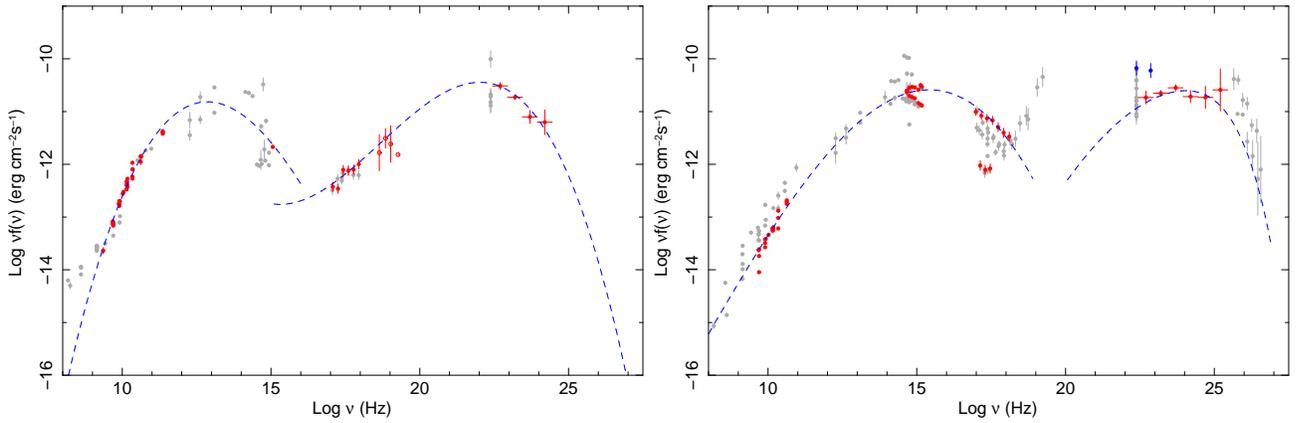

\epsscale{.80}
\includegraphics[height=8.5cm,angle=-90]{4C29.45_ASO0252.ps}
\includegraphics[height=8.5cm,angle=-90]{ON231_ASO0259.ps}
\caption{The SED of 0FGL J1159.2+2912 = 4C29.45 (left) and
of 0FGL J1221.7+2814 = ON231= W Comae (right)}
\label{fig:sed_aso0252}
\end{figure}

\begin{figure}
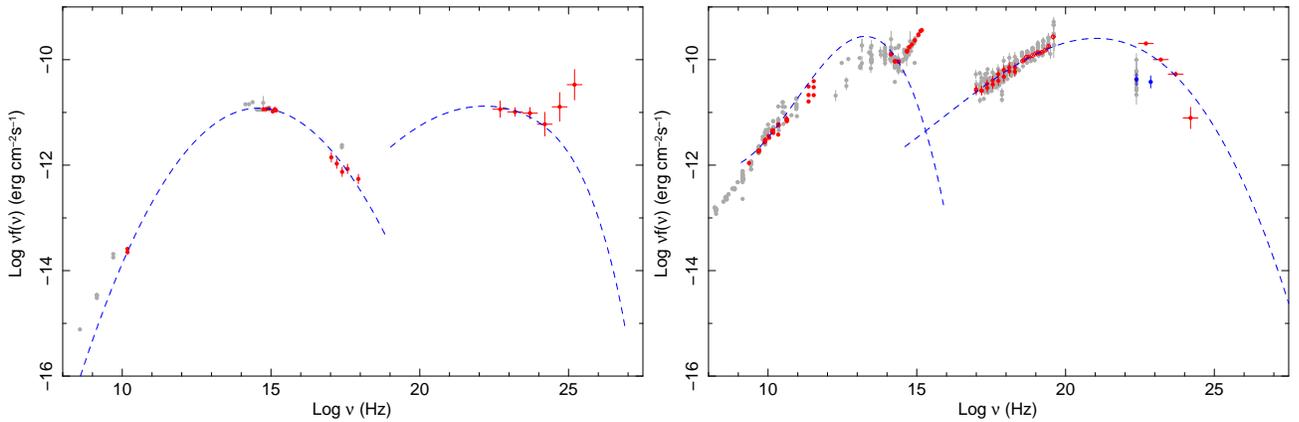

\epsscale{.80}
\includegraphics[height=8.5cm,angle=-90]{PG1246+586_ASO0270.ps}
\includegraphics[height=8.5cm,angle=-90]{3C273_ASO0263.ps}
\caption{The SED of 0FGL J1248.7+5811 = PG 1246+586 (left) and
of 0FGL J1229.1+0202 = 3C273  (right)}
\label{fig:sed_aso0270}
\end{figure}

\begin{figure}
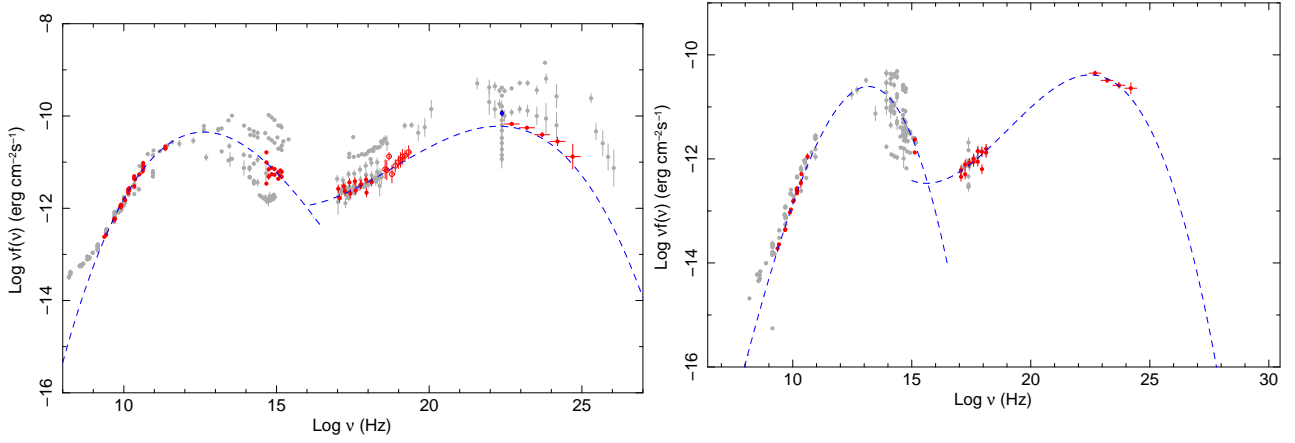

\epsscale{.80}
\includegraphics[height=8.5cm,angle=-90]{3C279_ASO0272.ps}
\includegraphics[height=8.5cm,angle=-90]{1Jy1308+326_ASO0280.ps}
\caption{The SED of 0FGL J1256.1-0547 = 3C279 (left) and
of 0FGL J1310.6+3220 = 1Jy1308+326 (right)}
\label{fig:sed_aso0272}
\end{figure}

\begin{figure}
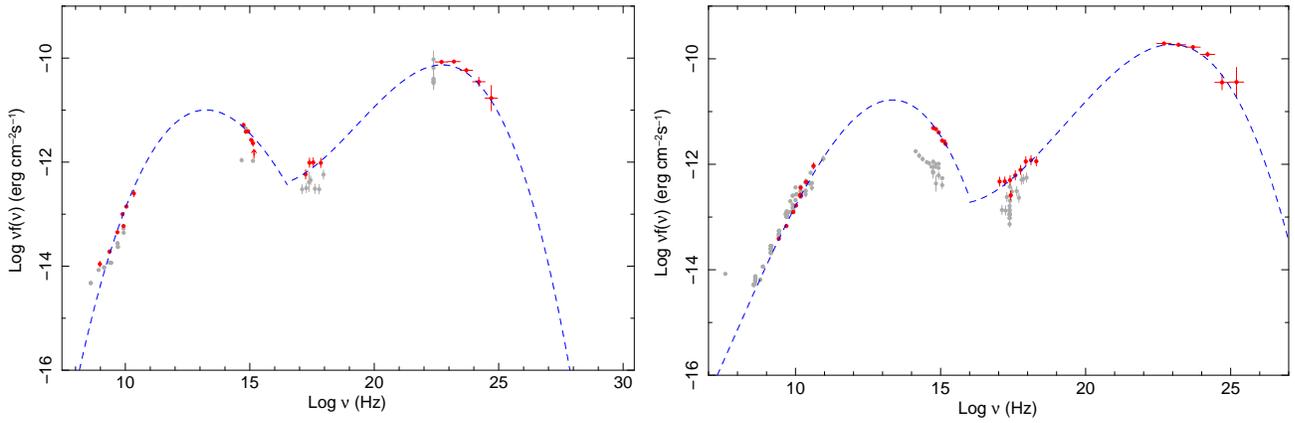

\epsscale{.80}
\includegraphics[height=8.5cm,angle=-90]{PKS1454-354_ASO0318.ps}
\includegraphics[height=8.5cm,angle=-90]{PKS1502+106_ASO0321.ps}
\caption{The SED of  0FGL J1457.6-3538 =  PKS 1454-354 (left) and
of 0FGL J1504.3+1030 = PKS1502+106 (right)}
\label{fig:sed_aso0287}
\end{figure}

\begin{figure}
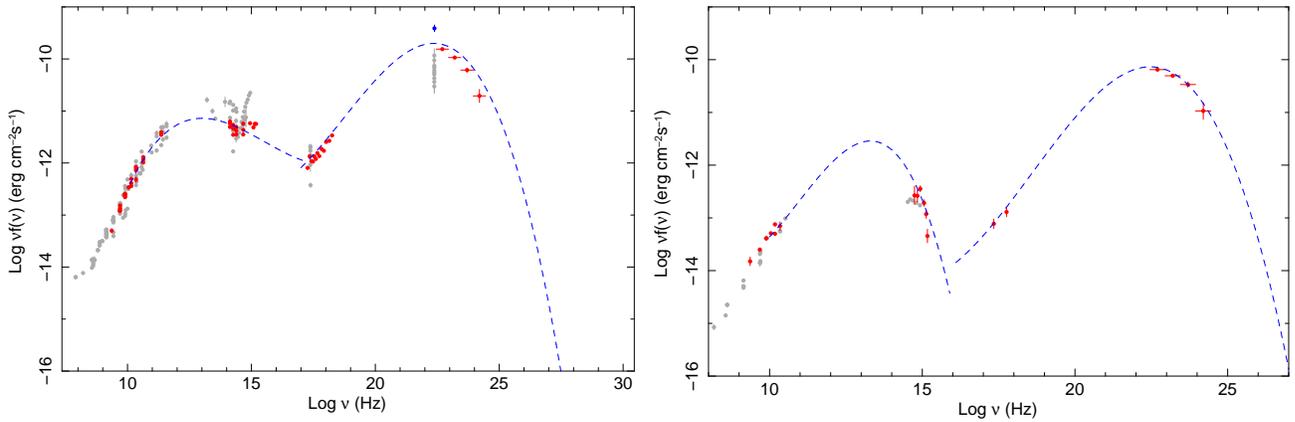

\epsscale{.80}
\includegraphics[height=8.5cm,angle=-90]{PKS1510-089_ASO0326.ps}
\includegraphics[height=8.5cm,angle=-90]{B21520+31_ASO0332.ps}
\caption{The SED of 0FGL J1512.7-0905 =  PKS 1510-089 (left) and
of 0FGL J1522.2+3143 = B2 1520+31 (right)}
\label{fig:sed_aso0326}
\end{figure}

\begin{figure}
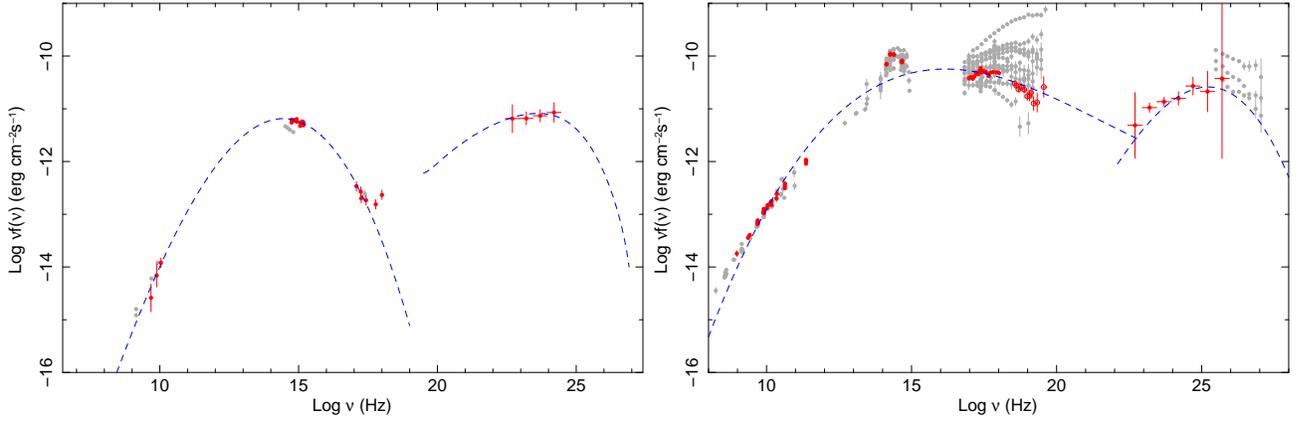

\epsscale{.80}
\includegraphics[height=8.5cm,angle=-90]{1RXS15429_ASO0336.ps}
\includegraphics[height=8.5cm,angle=-90]{MKN501_ASO0365.ps}
\caption{The SED of 0FGL J1543.1+6130 =  GB6 J1542+6129 (left) and
of 0FGL J1653.9+3946 =  Mkn 501 (right)}
\label{fig:sed_aso0336}
\end{figure}

\begin{figure}
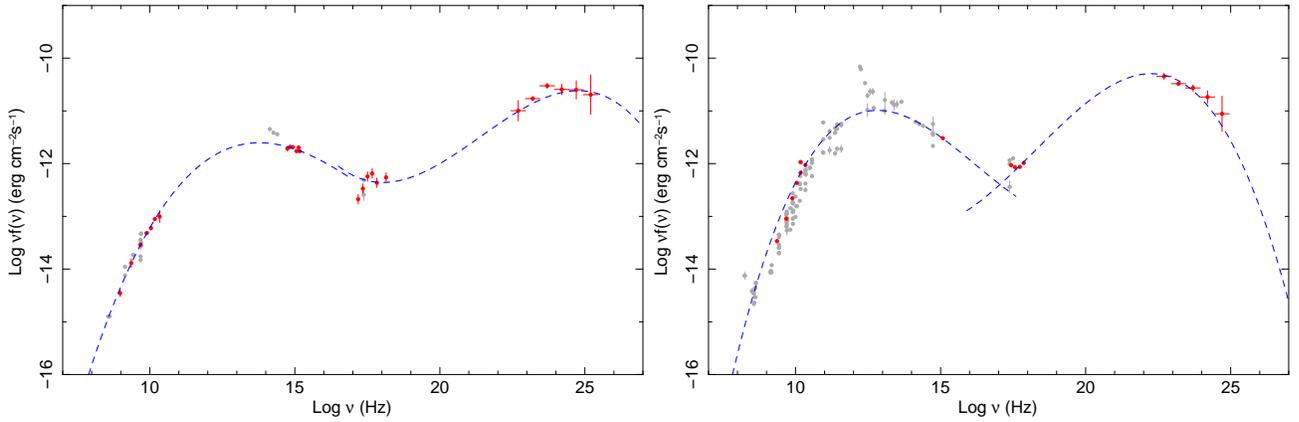

\epsscale{.80}
\includegraphics[height=8.5cm,angle=-90]{PKS1717+177_ASO0370.ps}
\includegraphics[height=8.5cm,angle=-90]{OT081_ASO0392.ps}
\caption{The SED of 0FGL J1719.3+1746 =  PKS 1717+177 (left) and
of 0FGL J1751.5+0935 = OT081}
\label{fig:sed_aso0370}
\end{figure}

\begin{figure}
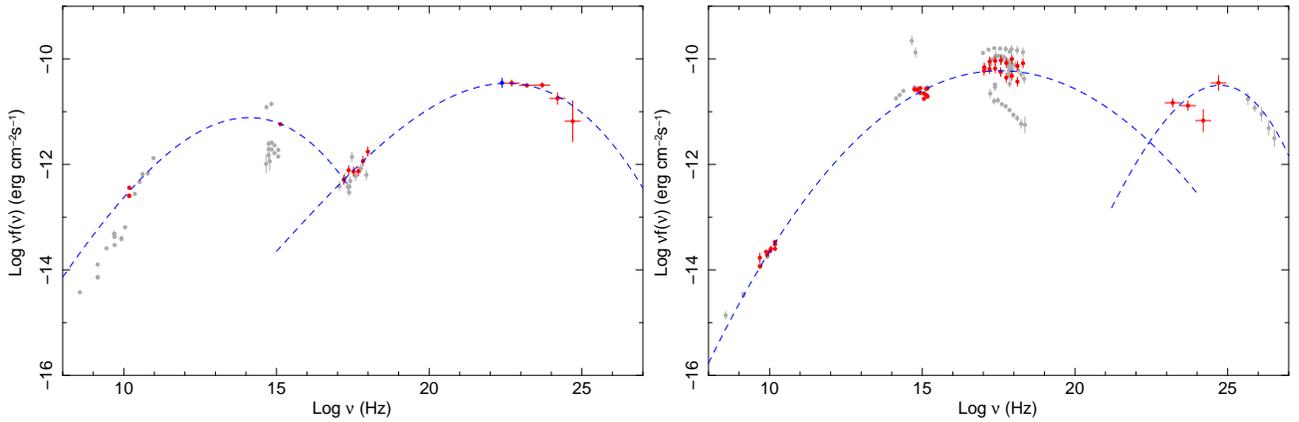

\epsscale{.80}
\includegraphics[height=8.5cm,angle=-90]{4C66.20_ASO0425.ps}
\includegraphics[height=8.5cm,angle=-90]{1ES1959+650_ASO0452.ps}
\caption{The SED of 0FGL J1849.4+6706 = 4C66.20 (left) and
of 0FGL J2000.2+6506 = 1ES1959+650 (right)}
\label{fig:sed_aso0425}
\end{figure}

\begin{figure}
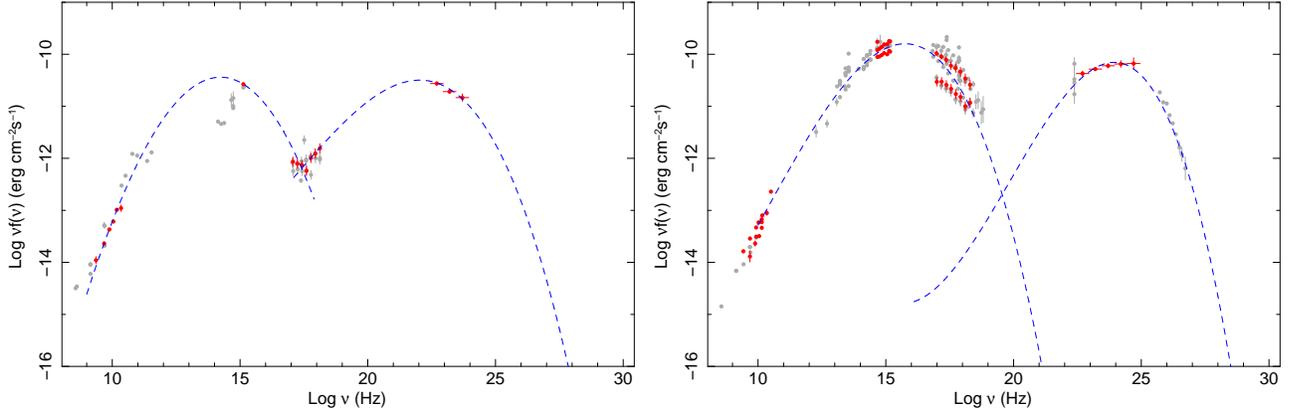

\epsscale{.80}
\includegraphics[height=8.5cm,angle=-90]{OX169_ASO0485.ps}
\includegraphics[height=8.5cm,angle=-90]{PKS2155-304_ASO0490.ps}
\caption{The SED of 0FGL J2143.2+1741 = S3~2141+17 (left) and
 of 0FGL J2158.8-3014 = PKS2155-304 (right)}
\label{fig:sed_aso0485}
\end{figure}

\begin{figure}
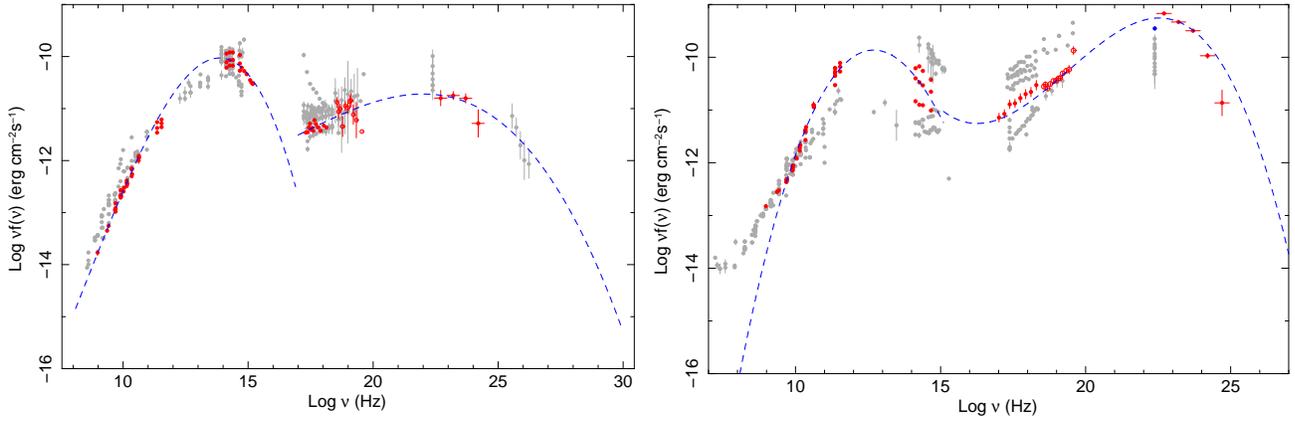

\epsscale{.40}
\includegraphics[height=8.5cm,angle=-90]{BLLAC_ASO0491.ps}
\includegraphics[height=8.5cm,angle=-90]{3C454.3_ASO0513.ps}
\caption{The SED of 0FGL J2202.4+4217 = BL~Lacertae (left) and
of 0FGL J2254.0+1609 = 3C454.3}
 \label{fig:sed_aso0491}
\end{figure}

\begin{figure}
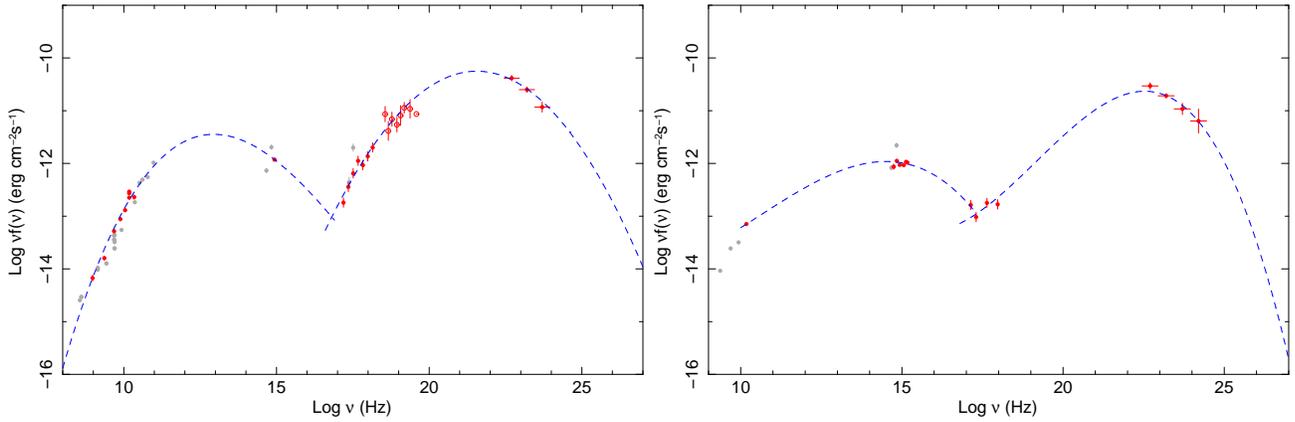

\epsscale{.80}
\includegraphics[height=8.5cm,angle=-90]{PKS2325+093_ASO0524.ps}
\includegraphics[height=8.5cm,angle=-90]{PMN2345m1555_ASO0528.ps}
\caption{The SED of 0FGL J2327.3+0947 =  PKS~2325+093 (left) and
of 0FGL J2344.5-1559 = PMN J2345-1555 (right)}
\label{fig:sed_last}
\end{figure}

The SEDs that we have built are shown from Fig. \ref{fig:sed_first} through Fig. \ref{fig:sed_last}, where the \fermi \gr ~data and the quasi-simultaneous multi-frequency measurements  appear as large filled red symbols. In all the SEDs we have also included non-simultaneous multi-frequency archival measurements  (small open grey points) to increase the data coverage in some energy bands and to illustrate the historical range of variability at different frequencies. Archival data points have been collected using the NED (NASA/IPAC Extragalactic Database) and ASDC (ASI Science Data Center) on-line services. The TeV data have been derived from the available literature as listed in Table \ref{tab:TeV}.

Fig. \ref{fig:sed_first} to Fig. \ref{fig:sed_last} show that in all cases the overall shape of the SEDs exhibit the typical broad double hump distribution, where the first bump is attributed to synchrotron radiation and the second one is likely due to one or more components related to inverse Compton emission.
The dashed lines represent the best fit to the data as described in the next Section.

Our SEDs show that there are considerable differences in the position of the peaks of the two components and on their relative peak intensities. Large variability is also present, especially at optical/UV and X-ray frequencies. Gamma-ray variability cannot be evaluated as the \fermi data that we are using are averaged over the entire LBAS data taking period.
The  \gr ~variability of \fermi LBAS blazars is discussed in detail in a separate paper (Abdo et al. 2009, in preparation).

A complete description of the \gr ~spectral shape of LBAS sources is given in Abdo et al (2009f). Here we note that in most cases the \fermi data
cannot be fit by a simple power law as significant curvature is detected.  Downward (convex) curvature is often observed in sources where synchrotron peak is located at low energies
(e.g., PKS0454-234, PKS1454-354 and PKS1502+106, 3C454.3 etc.)  whereas very flat  or even concave type curvature is exhibited by high synchrotron peaked objects (e.g., 3C66A, PKS 0447-439,
1ES 0502+675, and  PG 1246+586). A possible explanation of these features is discussed in Sect. \ref{sec:sedsparameters}

\section{Blazar SED observational parameters}
\label{sec:obsparameters}

 We now estimate some key observational parameters that characterize the SED of our blazars, namely, the radio spectral index ($\alpha_{r}$), the peak frequency and peak flux of the synchrotron component  (\nupS and \nupS F(\nupS)), and the peak frequency and flux of the inverse Compton part of the SED (\nupIC and \nupIC F(\nupIC)).

 \subsection{The radio spectral slope}
 \label{sec:alphar}

\begin{figure}
\epsscale{.40}
\includegraphics[height=16.cm,angle=-90]{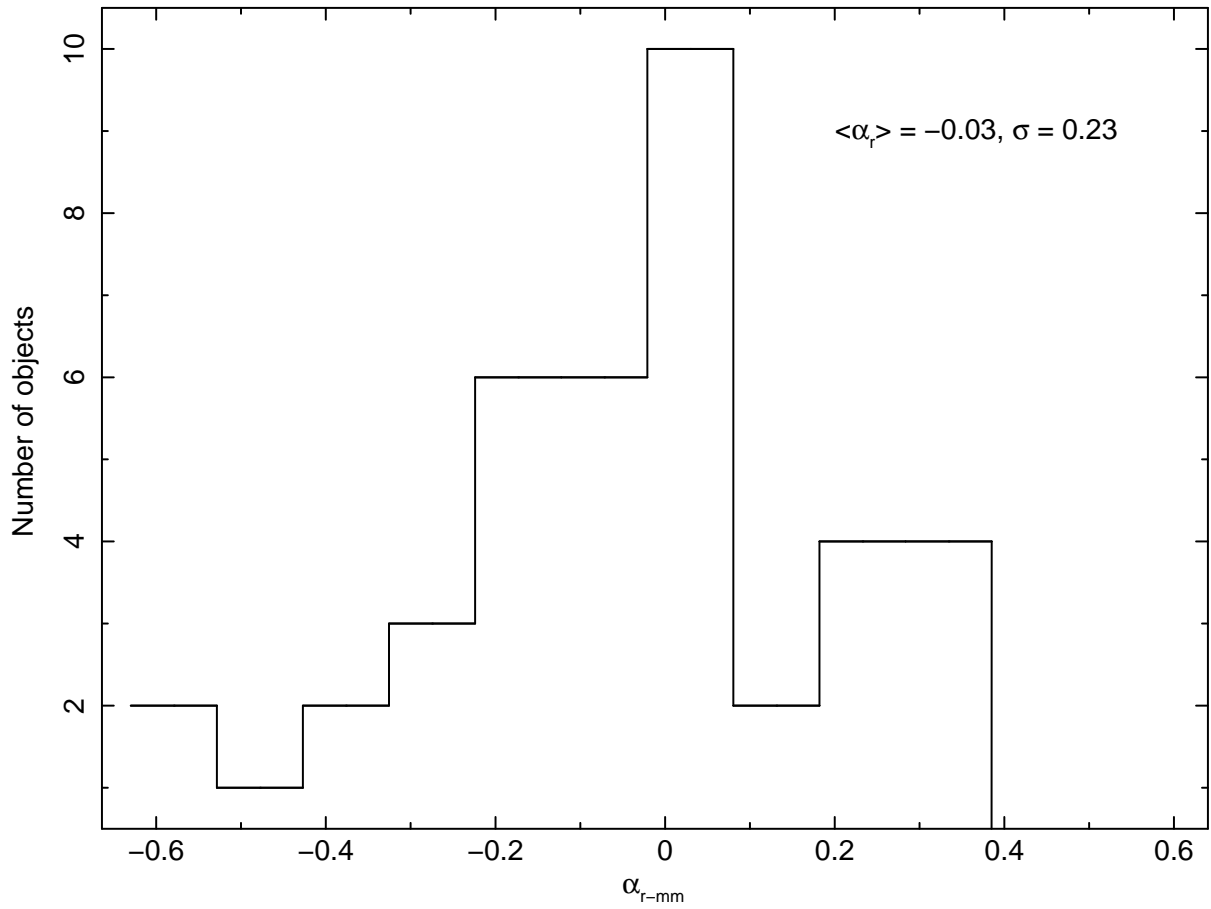}
\caption{The distribution of radio spectral index ( $f_r(\nu)\propto \nu^{\alpha_r}$) measured with the radio data of our 48 SEDs.}
\label{fig:alphaRdistrSED}
\end{figure}

\begin{figure}
\epsscale{.40}
\includegraphics[height=16.cm,angle=-90]{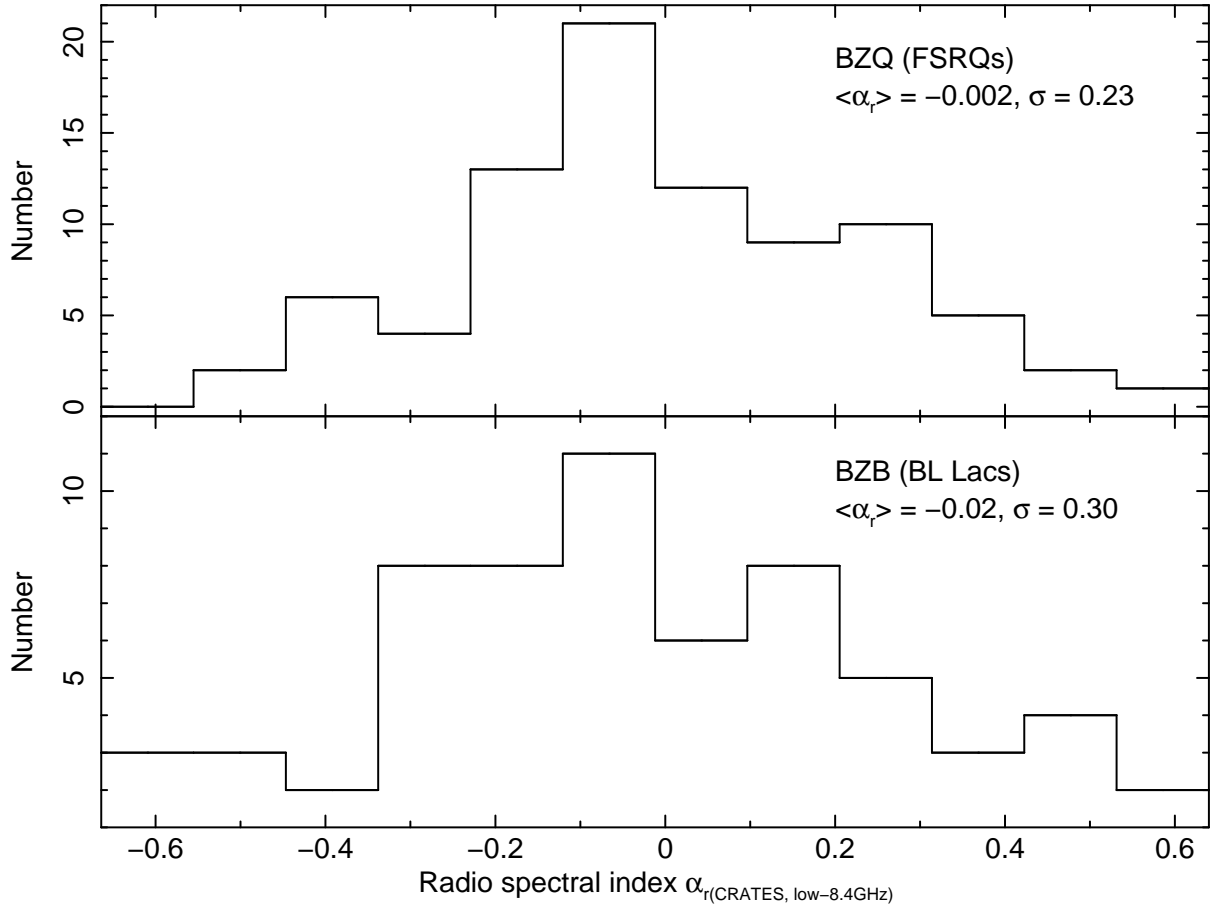}\\
\caption{The distribution of radio spectral index ( $f_r(\nu)\propto \nu^{\alpha_r}$) taken from the CRATES catalog, estimated between the CRATES low frequency, ($\sim $1GHz) and  8.4 GHz, for the sample of FSRQ (top panel) and BL Lacs (bottom panel).}
\label{fig:alphaRdistr}
\end{figure}

To estimate the blazar spectral slope (${\alpha_r}$, where $f_r(\nu)\propto \nu^{\alpha_r}$ ) in the radio/mm band we performed a linear regression of all the radio flux measurements
that have been used for the SEDs, including the non-simultaneous ones. The set of frequencies used for the linear regression in not the same for every source but
ranged from below 1 GHz up to about 100 GHz, for those sources for which microwave flux measurements are available.
The distribution of the radio spectral slopes $\alpha_{r}$ obtained with this method has an average value $\langle \alpha_{r}\rangle = -0.03$
and a standard deviation  $\sigma = 0.23$  (see Fig.  \ref{fig:alphaRdistrSED}). Figure \ref{fig:alphaRdistr} shows the distribution of the radio spectral slopes between $\sim $1GHz and 8.4 GHz taken from  the CRATES catalog \citep{crates} for the subsample of FSRQs and BL Lac objects respectively. The distributions shown in Figs. \ref{fig:alphaRdistrSED}  and \ref{fig:alphaRdistr} are all very similar
with an almost identical average value $\langle \alpha_{r}\rangle \sim  0.0$ and similar standard deviations $\sigma \sim $ 0.2/0.3.  In particular, for the $\alpha_{r}$ distributions
of FSRQs and BL Lacs shown in Fig.  \ref{fig:alphaRdistr} a Kolmogorov-Smirnov test gives a probablity of 0.43 that they come from the same parent population.
We conclude that the radio to micro-wave spectral slope in our SEDs is quite flat ($\langle \alpha_r \rangle \sim 0$) and consistent with being the same in all blazar types.

\subsection{The \aox--\aro ~plane}

The \aox-\aro ~plot of the LBAS sample is shown in Fig. \ref{fig:aoxaro} which also includes all blazars in the BZCat catalog for which we have radio, optical and X-ray measurements (small red dots). Note that \fermi FSRQs (filled circles), like all FSRQs discovered in any other energy band, are exclusively located along the top-left / bottom-right band, whereas
BL Lacs (open circles) can be found in all parts of the plane, albeit with a prevalence in the horizontal area defined by values of  \aro  ~between 0.2 and 0.4, which is where
HBL sources are located \citep{padgio95}.
The area of the  \aox-\aro ~space where the hypothetical population of UHBL (ultra high energy peaked) blazars \citep[that is sources where the synchrotron component is so energetic
to peak in the MeV region][]{ghisellini99,giommi01} could have been found, is empty, implying that these sources are either very rare, very weak or non existent \citep[see also][]{Costamante07}.

\subsection{The synchrotron peak energy (\nupS) and  peak intensity (\nupS F(\nupS))}

We estimated the peak energy (\nupS) and peak intensity (\nupS F(\nupS)) of the synchrotron power from the SEDs reported in Figs. \ref{fig:sed_first} - \ref{fig:sed_last} by fitting
the part of the SEDs that is dominated by Synchrotron emission. As fitting function we used a simple third degree polynomial:

\begin{equation}
\nu F_{\nu} = a\cdot \nu^3+b\cdot \nu^2+ c\cdot \nu+ d.
\label{poly}
\end{equation}

In the case of high redshift sources (e.g., J0229.5-3640, J0921.0+4437 and J1457.4$-$3538, J1522.2+3143), we excluded from the fitting procedure all points in the optical/UV bands that are likely to be significantly affected by Lyman-$\alpha$ forest absorption.

 \subsection{An empirical method to derive \nupS and \nupS F(\nupS)  from \aox ~and \aro}

As shown by  Padovani \& Giommi (1995)
 the peak of the synchrotron power \nupS in the SED of a blazar determines its position in the  \aox  - \aro ~plane
\citep[see Fig. 12 of][]{padgio95}, see also Padovani \etal ~(2003). 
Here we exploit this dependence showing that the value of \nupS  can be estimated from  \aox  - \aro ~through the following analytical relationship.

\begin{equation}
\mbox{Log($\nu_{peak_S}$) } = \left\{
\begin{array}{l}
13.85+2.30 X  ~ ~~~if ~~ {\rm ~X}< 0 ~{\rm and ~Y} < 0.3~  \\
13.15 + 6.58 Y ~~~~ {\rm otherwise}  \end{array} \right.
\label{eq.lognu}
\end{equation}
where  X = 0.565 $-~1.433\cdot$\aro +~0.155$\cdot$\aox ~and Y = 1.0 $-~0.661\cdot$\aro $-~0.339\cdot$\aox

We have calibrated this relationship using the  \nupS values directly measured from our 48 quasi-simultaneous SED and the corresponding \aox ~and \aro ~values.

Fig. \ref{fig:nupeaks} (top panel) shows the values of Log(\nupS) estimated from equation (\ref{eq.lognu}) plotted against  the values of Log(\nupS) measured by fitting a SSC model to the synchrotron part of the quasi-simultaneous SEDs of  figs. \ref{fig:sed_first} to \ref{fig:sed_last}. The distribution of the difference between the values estimated with the two methods has
a mean value of 0.04 and a standard deviation of 0.58, implying that the value of  Log(\nupS) can be derived even from non-simultaneous values of \aox ~and \aro ~within 0.6 decade at one sigma level and within 1 decade in almost all cases.

It must be noted, however, that this method assumes that the optical and X-ray fluxes are not contaminated by thermal emission from the disk or accretion. In blazars where thermal flux
components are not negligible (this should probably occur more frequently in low radio luminosity sources) the method described above may lead to a significant overestimation of the position of \nupS.

The peak flux   \nupS F(\nupS) can be estimated using the following relationship

\begin{equation}
{\rm Log}(\nu_{peak_S} F({\nu_{peak_S}}))   = 0.5\cdot{\rm Log} (\nu_{peakS}) -20.4 +0.9\cdot{\rm Log(R_{5GHz})},
\end{equation}

where ${\rm R_{5GHz}}$ is the radio flux density at 5GHz in units of mJy.

Fig. \ref{fig:nupeaks} (bottom panel) plots the value of   \nupS F(\nupS) estimated with the two methods. Also in this case the match is very good with an
average value of -0.01 for the difference between the two estimates and  a standard deviation of 0.33.

It is quite remarkable that one can derive the synchrotron peak flux simply from \nupS and from the radio flux as this implies that within a factor of 10
the radio emission represents a long-term calorimeter for the whole jet activity and the basic source power.

 \subsection{The peak frequency and peak intensity of the inverse Compton bump}

We have estimated the peak of the inverse Compton power in the SED (\nupIC) and the corresponding peak flux (\nupIC F({\nupIC})) by fitting the X-ray to \gr ~part of the SED,  which is dominated by inverse Compton emission using the polynomial function of Equation \ref{poly}.

There are some objects in which the soft X-ray band is still dominated by synchrotron radiation, and only the \fermi data can be used to constrain the inverse Compton component, so the above method is subject to large uncertainties. For this reason, in these cases, we have used the ASDC SED \footnote{http://tools.asdc.asi.it/SED/} interface to fit the simultaneous data points to a SSC model with a log-parabolic electron spectrum \citep{Tramacere09}.

For the whole sample we have determined  \nupIC as the value of  $\nu$ which maximizes $\nu F_{\nu}$ in Equation \ref{poly} or the predictions of the SSC model. The results are reported
in columns 6 and 7 of Table \ref{tab:SED_parameters}.
The best fit to both the synchrotron and inverse Compton components appear as dashed  lines in Figs.  \ref{fig:sed_first} - \ref{fig:sed_last}.

Fig. \ref{fig:slopeVSnup} (bottom panel) shows that the \nupIC, derived as described above for the 48 sources for which we have built the SEDs, is strongly correlated with their \gr ~spectral slope ($\Gamma$) taken from Table 3 of Abdo \etal ~(2009b).
We note that the scatter in the plots of Fig. \ref{fig:slopeVSnup} is largest in the regions of low \nupS/\nupIC - steep values of $\Gamma$, probably reflecting the presence of \gr ~spectral curvature (see Sect. \ref{sec:broadband}) .  The best fit to the \nupIC -- $\Gamma$ relationship is

\begin{equation}
Log (\nu_{\rm peak}^{IC}) =  -4.0\cdot\Gamma+31.6
\label{nupIC}
\end{equation}

Since the 48 objects for which we have quasi-simultaneous SEDs are representative of the entire LBAS sample the above equation can be used to estimate the \nupIC of the LBAS sources for which we have no simultaneous SEDs. We have done so and we have listed the results  in column 6 of Table \ref{tab:SED_parameters}.  The statistical uncertainty associated to \nupIC calculated via eq. \ref{nupIC} can be estimated from the distribution of the difference between
 \nupIC measured from the SED and that from eq. \ref{nupIC}. This distribution is centered on the value of 0 and has a sigma of 0.51; considering that the value of \nupIC from the SED is also subject to a a similar error we conservatively conclude that the Log of \nupIC values estimated through eq. \ref{nupIC} has an associated error of about 0.7.

\section{A SED-based classification scheme for blazars and other AGN}
\label{sec:classification}

Blazars, like other types of AGN, have been classified in the past according to heterogeneous criteria, often based on observational properties related to the energy band where they were first discovered. This lack of a stable and clear definition can lead to multiple classification of the same object and may cause subtle selection effects and biases in statistical analyses.
Now that \fermi has started producing large and homogeneous samples of blazars it is useful to re-asses the issue of blazar classification from a physical viewpoint taking into account the results of our SED study, so as to build a more robust base for future statistical/populations work.

We describe here a physical classification scheme based on the widely accepted AGN standard paradigm \citep[e.g.,][]{Urry95} and on well-known radiation emission processes.

The radiation emitted by an AGN is usually attributed to one (or both) of the following two physical processes:

\begin{enumerate}
\item[i)]
 thermal radiation originating from in-falling matter strongly heated in the inner parts of an accretion disk close to the black hole. This radiation is often assumed to be Comptonized by a hot corona producing the power law X-ray emission.
\item[ii)]
non-thermal emission emitted in a magnetic field by highly energetic particles that have been accelerated in a jet of material ejected from the nucleus at relativistic speed.
\end{enumerate}
The first process produces radiation mostly in the optical, UV and X-ray bands, whereas the radiation produced through the second mechanism encompasses the entire electromagnetic spectrum, from radio waves, to the most energetic $\gamma$-rays.
AGN that are energetically dominated by thermal radiation, (in the optical-X-ray band) can be classified as {\it thermal dominated}, or {\it disk dominated}  AGN, whereas AGN where the non-thermal processes are energetically dominant at all frequencies can be classified as {\it Non-Thermal radiation dominated} or {\it Jet dominated} AGN.  AGN can therefore be subdivided as follows.
\begin{itemize}
\item{\bf Thermal/Disk Dominated AGN} \\
These are objects usually called QSOs or Seyfert galaxies which do not show significant nuclear radio emission compared to the observed emission in the optical or X-ray band.
Although thermal-dominated AGN are the large majority ($\approx $90~\%) of AGN here we do not go into further detail about their sub-classification since none of the sources so far detected by \fermi {\it is} Thermal/disk dominated.

We feel however necessary to consider this type of AGN in this context as in some cases both the accretion (thermal) and the Jet (non-thermal) component may be present in the optical, UV or X-ray flux of the same object  \citep[e.g., 3C120, 3C 273][]{grandi04}. This mix of accretion and non-thermal radiation is rarely seen in the very bright $\gamma$-ray sources detected so far, but it will probably become more common as the sensitivity of the \fermi survey increases with time and a large number of fainter  and  less aligned sources are detected.

\item{\bf Non-Thermal/Jet dominated AGN} \\
The class of Non-Thermal/Jet Powered AGN, corresponds to the usual type of sources known as radio loud AGN. These can be subdivided into blazars and non-aligned non-thermal dominated AGN depending on the orientation of their jets with respect to the line of sight.

\begin{itemize}
\item  {\bf Blazars.}
These are core-dominated flat or inverted radio spectrum radio loud AGN. The radio core dominance
and the flat radio spectrum  together with strong and rapid variability (including
superluminal motion)  are the observational indicators that these objects point their
radio jet in a direction that is closely aligned to our line of sight. Because of this
very special perspective their light is strongly amplified by relativistic effects and the
time-scales of observed variations are significantly shortened.  \\
Blazars are divided into two main subclasses, FSRQs and BL Lacs depending on their optical spectral properties:

 \subitem  {\bf- FSRQs} or {\it Blazars of the QSO type} or {\bf BZQ} \citep[][]{Massaro09}. These are blazars showing broad emission lines in their optical spectrum
 just like normal QSOs. This category includes objects normally referred to as flat spectrum radio quasars (FSRQs) and
 broad-line radio galaxies.
 \subitem {\bf - BL Lacs} or {\it Blazars of the BL Lac type} or {\bf BZB}  \citep[][]{Massaro09}. These are objects normally called BL Lacs or
BL Lacertae objects. Their radio compactness and broad-band SED are very similar to that of
strong lined blazars but they have no strong and broad lines in their optical spectrum \citep[see e.g.,][]{marcha96}. \\

Sometimes, objects which show many of the hallmarks of blazars do not have optical spectra of sufficient quality to safely determine
the presence of broad emission lines or to accurately measure their equivalent width. In these cases the blazar subclass cannot be established
and therefore these objects have to be referred to as \subitem {\bf - BZU} or {\it- Blazars of the Unknown type}  \citep[see also][]{Massaro09}.

\item {\bf Non-aligned Non-Thermal dominated AGN.}  These sources are radio loud AGN with jets pointed at large or
intermediate \citep[$\approx 15-40 \degr $, see ][]{Urry95} angles with respect to the line of sight. For this reason they are sometimes called
non-aligned, misaligned or mispointed blazars. This category includes:
\subitem  {\bf - Radio galaxies} or {\it Non-aligned Non-Thermal dominated  AGN with no broad emission lines}
which are sources often showing extended, double--sided radio jets/lobes pointing in opposite directions in the plane
of the sky with respect to the central nucleus. The jet is clearly oriented at a very large angle with respect to the line of sight.
The nuclear emission is similar to that of blazars but it is not amplified and therefore it is usually fainter than the extended
emission, especially at low radio frequencies. The broad emission lines are not present in these sources because at such large angles
they are hidden by the torus.
\subitem {\bf - SSRQ} or {\it Non-aligned Non-Thermal dominated AGN with broad emission lines} which are sources usually known as
Steep Spectrum Radio Quasars (SSRQ), hence the orientation of the jet in these sources is thought to be intermediate between that of blazars and radio galaxies
 \citep[e.g.,][]{Urry95}).

\end{itemize}
\end{itemize}

In the literature  BL Lac objects are often subdivided into two or three subclasses depending on their SEDs. This classification was  first introduced by
Padovani \& Giommi (1995) who used the peak energy of the synchrotron emission, which reflects the maximum energy  the particles can be accelerated in the jet, to classify BL Lac into low energy and high energy synchrotron peak objects, respectively called LBL and HBL. In the following we extend this definition to all types of Non-Thermal dominated AGN using new acronyms (LSP, ISP and HSP) to avoid confusion.

\begin{itemize}
\item  {\bf LSP}  or {\it Low Synchrotron Peaked}  blazars.  These are sources where the synchrotron power peaks at
low energy (i.e. in the far IR or IR band or $\nu_{peak} \lsim10^{14}Hz$) and therefore their X-ray emission is flat (\ax $\approx $0.4-0.7) and due
to the rising part of the inverse Compton component (see Fig. \ref{fig:iblhbl}). At these relatively low energies the inverse Compton scattering occurs in the Thomson regime (see Sec. \ref{sec:sedsparameters} and Fig. \ref{fig:gammapeak}).

\item {\bf ISP}  or {\it Intermediate Synchrotron Peaked} blazars. Sources where the synchrotron emission peaks at intermediate energies ($10^{14} \lsim \nu_{peak} \lsim 10^{15}Hz$). In this case the X-ray band includes both the tail of the synchrotron emission and the rise of the inverse Compton component (see Fig. \ref{fig:iblhbl}).

\item  {\bf  HSP} or {\it High Synchrotron Peaked } blazars. Sources where the emitting particles are accelerated at much higher energies
than in LSPs so that the peak of the synchrotron power reaches UV or higher energies ($\nu_{peak} \gsim 10^{15}Hz$) (see Fig. \ref{fig:iblhbl}) \citep{padgio96}. Under these conditions the synchrotron emission dominates the observed flux in the X-ray band and the inverse Compton scattering occurs in the Klein Nishina regime (see Sec. \ref{sec:sedsparameters} and Fig. \ref{fig:gammapeak}).

\end{itemize}

Ideally, blazars should be classified on the basis of a complete SED built with simultaneous data. As in most cases this is not possible, LSP or HSP objects can still be recognized by estimating their \nupS from \aox ~and \aro ~and from their  X-ray spectral shape or by their radio to X-ray spectral slope \citep{pad03}.

In LSP sources the X-ray spectrum is flat (photon spectral index 1.5 $<  \gamma_{\rm x} < 1.8$) and dominated by the IC component.
In HSP sources the X-ray spectrum is instead still due to synchrotron emission and it is usually steep ($\gamma_{\rm x} > 2 $) if \nupS $\lsim 10^{17}Hz$  but it can still be flat in extreme
HSPs where \nupS is well into the X-ray band; the radio to X-ray spectral index, $\alpha_{rx}$ of these blazars is less than 0.7. In ISP objects both the (steep) tail of the synchrotron emission
and the (flat) rise of the IC component are within the X-ray band (see figure \ref{fig:iblhbl}), and $ 0.7 \lsim  \alpha_{rx} \lsim 0.8$.

\subsection{The distribution of synchrotron and inverse Compton peak frequencies}

Now that we have a new  SED-based classification of blazars and we have a reliable method of estimating \nupS we inspect the LBAS sample in terms of its content of LSP, ISP, and HSP objects and
we compare it with that of samples selected in other energy bands.

The distribution of the  synchrotron peak frequency (\nupS) of LBAS blazars (estimated using the \aox - \aro ~method) is plotted in Fig. \ref{fig:nupBbzBzq} for the FSRQ and the BL Lac subsamples (top and bottom panels respectively, solid histograms). While the \nupS distribution of FSRQs starts at
$\sim $10$^{12.5}$ Hz, peaks at  $\sim $10$^{13.3}$ Hz and it does not extend beyond $\approx$ 10$^{14.5}$ Hz, the distribution of BL Lacs is much flatter, starts at  $\sim $10$^{13}$ Hz and reaches much higher frequencies ( $\approx$ 10$^{17}$ Hz) than that of FSRQs. For comparison, in the same figure, we plot as a dotted histogram the distribution of \nupS of the sample of FSRQs and BL Lacs detected as foreground sources in the WMAP 3-year microwave anisotropy maps \citep{giommiWMAP3}.  In Fig. \ref{fig:nupBbzBzq_emss} we compare the  \nupS  distribution of the LBAS sample with that of the X-ray selected sample of blazars detected in the {\it Einstein} Extended Medium Sensitivity Survey \citep[EMSS][]{emss}.

From  Figs. \ref{fig:nupBbzBzq} and \ref{fig:nupBbzBzq_emss} we see that the \nupS distribution of FSRQs is consistent with being the same in the \gr, radio/microwave and in the X-ray band.
We note that the large majority of FSRQs are of the LSP type while no FSRQs of the HSP type have been found at any frequency. On the contrary, the \nupS distribution of BL Lac objects is very different in the three energy bands. It is strongly peaked at  $\sim $10$^{13.3}$ Hz in the microwave band, where HBL sources are very rare, whereas in the X-ray and \gr ~bands HSP sources are more abundant than LSPs.

Fig. \ref{fig:nupIC} shows the distribution of  the inverse Compton peak frequency, \nupIC, of the FSRQs (dot-dashed histogram) and the BL Lacs (solid histogram) in the LBAS sample.
The two distributions are quite different with the BL Lacs exhibiting much higher  \nupIC values, reproducing the case of the distribution of synchrotron \nupS shown in Fig. \ref{fig:nupeaks}. This is most likely due to the same reason that causes the different \nupS distributions in the two blazar subclasses.

\subsection {Summary of observational findings and sources classification}

The blazar observational parameters estimated from the quasi-simultaneous SEDs and from the broad band spectral indices \aox, \aro ~for the cases where no simultaneous SEDs are available are summarized in Table  \ref{tab:SED_parameters} where we also classify our blazars according to the scheme described in Sec. \ref{sec:classification}. Column 1 gives the source name, column 2 indicates if the quasi-simultaneous SED for the source is available, column 3 gives the radio spectral index  $\alpha_{r}$ as estimated in Sect. \ref{sec:alphar}, columns 4 and 5 give the synchrotron peak frequency  (\nupS) and intensity  (\nupS F(\nupS)) estimated from the SED
 and with the \aox-\aro ~method respectively, column 6 and 7 give the inverse Compton bump peak frequency (\nupIC) and intensity (\nupIC F(\nupIC)) estimated from the SEDs and from the correlation between \nupIC and the \gr ~spectral slope (see Fig. \ref{fig:slopeVSnup}) respectively, column 8 gives the particle peak energy (Lorentz factor) estimated assuming a simple SSC model ($\gamma_{peak}^{SSC}$ = $\sqrt{3/4 \cdot \nu_{\rm peak}^{IC}/\nu_{\rm peak}^{S}}$, see Eq. \ref{nupSSC} of Section \ref{sec:sedsparameters}), column 9 gives the Compton dominance (\nupIC F(\nupIC)/\nupS F(\nupS)), columns 10 and
 11 give the source classification based on the optical spectrum and on the shape of the SED according to the scheme described above.

\section{Implications for physical modeling}
\label{sec:sedsparameters}

The quasi-simultaneous SEDs reported in this paper show the typical two bump shape that is seen in radio or X-ray selected blazars.
According to current models the low energy bump is interpreted as synchrotron (S)
emission from highly relativistic electrons, and the high energy bump is related
to inverse Compton (IC) emission of various underlying radiation fields.\\
In the case of synchrotron self Compton model \citep[SSC,][]{Jones74,GhiselliniMaraschi89}
the seed photons for the IC process  are the  synchrotron photons produced by the
same population  of relativistic electron.\\
In the case of external radiation Compton (ERC) scenario \citep{Sikora94,Dermer02},
the  seed photons for the IC process  are typically UV photons generated by the accretion
disk surrounding the black  hole, and reflected toward the jet by the Broad Line
Region (BLR) within a typical distance from the accretion disk of the order of one pc.
If the emission occurs at larger distances, the external radiation is likely to
be provided by a dusty torus \citep{Sikora02}. In this case the photon field is
typically peaked at IR frequencies.\\

In this Section we follow a phenomenological approach to obtain information about the peak Lorentz factor
of the electron distribution ($\gamma_{peak}$)  most contributing to the synchroton emission
and to the inverse Compton process. To test the methods used to to estimate $\gamma_{peak}$, we employ an accurate numerical model
\citep{Tramacere09,Tramacere07a,Massaro06,Tramacere03} that can reproduce
both the SSC and ERC models. For the electron distribution we considered a log-parabola
of the form $n(\gamma)=K\cdot10^{~r~Log(\gamma/\gamma_{peak})^2}$  with $\gamma_{peak}$ ranging between
100 and $6\cdot10^5$ and with curvature parameter $r=0.4$  \citep{Massaro04,Tramacere07b}.
As input parameter for the benchmark SSC model we use a source size $R=10^{15}$ cm, a magnetic
field $B=0.1$ G, a beaming factor $\delta=10$, and an electron density N=1 $e^-/cm^{3}$
(N=$\int n(\gamma)d\gamma$). In the case of the benchmark ERC model, we use the same set of
parameters with the addition of  the external photon field produced by the accretion
disk and reflected by the BLR toward the emitting region with an efficiency
$\tau_{BLR}=0.1$. The accretion disk radiation is modelled by a multitemperature
black body, with a innermost disk temperature of $10^5$ K.

\subsection{The synchrotron peak frequency}

The dependence of the observed peak frequency of the synchrotron emission (\nupS) on
magnetic field intensity ($B$), electron Lorentz factor ($\gamma$), beaming factor ($\delta$) and redshift ($z$) is given by:
\begin{equation}
\nu_{peak}^{S}=3.2\times10^6(\gamma^S_{peak})^2 B \delta/(1+z)=\nu_{peak}^{S'}\delta/(1+z).
\label{nupS}
\end{equation}
where $\nu_{peak}^{S'}$ is the synchrotron peak frequency in the emitting region rest-frame.
A good estimate of $\gamma^S_{peak}$  in terms of the differential electron energy
distribution ($n(\gamma)=dN(\gamma)/d\gamma$) is given by the peak of  $\gamma^3 n(\gamma)$,
hereafter $\gamma_{3p}$ \citep{Tramacere09,Tramacere07b}.
In panel \textit{a} of Fig. \ref{fig:gammapeak} we plot the ratio of $\gamma_{peak}^S$
to $\gamma_{3p}$ as a function of $\gamma_{peak}^S$. The ratio is  steady and very
close to one over the whole range of $\gamma_{peak}^S$ values. The value of $\gamma_{peak}^S$ is
estimated by fitting the peak of the numerically  computed synchrotron SED
with a log-parabolic analytical function. Note, however, that there is a  degeneracy
on the value  of $\gamma_{peak}^S$  given by the product $B\delta$. We discuss this point in the next  subsection.

\subsection{The inverse Compton peak frequency}

In a simple SSC model, and under the Thomson regime (TH) of the IC scattering, the observed peak frequency of the synchrotron component (\nupS) is related
to the observed peak frequency of the inverse Compton one (\nupIC) by the following
relation:
\begin{equation}
{\nu_{peak}^{IC} \over \nu_{peak}^{S}} \simeq \frac{4}{3}(\gamma^{SSC}_{peak})^2
\label{nupSSC}
\end{equation}
where $\gamma_{peak}^{SSC}$ is of the same order of $\gamma^S_{peak}$.
Panels \textit{b} and \textit{c} of Fig.   \ref{fig:gammapeak} show that (for the choice of SSC parameters reported above ) this trend is valid only
for $\gamma_{peak}^{SSC} \lsim 2\cdot 10^4$ where the transition from Thomson to
 Klein Nishina (KN) regime occurs.  In the  KN regime Eq. (\ref{nupSSC}) is no longer valid: in fact, the kinematic limit for the maximum energy of the up-scattered
photons in the emitting region rest-frame is:
\begin{equation}
\nu_{max}^{IC} = \frac{4\gamma^2\nu_S}{1+4\gamma^2(h\nu_S/m_ec^2)}~~.
\end{equation}

As the energy of the seed photons in the electron rest-frames increases, the
maximum up-scattered photon energies approaches the energy of the up-scattering
electron ($\gamma m_ec^2$). This means that the peak energy of the IC emission
is no longer growing with $\gamma_{peak}^2$ according to Eq. (\ref{nupSSC}), but it starts
becoming smaller as shown in panels \textit{b} and \textit{c} of fig. \ref{fig:gammapeak}.
We note that this effect is particularly relevant for the case of HBL objects.\\
Other deviations from the trend given by Eq. (\ref{nupSSC}) occur when   further radiative
components add to a single zone SSC. In fact, for the case of External Compton scenario, the observed peak frequency of
the ERC component in  terms of the frequency of the external photon field in the disk
rest-frame ($\nu_{peak}^{' EXT}$) reads:
\begin{equation}
\frac{\nu_{peak}^{ERC} }{ \nu_{peak}^{' EXT}\Gamma} \simeq (\frac{4}{3}) (\gamma^{ERC}_{peak})^2\delta/(1+z)
\label{nupERC}
\end{equation}
where  $\nu_{peak}^{' EXT}\Gamma$ is the external photon field frequency transformed to the rest-frame
of the emitting region which is moving with a bulk Lorentz factor $\Gamma$, and assuming that the BLR radiation is isotropic.

If one uses Eq. (\ref{nupSSC}) in place of Eq. (\ref{nupERC}) (an assumption justified by the fact that the UV
and IR external radiation fields are usually dominated by the
non-thermal synchrotron emission of the source), a significant bias
on the value of $\gamma^{ERC}_{peak}$ is introduced in the ERC scenario.
In fact, the resulting value of $\gamma_{peak}$ is strongly overestimated in the case of external UV radiation field
($\gamma_{peak}^{SSC}>>\gamma^{ERC}_{peak}$ and $\gamma_{peak}^{SSC}>>\gamma^{S}_{peak}$ ). In the case of IR external
radiation field, the bias is smaller but the measured value of $\gamma^{SSC}_{peak}$ is still overestimating  both $\gamma^{ERC}_{peak}$
and  $\gamma^{S}_{peak}$.

In conclusion, when  $\gamma_{peak}$ is estimated through Eq. (\ref{nupSSC}) we expect two main biases:

\begin{enumerate}
    \item a bias related to the KN effect, affecting mostly HBL objects, which leads to an underestimation of  $\gamma_{peak}$
    \item a bias related to the ERC scenario, affecting FSRQs and IBL objects,  which yields  an overestimate of $\gamma_{peak}$
\end{enumerate}

These arguments provide an interesting diagnostic tool in the $\nu_{peak}^{S}$--$\gamma_{peak}^{SSC}$ plane. Objects radiating mainly
via the ERC mechanism are expected to lie above the $\nu_{peak}^{S} \propto  \gamma_{peak}^{S}$ line, and objects radiating \gr s mainly via
the SSC mechanism are expected to lie along the $\nu_{peak}^{S} \propto \gamma_{peak}^{S}$ line in the case of TH-IC regime, and below it in
the case of KN-IC regime.

To test this scenario  we use the value of $\gamma_{peak}^{SSC}$ obtained by Eq. (\ref{nupSSC})
applied to the numerically computed SSC/ERC SEDs , and we compare these trends with those
obtained applying  Eq. (\ref{nupSSC}) to the data of Table \ref{tab:SED_parameters}.
Fig. \ref{fig:nupvgammael} shows the location of HSP objects (blue solid boxes),  ISPs/LSPs
objects (orange solid boxes) and  FSRQs (red solid circles).\\
The values of $\gamma_{peak}^{SSC}$ estimated for the case of SSC
emission (dashed blue line with stars) show clearly the effect of the
transition from the TH to the KN regime. We note
that all but two of the HBLs, lie below the $\nu_{peak}^{S}
\propto  \gamma_{peak}^{S}$ line.  In particular all the HSP objects
below the $\nu_{peak}^{S} \propto \gamma_{peak}^{S}$ line have $\gamma_{peak}$
values below the prediction of the SSC scenario (solid blue line).
On the contrary, all the FSRQs and the LSP/ISP BL Lacs but one lie
above the $\nu_{peak}^{S} \propto \gamma_{peak}^{S}$ line. The majority of
the FSRQs objects have a value of $\gamma_{peak}$ in excess of
a factor $\sim 10^4$ and limited by the prediction from the ERC model (purple dashed line with stars).
The LBLs/IBLs sources are more uniformly distributed across the region delimited by the the
SSC TH prediction and by the ERC one. By further
dividing the sample in Compton Dominated (CD) objects
(\nupIC F(\nupIC) $>$ 2 \nupS F(\nupS)) and non Compton Dominated (NCD) objects
 (\nupIC F(\nupIC) $\le$ 2 \nupS F(\nupS)), we found that all the CD objects lie above the $\nu_{peak}^{S}
\propto \gamma_{peak}^{S}$ line and populate the region between the SSC
TH and the ERC regime, with the FSRQs clustering toward the ERC
region.

Our analysis shows that the ERC model could explain the
high CD values as well as the high values of $\gamma_{peak}^{SSC}$
estimated in the case of FSRQs and ISP/LSP BL Lacs.
In order to explain the high values of $\gamma_{peak}^{SSC}$
obtained in the case of FSRQs in the context of single
zone SSC emission model, a very small value of the magnetic field with
($B<0.01$ G) is required.\\


As a final step, we discuss two additional effects that have consequences for the source distribution is this parameter
space:

\begin{enumerate}
\item  The $B\delta$ degeneracy on $\gamma_{peak}^{S}$ can affect the transition
region from TH to KN regime, since high values of $\delta$ allow the TH regime
to propagate towards higher frequencies.
\item The values of $\gamma_{peak}$ in  the case of an UV external radiation field (purple line Fig.\ref{fig:nupvgammael} )
constitutes an upper limit to the observed values of $\gamma_{peak}$, meaning
that objects in the  region below the ERC prediction line require
a wider range of external photon energies, extending down to the IR band.
\end{enumerate}

To take into account both these effects we perform Monte Carlo (MC) simulations.
Specifically, we generate both the SSC
and ERC numerical computation of the SEDs extracting $\delta$, $B$
and the temperature of the accretion disk $T$ from a random
uniform distribution
in order to cover a larger volume of the parameter space.
We generate 1000 realizations, with $\delta$ ranging in the interval [10-15],
$B$ in the interval [0.01-1] G and $T$ in the interval [$10-10^{4.5}$] K.
In  Fig.  \ref{fig:nupvgammael}  the MC results for the case of SSC fall within the area delimited by the blue contour line, while
the results in the case of ERC model, are delimited by the light red contour line. \\

We note that the MC simulations, compared to the ERC one for the only case of UV external photons (purple line), cover a much wider region of the parameter space. In the case of the
 MC SEDs, the range of temperatures of the BB emission allows us to take into account external photon fields peaking at IR frequencies. The resulting MC realizations  populate the whole
 parameter space delimited by the ERC/UV (purple line) and the SSC/TH case (solid blue line, below about $10^{15}$ Hz). This suggests, that in the ERC paradigm, the observed data, FSRQs (red circles) and ISP/LSP BL Lacs (orange square symbols), require external photon fields ranging form the UV down to the IR.

An alternative scenario that can explain the distribution of LBAS blazars in the plot of Fig. \ref{fig:nupvgammael}
advocates the superposition of two or more SSC components with different intrinsic energetics reflecting different conditions of the associated components.
Such composition of multiple relativistic plasmoids predicts that the large $\gamma$-ray excess, over a simple SSC model, observed in many LSP blazars in Fig. \ref{fig:nupvgammael}, and the flat or concave shape of the \gr ~SED of a number of ISP/HSP blazars  (see e.g., 3C66A, Fig. \ref{fig:sed_56}, PKS 0447-439, Fig. \ref{fig:sed_1112}, 1ES 0502+675, Fig. \ref{fig:sed_aso0103}, and  PG 1246+586, Fig. \ref{fig:sed_aso0270}) is the result of the presence of a second (or higher order) SSC component that is subdominant in the low-$\nu$ (radio-IR) range but emerges at higher energies, after the synchrotron peak of the first, less energetic, component
\citep[see e.g., the case of S5 0716+714][] {giommi08}. The combination of these multiple components is consistent with the intensity and spectral variability observed in these blazars.
In such a model, the SED of HBL-type sources can be fitted with a primary SSC component which peaks at the IR/optical (S) and \gr ~band (IC), and with a second more energetic and usually more variable component, which peaks in the UV or X-ray band (S) and at $\approx $ GeV energies (IC) thus explaining the widespread variability of these sources at TeV energies.

The predictions of this last model are quite different from those of the ERC model and can therefore be tested by future multifrequency observation campaigns.
A specific discussion of the details of such a multi-component SSC model will be presented in a dedicated paper.

\section{Summary and conclusion}
\label{sec:discussion}

\begin{figure}
\epsscale{.40}
\includegraphics[height=16.cm,angle=-90]{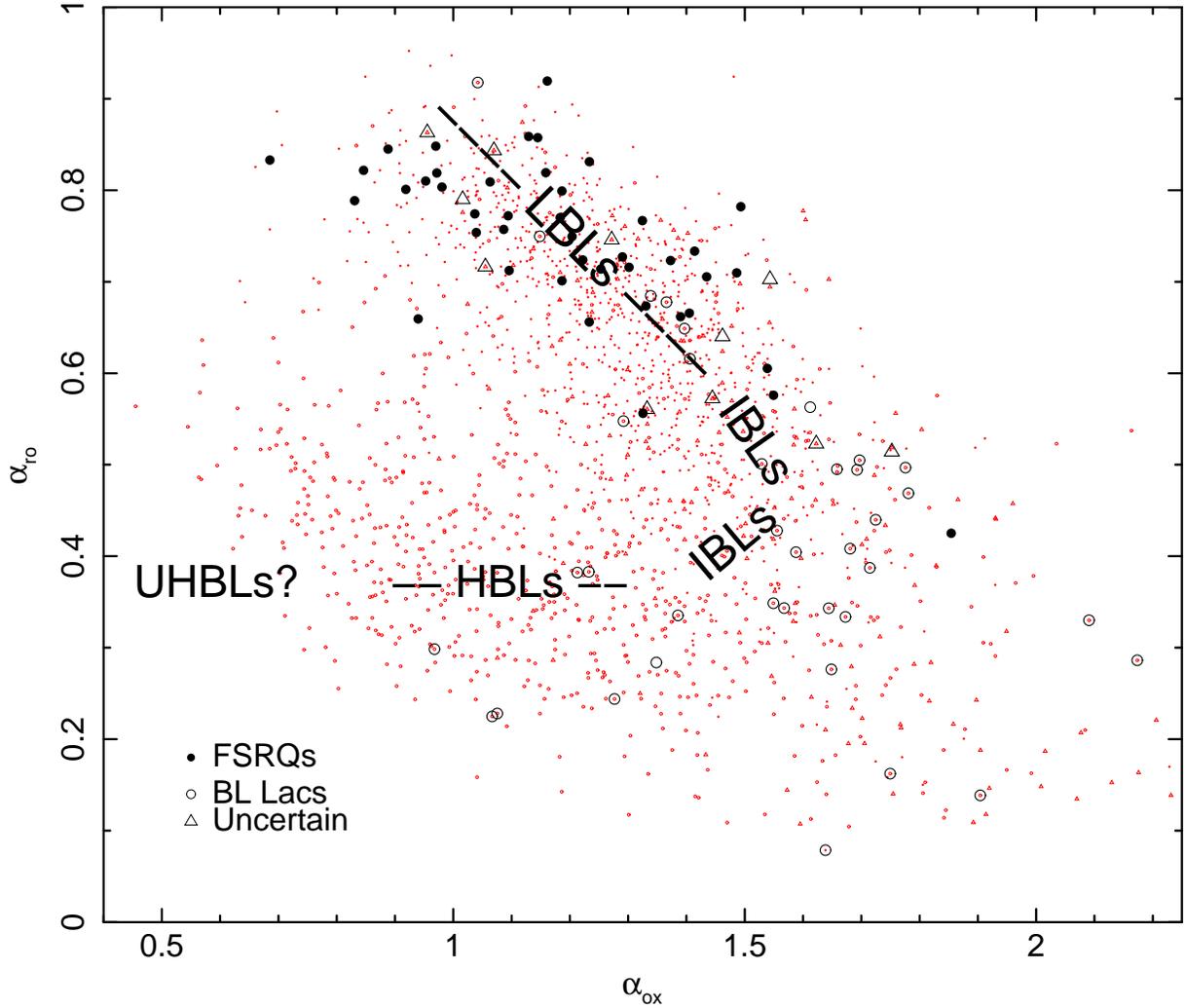}
\caption{The \aox- \aro ~plot of the LBAS blazars (large symbols) compared to the sample of blazars in the BZCAT catalog for which there is radio optical and X-ray information (small red symbols). All gamma-ray selected blazars are located in regions covered  by previously known blazars. No new \gr ~type of blazars has been found, in particular there is no evidence for thy hypothetical population of Ultra High energy peaked blazars (UHBLs), with synchrotron peak in the \gr ~band (Log(\nupS) $> 10^{20}$ Hz).}
\label{fig:aoxaro}
\end{figure}

We have carried out a detailed investigation of the broad-band (radio to high-energy \gr) spectral properties of the LBAS sample of \fermi bright blazars using a large number of
multi-frequency simultaneous observations as well as literature and archival data.  Using data obtained with \frmi, \sw, radio/mm telescopes, infra-red, and optical facilities
we have been able to assemble simultaneous or quasi-simultaneous  SEDs of a sizable and representative fraction of an homogeneous sample of blazars detected during a \gr ~all sky survey and not under special circumstances such as strong flaring activity.  This collection of high-quality, well sampled, nearly simultaneous, broad-band SEDs for a large number of blazars is unprecedented and allowed us
to estimate a number of important parameters characterizing the SED of \gr ~selected blazars and to address some key aspects of blazar demographics and physics.
Our main results are as follows:

\begin{enumerate}

\item
We derived reliable estimates of the frequency of the synchrotron  (\nupS ) and of the inverse Compton peaks (\nupIC ) for over 100 LBAS blazars. This was done directly from the simultaneous data for the 48 sources for which we have  the SEDs (see Figs \ref{fig:sed_first} - \ref{fig:sed_last}). For the remaining ones, \nupS and \nupIC were estimated indirectly using a refined version of the method of Padovani \& Giommi (1995)
based on the position in the \aox - \aro ~plane, for the former, and on the slope of the \gr ~spectrum for the latter,  as the \gr ~spectral slope and \nupIC are strongly correlated (see Fig.\ref{fig:slopeVSnup}).
The determination of \nupS for the large majority of the sources in the sample prompted us to develop a new SED-based classification scheme for {\it all} non-thermal dominated AGN  based on an extension of the  of the classification previously used for BL Lac objects only (see Sect. \ref{sec:classification}).
We also find that the \gr ~spectral slope is strongly correlated with the slope of the X-ray spectrum  (see Fig. \ref{fig:slopexslopeg}). Such a correlation is expected at first order in  synchrotron - inverse Compton scenarios, however the expected spectral slopes in the two energy bands depend on the position of the Synchrotron (e.g. Padovani \& Giommi 1996) and Inverse Compton peaks
broadening the correlation.

\item
Considering that a) all the  \gr ~sources in the bright sample of \fermi blazars that have been associated to radio loud AGN \cite{AbdoLATpaper, AbdoAGNpaper} have \aox ~and \aro ~similar to those of previously known blazars (see Fig. \ref{fig:aoxaro});  b) that among the only 7 still unidentified sources with Galactic latitude $|b| > 10^\circ$, 2 are likely blazars (similar to the ones already identified as the \gr ~error region includes radio-optical candidates with \aox $\gsim$ 1.4 and \aro $\sim 0.5$) and that the error region of the remaining 5 do not include any radio candidates brighter than 3 mJy,  we can conclude that gamma-ray selected blazars have broad-band spectral properties similar to those of radio and X-ray discovered blazars implying that they are all drawn from the same underlying population.
No evidence was found for the hypothetical class of Ultra High Energy peaked (UHBL) blazars \citep[see][]{ghisellini99,giommi01,nieppola} characterized by a synchrotron emission that is so energetic to reach the \gr ~band, and thus populate the extreme part of the \aox--\aro ~diagram defined by 0.2 $<$ \aro $~< $ 0.4 and \aox $~\lsim$ 0.7 (see Fig. \ref{fig:aoxaro}).
These sources, if bright and existed in good numbers, should have been found in a \gr ~survey such as LBAS, just as the population of HBL BL Lacs was discovered when X-ray surveys became available.  Alternatively, UHBLs could be intrinsically weak \gr ~sources and/or mis-identified  \citep{Costamante07} and their discovery must await the availability of much deeper samples than LBAS.

\item The distribution of the synchrotron peak frequency is very different for the FSRQ and BL Lac subsamples with values of \nupS located between 10$^{12.5}$ and  10$^{14.5}$ Hz in FSRQ  and between $10^{13}$ and  $10^{17}$ Hz in BL Lacs (see Fig. \ref{fig:nupBbzBzq}). This result rules out the existence of FSRQs of the HSP type (HBL in the old BL Lac nomenclature), consistent with what also observed in radio, microwave and  X-ray surveys.
The much larger \nupS values that can be reached by BL Lacs explain their observed harder \gr ~spectral slopes and hence the much better sensitivity of the LAT instrument to these sources \citep[see Fig. 7 of ][]{AbdoAGNpaper}. This selection effect will be even stronger above a few GeV  and fits with the well-known fact that TeV detected blazars are almost exclusively of the HSP (HBL) type. This also reproduces the case of the soft X-ray band where HSB BL Lacs (HBLs) are the dominant type of blazars.

\item A remarkable difference between LSP and HSP sources (see Sect.  \ref{sec:classification}) is that more than  50\% (10/16) of the HSP blazars with radio flux larger than 300 mJy at 1.4GHz,  in the BZCat catalog are detected in the LBAS sample while this fraction goes down to only $\lsim$13\% (58/452) for LSP blazars with radio flux larger than 500 mJy at 1.4GHz. Note that the sample of
undetected LBL blazars has similar overall properties than that of the detected ones, e.g.,  $<z>_{detected}$=1.0, $<z>_{undetected}$= 1.1,  $ <$Vmag$>_{detected}$= 17.1 and
 $ <$Vmag$>_{undetected}$=17.7. However, some authors  \citep[e.g.,][]{Kov09,Lister09,Pushkarev09,Kov09a,Savolainen09}
 showed that the LAT detected blazars might have larger doppler boosting factors than undetected ones.
 A detailed comparison of all important parameters of \gr ~detected and undetected blazars will be done when the much larger catalog of \gr ~sources based on approximately 1 year of LAT data is available.

\item The minimum  \nupS of BL Lac objects of $\sim 10^{13}$ Hz is consistent with the results of Maselli \etal ~(2009)
who conducted a careful search for very low synchrotron peaked BL Lac objects among the over 2000 blazars of the BZcat list and found them to be very rare or non-existent. The fact that the BL Lac minimum \nupS appears to be larger than in FSRQs could be due to some intrinsic difference in the mechanism of particle acceleration in the two types of blazars or to a mere selection effect. In fact, the non-thermal emission of very low \nupS BL Lacs would be minimal in the optical band (see Fig. \ref{fig:iblhbl}) causing them to be classified more easily as FSRQs rather than BL Lacs if low intensity broad lines (which would normally be below the non-thermal continuum) are  present in this type of objects.

We note that for LSP sources (\nupS $< 10^{14}$ Hz), the ratio of \gr ~detected FSRQs compared to BL Lacs is  approximately $4$, i.e. a value similar to that seen in the radio and microwave bands  \citep[$\sim $6 both in the 1Jy and in the WMAP3 samples,][]{stickel91,giommiWMAP3}. This  strongly suggests that the mechanism that produces $\gamma$-rays
is, at first order, the same in both LBL FSRQs and BL Lacs \citep[see also][]{giommiWMAP3}.

\item The results of this study lead to the conclusion that a simple homogeneous, one-zone, SSC model cannot explain the SED of the majority of the detected sources, especially of the LBL type (see Fig. \ref{fig:nupvgammael}). In addition, differential variability in the simultaneous optical and X-ray data observed in IBL and HBL objects (that is close to the peak of the synchrotron component) suggests that multiple components are present in non LBL blazars \citep[e.g., S5 0716+714, ][]{giommi08}  as also clearly shown by simultaneous X-ray/TeV campaigns  \citep[e.g., PKS 2155-304,][]{Aharonian09}.
Our results also show that ERC models can easily fit the data as they can cover a very wide part of the parameter space of Fig. \ref{fig:nupvgammael} (orange squares). However, models that are based on the presence of external radiation fields that are significantly different in FSRQs and BL Lacs, such as the broad-line region, accretion disk etc., must
explain why a)
the ratio of the number of FSRQs  and BL Lacs of the LBL type (which have similar \gr ~spectral slopes and therefore are affected in the same way by the higher LAT sensitivity to hard sources)  is  similar in radio/microwave selected samples (e.g., 1 Jy, WMAP) and in the LBAS \gr ~selected sample, and b)
why BL Lacs  appear to show equal, or even larger, values of  $\gamma_{peak}^{SSC}$ (that is larger \gr ~excess above SSC) than FSRQs in Fig.  \ref{fig:nupvgammael}.
Finally, any emission model should explain why only less than 13\% of bright radio sources  (F $>$ 0.5 Jy @ 1.4 GHz) of the LBL type are in the LBAS sample, while the other 87\% with similar observational properties are below the LBAS detection threshold and may well be radiating close to simple SSC.  We intend to address these topics in future papers.

\end{enumerate}

\newpage
\appendix
\begin{center}
  {\bf APPENDIX}
\end{center}
\section{Unfolding analysis}
\label{app}

The purpose of the unfolding method is to estimate the true distribution (in this case the true source energy spectrum), given the observed one and assuming the knowledge of the smearing matrix, which describes the migration effects among the energy bins as well as the efficiencies (Mazziotta 2009). The smearing matrix is evaluated using the Monte Carlo package \textit{Gleam}, a \textit{Geant4}  based simulation code of the instrument ~\cite{Atwood2009},
 and taking into account the pointing history of the source under investigation.
The unfolding analysis is performed selecting, from the initial data samples, events in an energy dependent RoI centered on the position of the source under investigation. The maximum allowed angular separation of the events selected from the source position is a decreasing function of energy that reproduces the behavior of the Point Spread Function (PSF) of the LAT. Events entering the LAT with a zenith angle larger than $105^\circ$ with respect to the Earth reference frame  and with an angle larger than $66.4^\circ$ with respect to the Z-axis in the instruments reference frame have been also excluded from this analysis.
The observed spectrum built from the data selected according the procedure described above includes the background contributions, that have to be subtracted before performing the unfolding.
In the examples shown in figures  ~\ref{fig:unf_3C454.3} and ~\ref{fig:unf_ASO0235} the background counts have been evaluated from real data, considering the photons in an annulus external to the analysis RoI and rescaling them in each observed energy bin for the ratio between solid angles and live times.
Once the source spectrum has been unfolded from the observed one, both statistical and systematic errors on the observed energy distribution can be easily propagated to the unfolded spectrum.
In figures ~\ref{fig:unf_3C454.3} and ~\ref{fig:unf_ASO0235} a comparison between the SEDs obtained  with the unfolding and the spectra obtained with \textit{gtlike}
is shown for the blazars 3C454.3 and ASO0235+164. The unfolded spectra are consistent with the ones obtained from \textit{gtlike}.

{\it Acknowledgments}
The \fermi LAT Collaboration acknowledges the generous
support of a number of agencies and institutes that have supported the $Fermi$
LAT Collaboration. These include the National Aeronautics and Space
Administration and the Department of Energy in the United States, the
Commissariat \`a l'Energie Atomique and the Centre National de la Recherche
Scientifique / Institut National de Physique Nucl\'eaire et de Physique des
Particules in France, the Agenzia Spaziale Italiana and the Istituto Nazionale
di Fisica Nucleare in Italy, the Ministry of Education, Culture, Sports,
Science and Technology (MEXT), High Energy Accelerator Research Organization
(KEK) and Japan Aerospace Exploration Agency (JAXA) in Japan, and the K.\ A.\
Wallenberg Foundation, the Swedish Research Council and the Swedish National
Space Board in Sweden.
Additional support for science analysis during the operations phase from the following agencies is also gratefully acknowledged: the
Istituto Nazionale di Astrofisica in Italy and the K.~A. Wallenberg Foundation in Sweden.
 This research is based also on observations with the
100-m telescope of the MPIfR (Max-Planck-Institut f\"ur Radioastronomie) at Effelsberg.
\mbox{RATAN-600} observations are supported in part by the Russian
Foundation for Basic Research (projects 01-02-16812 and 08-02-00545).
Part of this work was supported by Georgian National Science Foundation grant GNSF/ST-08/4-404
The Mid-infrared VISIR results are based on observations carried out at the European Southern Observatory
under programmes ID 078.B-0366, 079.B-0448 and 081.B-0404.
The Submillimeter Array is a joint project between the Smithsonian
Astrophysical Observatory and the Academia Sinica Institute of Astronomy and
Astrophysics and is funded by the Smithsonian Institution and the Academia
Sinica.
We acknowledge the use of data and software facilities from the ASI
Science Data Center (ASDC), managed by the Italian Space Agency (ASI).  Part of this work is based on
archival data and on bibliographic information obtained from the NASA/IPAC Extragalactic Database (NED) and
from the Astrophysics Data System (ADS).

{\it Facilities:} {Fermi LAT, \swift XRT, UVOT, OVRO, RATAN, GASP-WEBT, SMA, AGILE}.

\newpage



%
\newpage

\begin{deluxetable}{llcccccccccc}
\tablecolumns{11}
 \tablewidth{0pc}
 \tabletypesize{\tiny}
 \tablecaption{Sources list and basic parameters}
\rotate
\tablehead{
\colhead{LAT Name} & \colhead{Counterpart name(s)}  & \colhead{R.A.} & \colhead{Dec} & Z  &Radio flux & Vmag & X-ray flux \tablenotemark{\ast} & \colhead{\aro}& \colhead{\aox} & \colhead{\arx}& \colhead{\axg}\\
\colhead{0FGL}& & (J2000.0) & (J2000.0) & &6cm, mJy && erg cm$^{-2}$s$^{-1}$ & & &  &\\
 \colhead{(1)} &
 \colhead{(2)} &
 \colhead{(3)} &
 \colhead{(4)} &
 \colhead{(5)} &
 \colhead{(6)} &
 \colhead{(7)} &
 \colhead{(8)} &
 \colhead{(9)} &
 \colhead{(10)} &
  \colhead{(11)} &
 \colhead{(12)}
}
\startdata
J0017.4$-$0503&1RXS J001736.1-0/BZQJ0017-0512   &00 17 35.8&$-$05 12 41.6&0.227&274&18.2&8.03e-13&0.59&1.28&0.83&0.57\\
J0033.6$-$1921&1RXS J003334.6-1/BZBJ0033-1921   &00 33 34.3&$-$19 21 33.6&0.61&-&16.1&8.43e-12&0.23&1.08&0.53&1.04\\
J0050.5$-$0928&PKS0048-09      /BZBJ0050-0929   &00 50 41.2&$-$09 29 05.2&-&931&17.3&2.33e-12&0.52&1.46&0.84&0.72\\
J0051.1$-$0647&PKS0048-71      /BZQJ0051-0650   &00 51 08.2&$-$06 50 02.1&1.975&841&20.1&4.18e-13&0.85&1.02&0.93&0.59\\
J0100.2+0750&GB6J0100+0745             &01 00 20.8&+07 45 50.4&-&101&-&-&-&-&-&0\\
J0112.1+2247&GC 0109+224     /BZBJ0112+2244   &01 12 05.8&+22 44 38.7&-&304&15.3&1.93e-12&0.33&1.67&0.79&0.74\\
J0118.7$-$2139&PKS 0116-219    /BZQJ0118-2141   &01 18 57.1&$-$21 41 30.0&1.165&559&20&1.42e-13&0.79&1.23&0.97&0.47\\
J0120.5$-$2703&1Jy0118-272     /BZBJ0120-2701   &01 20 31.6&$-$27 01 24.7&0.557&1000&16.5&7.22e-13&0.49&1.66&0.91&0.74\\
\enddata
\tablenotetext{*}{0.1-2.4 keV band}
\label{tab:sample}
\end{deluxetable}

\newpage

\begin{deluxetable}{lcccccccc}
\tablecolumns{9}
\tablewidth{0pc}
\tabletypesize{\tiny}

\tablecaption{Results of Fermi LAT data analysis (Flux in units of $\pdflux$) }
\rotate
\tablehead{
\colhead{LAT Name} &
\colhead{Band 1}&
\colhead{Band 2}&
\colhead{Band 3}&
\colhead{Band 4}&
\colhead{Band 5}&
\colhead{Band 6}&
\colhead{Band 7}\\

\colhead{0FGL}&
\colhead{$ 100 \div 316.2 $ MeV}&
\colhead{$ 316.2 \div 1000 $ MeV} &
\colhead{$ 1000 \div 3162.3 $ MeV}&
\colhead{$ 3162.3 \div 10000 $ MeV} &
\colhead{$ 10000 \div 31623 $ MeV}&
\colhead{$ 31623 \div 100000 $ MeV}&
\colhead{$100000 \div 316230 $ MeV}\\
}

\startdata
J0033.6$-$1921 & --- & (4.45 $\pm$ 1.06) $10^{-12}$ & (1.23 $\pm$ 0.28) $10^{-12}$ & (1.16 $\pm$ 0.47) $10^{-13}$ & (1.67 $\pm$ 0.96) $10^{-14}$ & ---&--- \\
J0050.5$-$0928 & (3.83 $\pm$ 0.64) $10^{-10}$ & (3.83 $\pm$ 0.31) $10^{-11}$ & (3.04 $\pm$ 0.49) $10^{-12}$ & (2.02 $\pm$ 0.65) $10^{-13}$ & --- & --- & --- \\
J0137.1+4751    & (3.33 $\pm$ 0.72) $10^{-10}$ & (3.14 $\pm$ 0.44) $10^{-11}$ & (3.29 $\pm$ 0.50) $10^{-12}$ & (2.64 $\pm$ 0.72) $10^{-13}$ & --- & --- & --- \\
J0210.8$-$5100  & (9.20 $\pm$ 0.80) $10^{-10}$ & (7.23 $\pm$ 0.55) $10^{-11}$ & (5.99 $\pm$ 0.68) $10^{-12}$ & (2.87 $\pm$ 0.79) $10^{-13}$ & --- & --- & ---\\
J0222.6+4302  & (9.70 $\pm$ 0.47) $10^{-10}$  & (9.89 $\pm$ 0.49) $10^{-11}$ & (8.77 $\pm$ 0.72) $10^{-12}$ & (9.26 $\pm$ 1.35) $10^{-13}$ & (1.16 $\pm$ 0.25) $10^{-13}$ & (1.47 $\pm$ 0.49) $10^{-14}$ & (6.57 $\pm$ 6.14) $10^{-16}$\\
J0229.5$-$3640 & (6.18 $\pm$ 0.68) $10^{-10}$ & (3.26 $\pm$ 0.40) $10^{-11}$ & (2.08 $\pm$ 0.42) $10^{-12}$ & (7.60 $\pm$ 4.00) $10^{-14}$ &---&---&---\\
J0238.4+2855   & (3.01 $\pm$ 0.65) $10^{-10}$& (1.95 $\pm$ 0.37) $10^{-11}$ & (1.56 $\pm$ 0.38) $10^{-12}$ & (6.91 $\pm$ 4.04) $10^{-14}$ & --- & --- & --- \\
J0238.6+1636   & (2.63 $\pm$ 0.11) $10^{-9}$ & (2.64 $\pm$ 0.09) $10^{-10}$ & (2.74 $\pm$ 0.13) $10^{-11}$ & (2.27 $\pm$ 0.20) $10^{-12}$ & (9.74 $\pm$ 2.27) $10^{-14}$ & (2.12 $\pm$ 2.04) $10^{-15}$ &---\\
\enddata
\end{deluxetable}

\begin{deluxetable}{clcccccc}
\tablecolumns{8} \tablewidth{0pc} \tabletypesize{\tiny} \tablecaption{Results of \swift UVOT analysis}
\tablehead{
\colhead{Source Name} & \colhead{Observation date}& \colhead{V mag \tablenotemark{\ast}}  & \colhead{B mag\tablenotemark{\ast}} &
 \colhead{U mag\tablenotemark{\ast}} &\colhead{UVW1\tablenotemark{\ast}} &\colhead{UVM2\tablenotemark{\ast}}&\colhead{UVW2\tablenotemark{\ast}} \\
\colhead{(1)} & \colhead{(2)} & \colhead{(3)} & \colhead{(4)} & \colhead{(5)} & \colhead{(6)} & \colhead{(7)}& \colhead{(8)}
}
\startdata
J0033.5$-$1921   & 11-Nov-2008 &  16.30$\pm$0.04& 16.59$\pm$0.03 & 15.67$\pm$0.03 & 15.59$\pm$0.04 &  15.44$\pm$0.04  &   15.60$\pm$0.03    \\
-         & 15-Nov-2008 & 16.33$\pm$0.07& 16.60$\pm$0.05&  15.65$\pm$0.04&15.59$\pm$0.05  &15.44$\pm$0.05   &15.57$\pm$0.04      \\
J0050.5$-$0928   &  04-Jun-2008 & -&-&-&-&-& 15.074$\pm$0.033\\
J0137.1+4751   &  18-Nov-2008 &15.43$\pm$0.03& 15.84$\pm$0.02& 15.14$\pm$0.04&15.38$\pm$0.05 &15.18$\pm$0.04 &15.43$\pm$0.04\\
-         &  11-Feb-2008&  -&- &-  & - & -  &-      \\
-         &  22-Nov-2007 &15.007$\pm$0.02 & 15.554$\pm$0.019&14.966$\pm$0.023&13.354$\pm$0.033& 15.632$\pm$0.036& 15.772$\pm$0.033\\
J0210.7$-$5100   &  4-May-2005 &   -&- &-  & - & -  &-      \\
-         &  26-Dic-2008 &17.553$\pm$0.062&17.037$\pm$0.039&18.059$\pm$0.046& 16.784$\pm$0.043& 16.679$\pm$0.044&17.058$\pm$0.039\\
\enddata
\tablenotetext{\ast}{All magnitudes are corrected for Galactic extintion}
\label{tab:uvot_analysis}
\end{deluxetable}

%
\begin{deluxetable}{clcccccc}
\tablecolumns{8}
 \tablewidth{0pc}
 \tabletypesize{\tiny}
 \tablecaption{Results of \swift XRT data analysis}
\tablehead{ \colhead{Source Name} & \colhead{Observation date}& \colhead{XRT exposure (s)}
 & \colhead{X-ray flux (2-10keV) \tablenotemark{\ast} } &\colhead{$\Gamma_x /a$ \tablenotemark{\ast\ast}} & \colhead{$b$} &\colhead{d.o.f.}&\colhead{$\chi^2_{reduced}$} \\ \colhead{(1)} & \colhead{(2)} & \colhead{(3)}
& \colhead{(4)} & \colhead{(5)} & \colhead{(6)} & \colhead{(7)}& \colhead{(8)}
}
\startdata
J0033.5$-$1921  &11-Nov-2008 &2946.2&$2.32\times10^{-12}$&2.29$\pm$0.07 &-0.15$\pm$0.27&23&1.03\\
-        &15-Nov-2008 &3939.8&$1.86\times10^{-12}$&2.4$\pm$0.06& -0.29$\pm$0.24&31&1.3\\
J0050.5$-$0928  & 04-Jun-2008 &865&1.27$\times 10^{-12}$&-&-&-&-\\
J0137.1+4751  & 18-Nov-2008 & 6520& 2.4$\times 10^{-12}$&1.38$\pm$0.11&-&11&1.51\\
-        & 11-Feb-2008& 4674& 1.7$\times10^{-12}$ & 1.59$\pm$0.16&-&4&1.6\\
-        & 22-Nov-2007 & 4643& 2.2$\times 10^{-12}$&1.58$\pm$0.01&-& 11&1.19\\
J0210.7$-$5100  &  4-May-2005 &2102& 1.89$\times 10^{-12}$& 1.85$\pm$ 0.12&-& 6&0.53\\
-        &  26-Dic-2008& 4595& 1.28$\times 10^{-12}$ &1.66$\pm$0.13 & -& 7& 1.93\\
\enddata

\tablenotetext{*}{When the photon statistics was too poor to allow a reliable best fit, the flux was estimated converting the observed count rate assuming a power law
model with photon index of 1.9 and low energy absorption due to Galactic N$_H$.}
\tablenotetext{**}{This column gives the power law photon index  $\gamma$ when a simple power law model could be used. In the case where the Log Parabola model was used
this column gives the $a$ parameter which represents the photon index at 1 keV.}
\label{tab:xrt_analysis}
\end{deluxetable}

\begin{deluxetable}{cccc}
\tablecolumns{4}
\tablewidth{0pc}
\tabletypesize{\tiny}
\tablecaption{Effelsberg radio data}
%
\tablehead{
\colhead{Source Name} &
\colhead{Observation date} &
\colhead{Frequency (GHz)} &
\colhead{Flux Density (Jy)} \\
\colhead{0FGL} &
\colhead{} &
\colhead{} &
\colhead{} \\
\colhead{(1)} &
\colhead{(2)} &
\colhead{(3)} &
\colhead{(4)}
}
\startdata
J0222.6+4302 &08-08-2008 &4.85 &1.434\\
 &08-08-2008 &8.35 &1.247\\
 &08-08-2008 &10.45 &1.159\\
 &08-08-2008 &14.60 &1.087\\
 &08-08-2008 &23.05 &1.086\\

J0238.4+2855 &17-09-2008 &2.64 &3.221\\
 &17-09-2008 &4.85 &3.609\\
 &17-09-2008 &8.35 &3.614\\
\enddata
 \label{tab:effelsberg}
\end{deluxetable}

\begin{deluxetable}{ccccc}
 \tabletypesize{\tiny}
  \tablecaption{OVRO radio data (15~Ghz) \label{table:ovro_data}}
  \tablehead{
    \colhead{Name} & \colhead{Date of Max.} & \colhead{Max. Flux Density} & \colhead{Date of Min.} & \colhead{Min. Flux Density} \\
    \colhead{0FGL} & & \colhead{(Jy)} & & \colhead{(Jy)} \\
    \colhead{(1)} & \colhead{(2)} & \colhead{(3)} & \colhead{(4)} & \colhead{(5)} }
  \tablewidth{0pt}
  \startdata
  J0050.5-0928 & 23-10-2008 & $1.079\pm 0.057$ & 19-12-2008 & $1.761\pm 0.092$\\
  J0137.1+4751 & 16-08-2008 & $3.05\pm 0.16$   & 24-10-2008 & $3.73\pm 0.20$\\
  J0222.6+4302 & 20-12-2008 & $0.861\pm 0.045$ & 16-08-2008 & $1.070\pm 0.056$\\
  J0238.4+2855 & 22-10-2008 & $2.99\pm 0.16$   & 30-08-2008 & $3.34\pm 0.18$\\
  J0423.1-0112 & 04-11-2008 & $3.90\pm 0.20$   & 19-11-2008 & $4.40\pm 0.23$\\
  J0531.0+1331 & 23-08-2008 & $2.87\pm 0.15$   & 05-11-2008 & $3.26\pm 0.17$\\
  J0722.0+7120 & 06-12-2008 & $1.511\pm 0.079$ & 26-08-2008 & $3.16\pm 0.17$\\
  J0855.4+2009 & 12-11-2008 & $2.37\pm 0.13$   & 24-12-2008 & $3.21\pm 0.17$\\
\enddata
 \label{tab:OVRO}
\end{deluxetable}

\begin{deluxetable}{lcc}
\tablecolumns{3}
\tablewidth{0pc}
\tabletypesize{\tiny}
\tablecaption{RATAN-600 flux density measurements in September 10 -- October 3, 2008}
\tablehead{
\colhead{Source Name} &
\colhead{Central frequency} &
\colhead{Flux Density} \\
\colhead{0FGL} &
\colhead{GHz} &
\colhead{Jy} \\
\colhead{(1)} &
\colhead{(2)} &
\colhead{(3)}
}
\startdata
0FGL J0050.5-0928       &        21.7   &         0.98  $\pm$  0.06 \\
       &        11.2   &         1.10  $\pm$  0.04 \\
       &         7.7   &         1.09  $\pm$  0.03 \\
       &         4.8   &         0.97  $\pm$  0.01 \\
       &         2.3   &         0.95  $\pm$  0.11 \\
       &         1.0   &         0.91  $\pm$  0.15 \\
0FGL J0137.1+4751       &        21.7   &         3.43  $\pm$  0.06 \\
       &        11.2   &         3.75  $\pm$  0.07 \\
\enddata
 \label{tab:RATAN}
\end{deluxetable}

\begin{deluxetable}{lrrrrr}
\tablecolumns{6}
 \tablewidth{0pc}
 \tabletypesize{\tiny}
 \tablecaption{GASP-WEBT data}
\tablehead{
\colhead{Name}  & \colhead{ Log ($\nu$)} & \colhead{Log ($\nu F_\nu$)} & \colhead{ Log ($\nu F_\nu$)} & \colhead{Log ($\nu F_\nu$) } & \colhead{$N_{\rm data}$}\\
                             & \colhead{} & \colhead{(average)}   & \colhead{(max)}  & \colhead{min}  & \\
 \colhead{(1)} &
 \colhead{(2)} &
 \colhead{(3)} &
 \colhead{(4)} &
 \colhead{(5)} &
 \colhead{(6)}
}
\startdata
J0222.6+4302   &  14.670  & $-$10.446  & $-$10.246  & $-$10.668 & 521 \\
     &  14.390  & $-$10.467  & $-$10.300  & $-$10.638 &  21 \\
      &  14.265  & $-$10.477  & $-$10.324  & $-$10.643 &  21 \\
     &  14.136  & $-$10.496  & $-$10.365  & $-$10.661 &  21 \\
       &  10.633  & $-$12.394  & $-$12.371  & $-$12.417 &   2 \\
   &  10.342  & $-$12.694  & $-$12.623  & $-$12.778 &   2 \\
      &  10.161  & $-$12.850  & $-$12.820  & $-$12.900 &   3 \\
      &   9.903  & $-$13.011  & $-$13.007  & $-$13.017 &   4 \\
\enddata
\label{tab:webt}
\end{deluxetable}

\begin{deluxetable}{cccccc}
\tablecolumns{6}
 \tablewidth{0pc}
 \tabletypesize{\tiny}
 \tablecaption{Mid-infrared photometry of blazars obtained with the VISIR instrument on VLT/UT3}
\tablehead{
\colhead{Source Name}  & \colhead{Observation date} & \colhead{UT} & \colhead{Filter}  & \colhead{Flux}  & \colhead{Error} \\
 0FGL&  & &  & \colhead{mJy}  & \colhead{mJy} \\
 \colhead{(1)} &
 \colhead{(2)} &
 \colhead{(3)} &
 \colhead{(4)} &
 \colhead{(5)} &
 \colhead{(6)}  \\
}
\startdata
J0238.6+1636 &  2008-07-06& 10:08&  PAH1&   82.44&   2.02   \\
    &   2008-07-07& 09:10&  PAH1&   108.02&  2.03\\
    &   2008-07-07& 09:36&  PAH2&   155.15&  3.59\\
    &   2008-07-07& 09:48&  Q2& 180.94&  9.59\\
J0457.0$-$2325& 2006-12-01& 12:00&  PAH2& 67.7&    1.8 \\
        &       2006-12-01&      12:00&   Q2&   73.0&    5.3 \\
J1256.1$-$0547  &   2007-07-15& 00:48&  PAH1&   292.09&  2.63   \\
    &   2007-07-15& 01:13&  Q2& 335.81&  11.66\\
\enddata
\label{table:MIR}
\end{deluxetable}

\begin{deluxetable}{lccc}
\tablecolumns{4}
 \tablewidth{0pc}
 \tabletypesize{\tiny}
 \tablecaption{MIPS (Spitzer) data}
\tablehead{
\colhead{Source Name}  & \colhead{Observation date} & \colhead{Log ($\nu$)} & \colhead{ Log ($\nu F_\nu$)} \\
 \colhead{(1)} &
 \colhead{(2)} &
 \colhead{(3)} &
 \colhead{(4)}  \\
}
\startdata
J1653.9+3946&      11-04-2004   & 13.097     & -11.080 \\
J1229.1+0202     &     15-01-2007    &13.097      &-10.121\\
          &15-01-2007    &12.632      &-10.391\\
          &15-01-2007    &12.273      &-10.680\\
          &11-07-2007    &13.097      &-10.160\\
          &11-07-2007    &12.632      &-10.499\\
J2202.4+4217    &      22-07-2006   & 13.097    &  -10.343\\
          &22-07-2006    &12.632      &-10.503\\
\enddata
\label{tab:spitzer}
\end{deluxetable}

\begin{deluxetable}{lcc}
\tablecolumns{4}
 \tablewidth{0pc}
 \tabletypesize{\tiny}
 \tablecaption{AGILE mean fluxes
 \label{tab:agilecomp}}
\tablehead{
\colhead{Name}  & \colhead{AGILE flux} & \colhead{AGILE Mean Exposure}  \\
                             & \colhead{$10^{-8}$ ph cm$^{-2}$s$^{-1}$} & \colhead{$10^8$ cm$^{2}$ s}    \\
 \colhead{(1)} &
 \colhead{(2)} &
 \colhead{(3)}
}

\startdata
0FGL J0538.8$-$4403   & 43+/-10 & 0.81    \\
0FGL J0722.0+7120       & 69+/-9 &  1.39     \\
0FGL J1104.5+3811                 &  42+/-13 & 0.51   \\
0FGL J1221.7+2814 & 38+/-11 &  0.50    \\
0FGL J1229.1+0202                   &  24+/-6 &  1.98  \\
0FGL J1256.1$-$0547              &  65+/-9 &  1.98    \\
0FGL J0538.8$-$4403  & 43+/-10& 0.81     \\
0FGL J1849.4+6706               &  20+/-4 &  5.52     \\
0FGL J2254.0+1609          & 200+/-14 &  1.16  \\
\enddata

\label{tab:agilecomp}
\end{deluxetable}

\begin{deluxetable}{lccc}
\tablecolumns{4}
 \tablewidth{0pc}
 \tabletypesize{\tiny}
 \tablecaption{TeV literature data}
%
 \tablehead{
 \colhead{Source Name} &
 \colhead{Periodo of observations} &
 \colhead{Instrument}&
 \colhead{reference } \\
 \colhead{0FGL} &
 \colhead{} &
 \colhead{} &
 \colhead{} \\
 \colhead{(1)} &
 \colhead{(2)} &
 \colhead{(3)} &
 \colhead{(4)}
 }
\startdata
J0222.6+4302       & 2007-2008                  & VERITAS   & \cite{Acciari09a}        \\
J0722.0+7120     & Apr 22-24 2008             & MAGIC       & \cite{Teshima08}                         \\
J1104.5+3811        & Nov 2004-Apr 2005          & MAGIC     & \cite{Albert07a}         \\
J1221.7+2814        & Jan-Apr 2008               & VERITAS   & \cite{Acciari09b}  \\
J1256.1$-$0547       & Jan-Apr 2006               & MAGIC     &  \cite{Errando08}     \\
J1653.9+3946       & May-Jul 2005               & MAGIC     & \cite{Albert07b}          \\
J2000.2+6506     & Sept-Oct 2004              & MAGIC     &  \cite{Albert06}       \\
J2158.8$-$3014    & Jul, Oct 2002, Jul-Sep 2003& HESS      &  \cite{Aharonian09}       \\
J2202.4+4217   & Aug-Dec 2005,  Jul-Sep 2006& MAGIC     &  \cite{Albert07b}         \\
\enddata
 \label{tab:TeV}
\end{deluxetable}

%
\begin{deluxetable}{lcccccccccl}
\tablecolumns{11}
 \tablewidth{0pc}
 \tabletypesize{\tiny}
 \tablecaption{Blazar SED parameters}
 \rotate
%
 \tablehead{
 \colhead{Name} &
  \colhead{SED} &
 \colhead{$\alpha_{r}$} &
 \colhead{Log(\nupS) }&
 \colhead{Log(\nupS F(\nupS))  } &
 \colhead{Log(\nupIC) }&
 \colhead{Log(\nupIC F(\nupIC))}&
 \colhead{Log($\gamma_{peak}^{SSC}$) } &
 \colhead{Compton}&
 \colhead{SED} &
 \colhead{Optical} \\
 \colhead{0FGL} &
 \colhead{available} &
  \colhead{} &
 \colhead{} &
 \colhead{} &
 \colhead{} &
 \colhead{} &
 \colhead{} &
 \colhead{dominance} &
  \colhead{classification} &
 \colhead{classification} \\
 \colhead{(1)} &
 \colhead{(2)} &
 \colhead{(3)\tablenotemark{\ast} } &
 \colhead{(4)\tablenotemark{\ast \ast}} &
 \colhead{(5)\tablenotemark{\ast \ast }} &
 \colhead{(6)\tablenotemark{\ast \ast \ast } } &
 \colhead{(7)}&
 \colhead{(8)\tablenotemark{\ast \ast \ast \ast}} &
 \colhead{(9)}&
 \colhead{(10)}&
 \colhead{(11)}
 }
\startdata
J0017.4$-$0503& - &0.127&$--$/13.6&$--$/-11.4&-/20.7&-&3.4&-&LSP&FSRQ \\
J0033.6$-$1921& Yes &0&16.1/16.3&-11.1/-11.2&24.3/24.8&-11.1&4&1&HSP&BL Lac \\
J0050.5$-$0928& Yes &0.205&14.3/14.4&-10.8/-10.6&22.4/23&-10.6&4&1.8&ISP&BL Lac \\
J0051.1$-$0647& - &-0.103&$--$/12.8&$--$/-11.4&-/22.7&-&4.9&-&LSP&FSRQ \\
J0100.2+0750& - &-0.076&$--$/$--$&$--$/$--$&-/24.4&-&-&-&-&Unidentified \\
J0112.1+2247& - &0.121&$--$/14.6&$--$/-10.8&-/23.1&-&4.1&-&ISP&BL Lac \\
J0118.7$-$2139& - &0.089&$--$/13&$--$/-11.5&-/22.3&-&4.5&-&LSP&FSRQ \\
J0120.5$-$2703& - &-0.114&$--$/13.9&$--$/-10.8&-/23.6&-&4.7&-&LSP&BL Lac \\
\enddata
\tablenotetext{*}{The radio power-law spectral index $\alpha_r$ is evaluated in the range 1-100 GHz.}  \\
\tablenotetext{**}{The value to the left is estimated directly from SED, while the value reported to the right has been estimated from \aox - \aro }
\tablenotetext{***}{The value to the left is estimated directly from SED, while the value reported to the right has been estimated from eq. \ref{nupIC}}
\tablenotetext{****}{Calculated assuming a simple SSC emission mechanism, i.e.  $\gamma_{peak}^{SSC}$ = $\sqrt{3/4 \cdot \nu_{peak}^{IC}/\nu_{peak}^S}$}

 \label{tab:SED_parameters}
\end{deluxetable}


\clearpage


\clearpage

\clearpage

\clearpage

\clearpage

\clearpage

\clearpage

\clearpage

\clearpage

\begin{figure}
\epsscale{.40}
\center{
\includegraphics[height=14.0cm,angle=-90]{nu_peakVSnu_peak.ps}\\
\includegraphics[height=14.0cm,angle=-90]{nufnuvsnufnu.ps} }
\caption{The synchrotron peak frequency (\nupS, top panel) and its corresponding peak flux, \nupS F(\nupS), bottom panel) value estimated from the SEDs of  Figs.  \ref{fig:sed_first} through \ref{fig:sed_last} is plotted against the value estimated using the method based on \aox and \aro values (see text for details). Despite the fact that \aox and \aro are based on non-simultaneous literature data, the scatter around the solid lines, representing perfect match, is $\approx$ 0.6 and  $\approx$ 0.4 in log space for \nupS and
\nupS F(\nupS) respectively.}
\label{fig:nupeaks}
\end{figure}

\begin{figure}
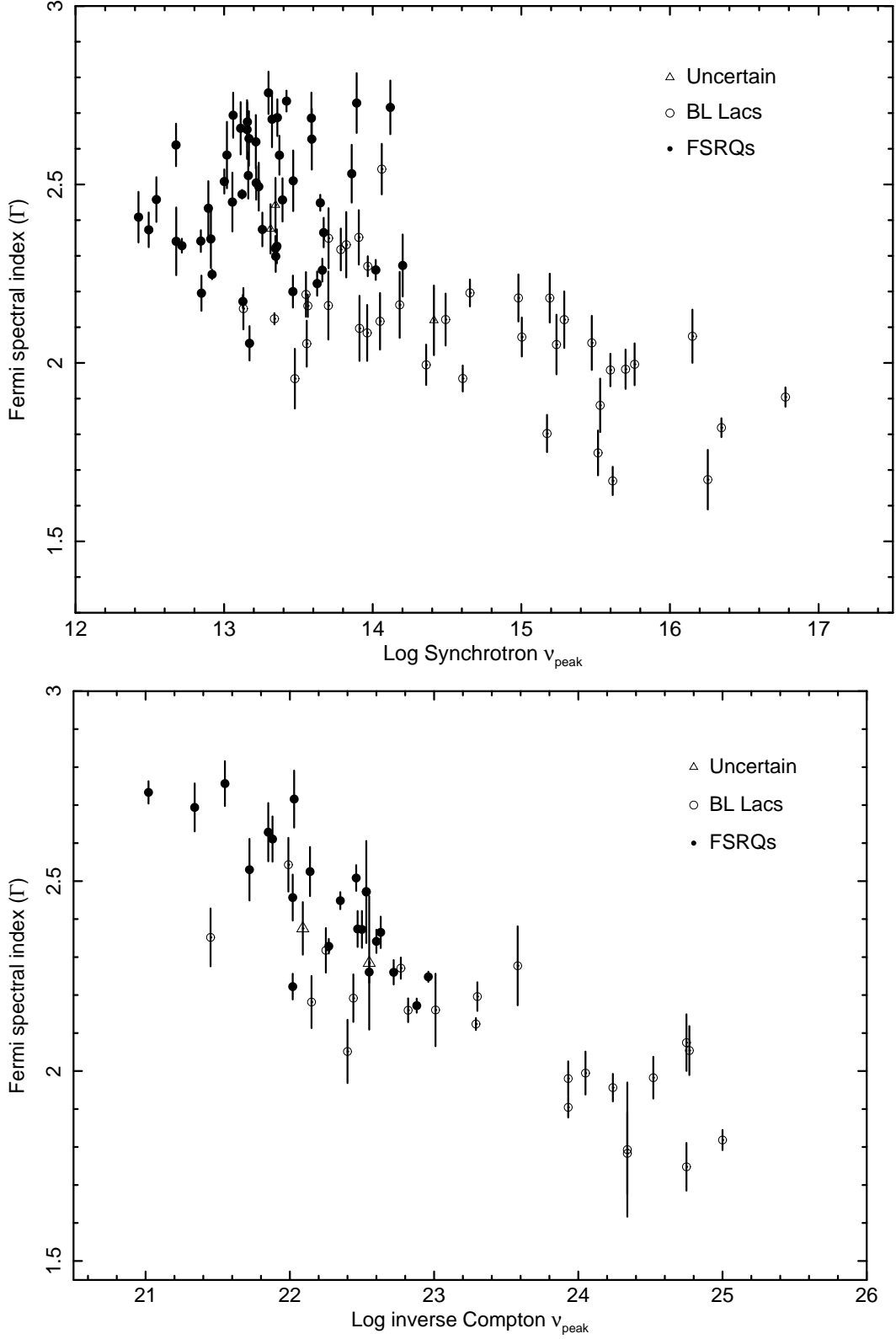

\epsscale{.40}
\includegraphics[height=14.cm,angle=-90]{slopeVSnup.ps} \\
\includegraphics[height=14.cm,angle=-90]{slopeVSnup_ic.ps}
\caption{The \gr ~power law photon spectral index ($\Gamma$) is plotted against the log of synchrotron peak energy (top panel) and the Log of inverse Compton peak energy (bottom panel). A clear correlation is present in both cases. Note that BL Lacs behave differently than FSRQs spanning a wider range of both \nupS and spectral slopes.}
 \label{fig:slopeVSnup}
\end{figure}

\begin{figure}
\epsscale{.40}
\includegraphics[height=20.cm,angle=-90]{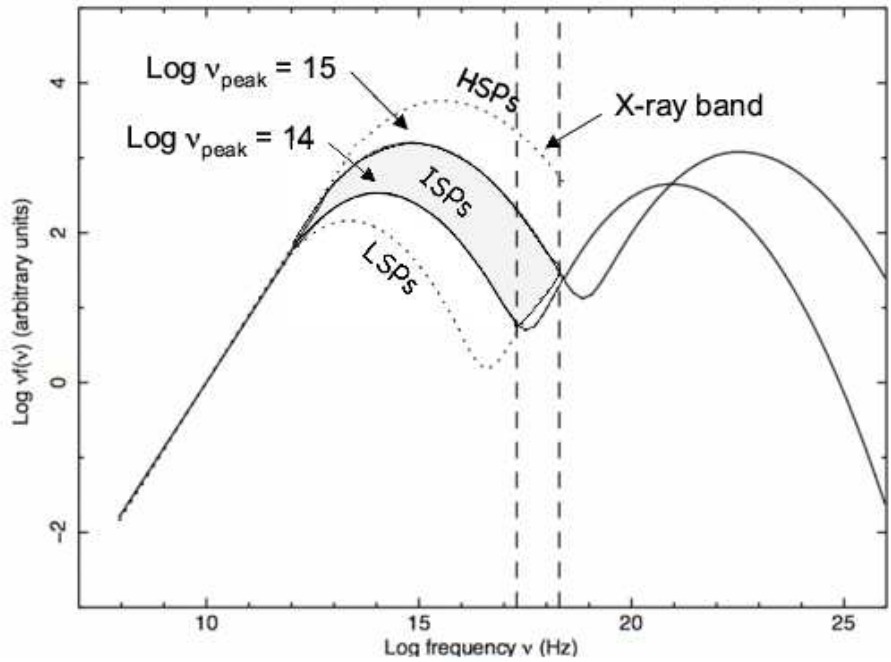}
\caption{The definition of different blazar types based on the peak of the synchrotron component  (\nupS) in their SED. Low Synchrotron Peaked blazars, or LSP are those
where \nupS  is located at frequencies lower then $10^{14}$ Hz (e.g., lower dotted line), for Intermediate Synchrotron Peaked sources, or IPB,  $10^{14}$ Hz $< \nu_{peak_S} <  10^{15}$ Hz, (SEDs with peak within the grey area) while for  High Synchrotron Peaked blazars, or HPS,  \nupS  $> 10^{15}$ Hz (e.g., upper dotted line).}
\label{fig:iblhbl}
\end{figure}

\begin{figure}
\epsscale{.40}
\includegraphics[height=16.cm,angle=-90]{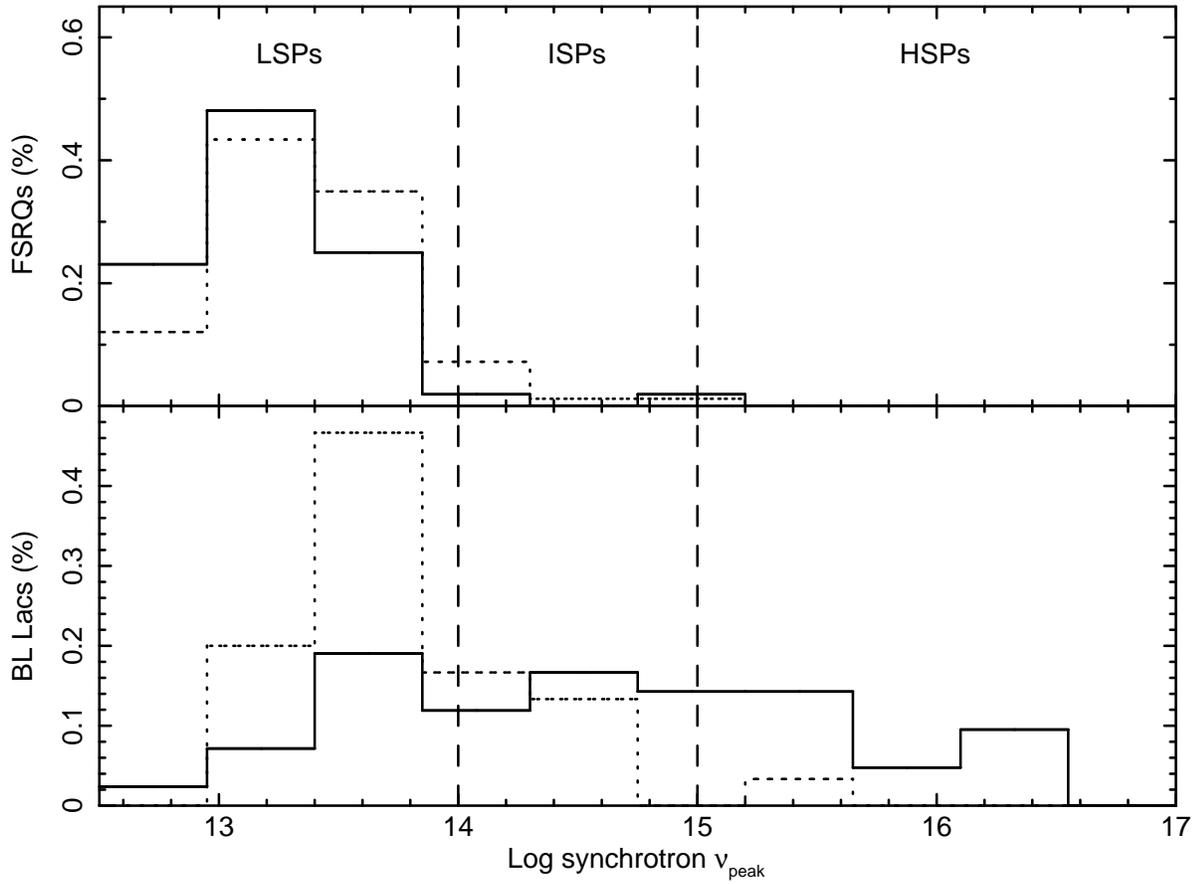}
\caption{The distribution of synchrotron peak energy for the sample of LBAS FSRQ (solid line, top panel) and BL Lacs (solid line, bottom panel)
compared to that of microwave selected blazars listed in the WMAP foreground sources catalog (dotted histograms).}
\label{fig:nupBbzBzq}
\end{figure}

\begin{figure}
\epsscale{.40}
\includegraphics[height=16.cm,angle=-90]{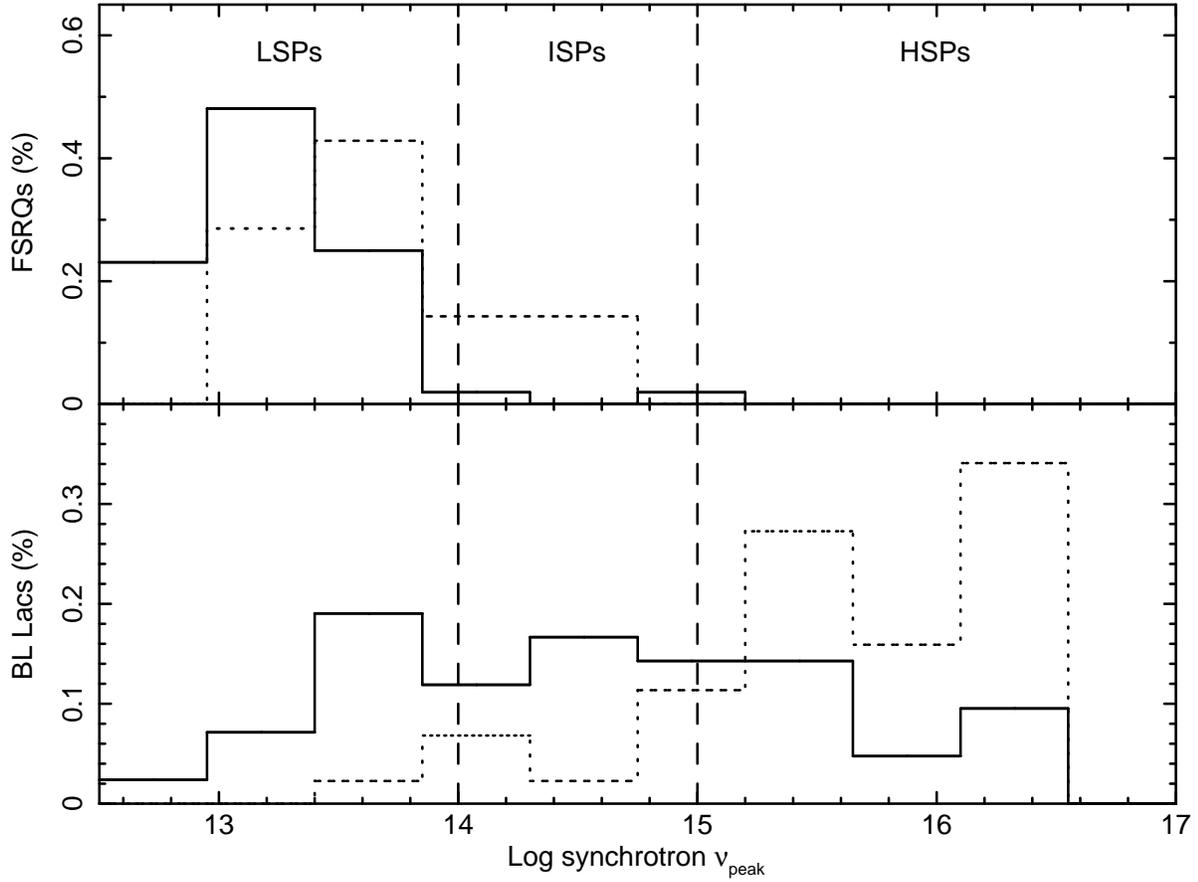}
\caption{The distribution of synchrotron peak energy for the sample of LBAS FSRQ (solid line, top panel) and BL Lacs (solid line, bottom panel) compared
to that of the sample of X-ray selected blazars of the {\it Einstein} Extended Medium Sensitivity Survey (EMSS, dotted histograms).}
\label{fig:nupBbzBzq_emss}
\end{figure}

\begin{figure}
\epsscale{.40}
\includegraphics[height=16.cm,angle=-90]{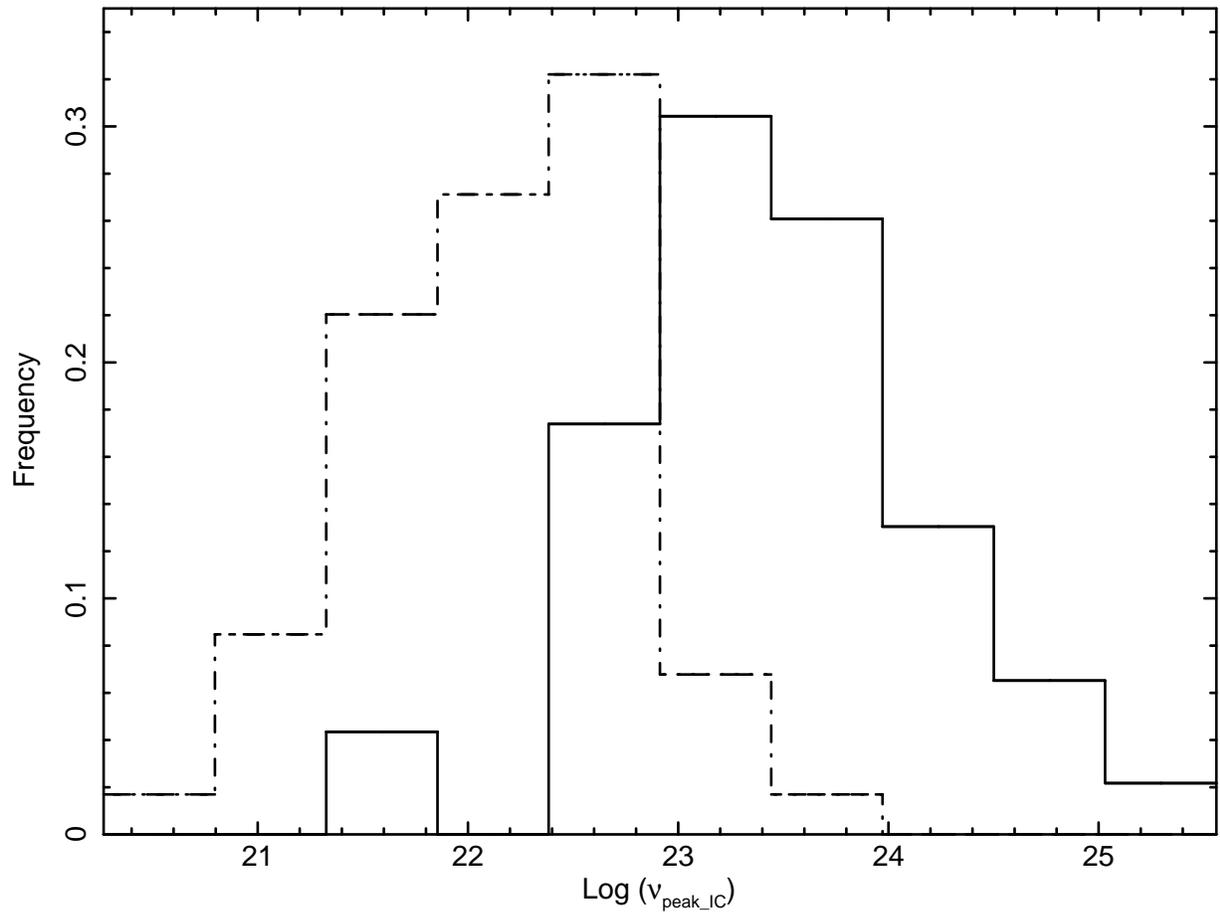}
\caption{The distribution of inverse Compton peak frequency for the sample of LBAS FSRQ (dot-dashed line) and BL Lacs (solid line).}
\label{fig:nupIC}
\end{figure}

\clearpage

\begin{figure}
\epsscale{.40}

\includegraphics[height=16.cm]{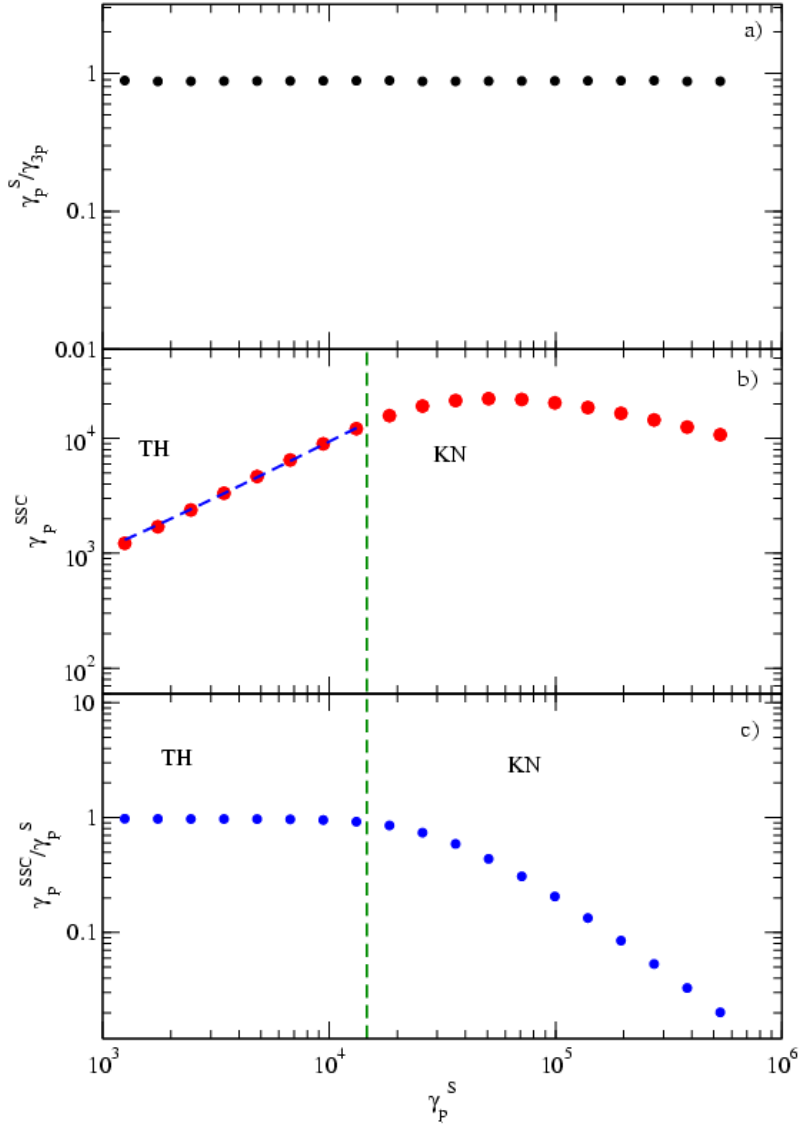}
\caption{Estimate of $\gamma_{peak}^{S}$ and $\gamma_{peak}^{SSC}$ for numerically computed SEDs in
the case of a SSC model and using as electron distribution  a log-parabola
$n(\gamma)=K\cdot10^{~r~Log(\gamma/\gamma_{peak})^2}$ with $\gamma_{peak}$ ranging between
100 and $6\cdot10^5$, and the curvature parameter $r=0.4$.
The other model parameters are: source size $R=10^{15}$ cm, a magnetic
field $B=0.1$ G, a beaming factor $\delta=10$, and an electron density N=1 $e^-/cm^{3}$
(N=$\int n(\gamma)d\gamma$). From to top to bottom: a) the ratio of $\gamma_{peak}^S$
to $\gamma_{3p}$ as a function of $\gamma_{peak}^S$. b) $\gamma_{peak}^{SSC}$ as a function
of   $\gamma_{peak}^S$, the transition from the TH trend (blue dashed line) to the KN
region is evident for $\gamma>2\cdot 10^4$. c) The ratio of $\gamma_{peak}^{SSC}$
to $\gamma_{peak}^S$, also in this case, above the TH region (vertical dashed green line)
is evident the effect of the KN suppression, $\gamma_{peak}^{SSC}$ gets to increasingly
underestimate $\gamma_{peak}^S$ as $\gamma_{peak}$ is increasing.
}
\label{fig:gammapeak}
\end{figure}

\begin{figure}
\epsscale{.40}
\includegraphics[height=16.cm,angle=-90]{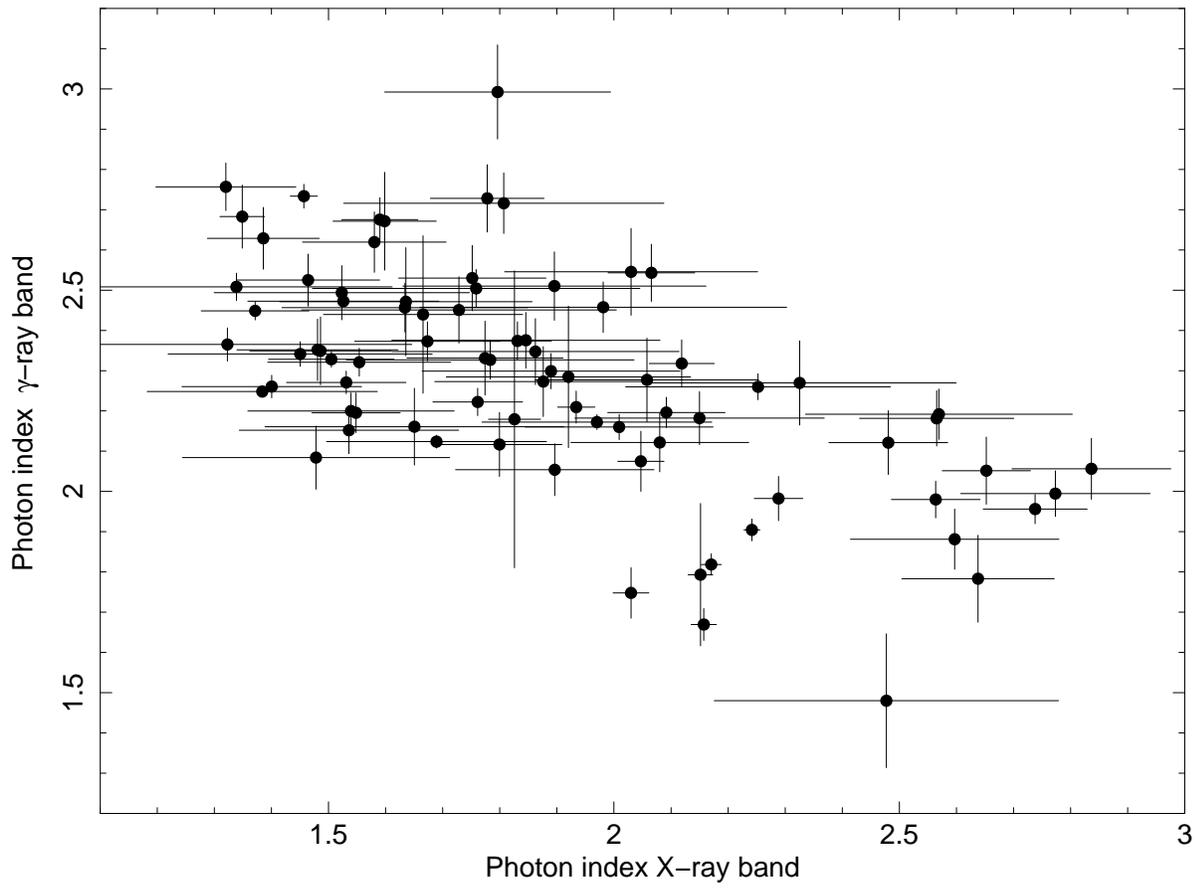}
\caption{The power law spectral slope in the \gr ~band is plotted against the spectral slope in the X-ray band. A clear correlation is present.}
\label{fig:slopexslopeg}
\end{figure}

\begin{figure}
\epsscale{.40}
\includegraphics[height=16.cm,angle=-90]{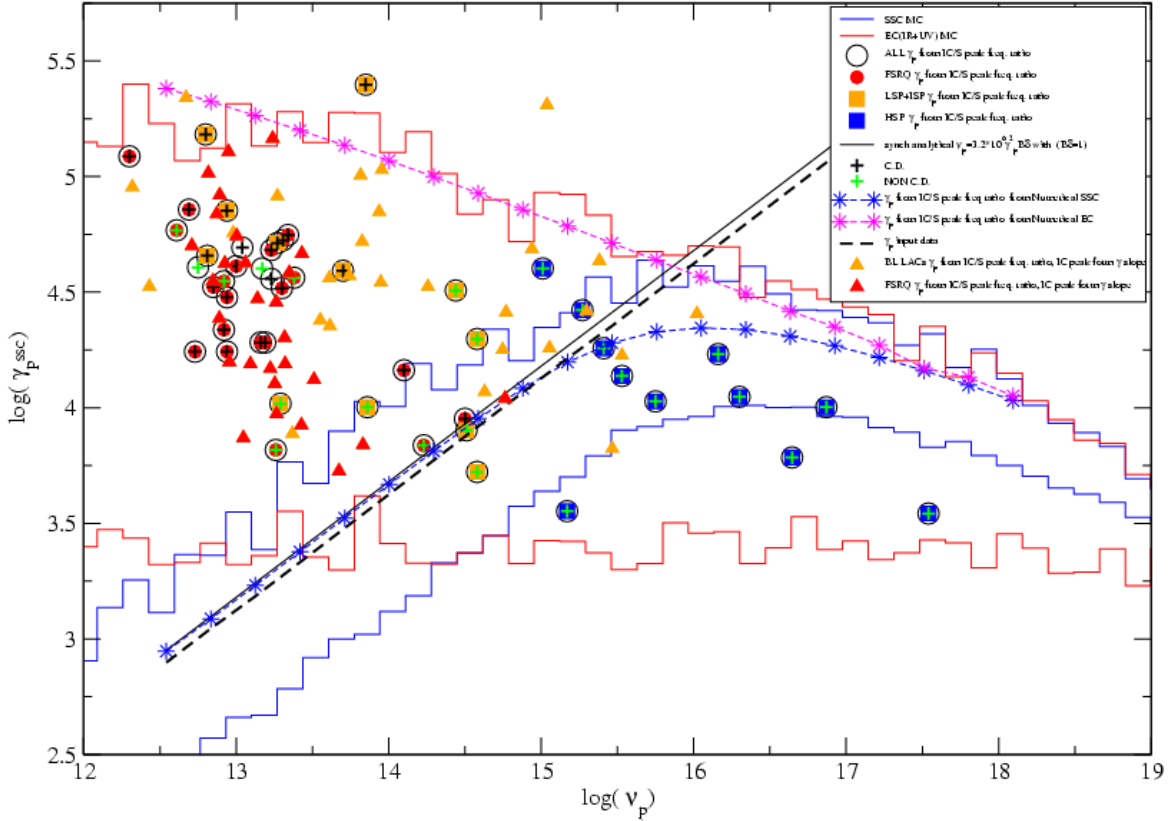}
\caption{ $\gamma_{peak}^{SSC}$ obtained by Eq. \ref{nupSSC}  for
the objects reported in Table 10. blue solid boxes represent HPB
objects, orange solid boxes represent IPBs/LPBs objects and red
solid circles represent FSRQs. The black solid line represents
$\nu_{peak}^S$ estimated by Eq. (\ref{nupS}) for both the ERC and
the SSC numerical SEDs, the blue solid line represents
$\gamma_{peak}^{SSC}$ estimated from Eq. (\ref{nupSSC}) applied to
numerically computed the SSC SEDs, and the solid purple line
represents the same for the case of ERC emission. The true value
of the simulation  is represented by the black dashed line.
Parameters of the model are given in Sect. 9. The blue and red
contours delimit the area covered by the estimate of
$\gamma_{peak}^{SSC}$ for the case of SSC and ERC models
respectively and for a Monte Carlo simulation with  values of
$\delta$ ranging between 10 and 15,  $B$  ranging between  0.01
and 1G and $T$ ranging between 10 and 10$^{4.5}$ K. }
\label{fig:nupvgammael}
\end{figure}

\begin{figure}
\begin{center}
\includegraphics[angle=0,scale=0.50]{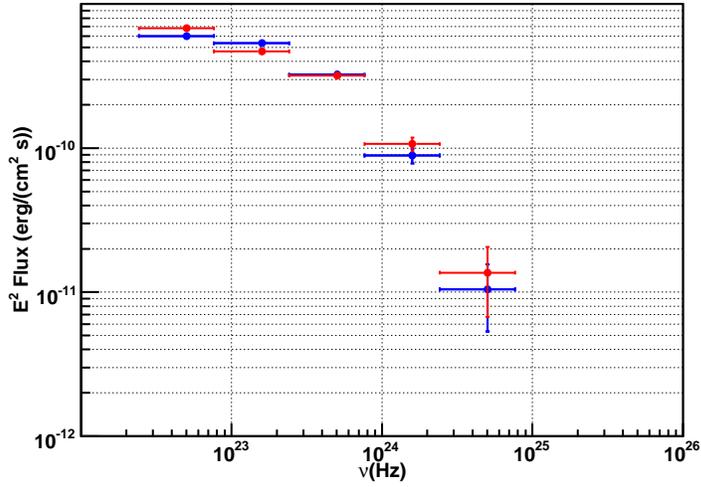}
\end{center}
\caption{Comparison of the SEDs of the blazar 3C454.3 obtained with
the maximum likelihood approach (red points) and with the unfolding
technique (blue points). }
\label{fig:unf_3C454.3}
\end{figure}

\begin{figure}
\begin{center}
\includegraphics[angle=0,scale=0.50]{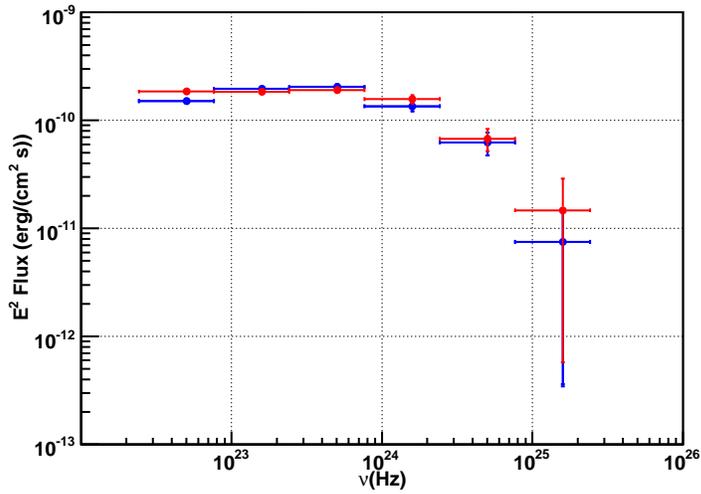}
\end{center}
\caption{Comparison of the SEDs of the blazar ASO0235+164 obtained with
the maximum likelihood approach (red points) and with the unfolding
technique (blue points). The horizontal error bars represent the bin
width.}
\label{fig:unf_ASO0235}
\end{figure}


\begin{thebibliography}{}

\bibitem[Abdo \etal ~2009a]{AbdoLATpaper} Abdo, A.A. et al, 2009a, \apjs,183,  46
\bibitem[Abdo \etal ~2009b]{AbdoAGNpaper} Abdo, A.A. et al, 2009b,  \apj, 700, 597
\bibitem[Abdo \etal ~2009c]{lsi61} Abdo, A.A. et al., 2009c,  \apj ~letters, submitted
\bibitem[Abdo \etal ~2009d]{vela1} Abdo, A.A. \ et al. \ 2009d \apj ~696, 934 arXiv:0812.2960
\bibitem[Abdo \etal ~2009e]{Abdo454.3}  Abdo A.A. \ et al. \ 2009e \apj, 699, 861
\bibitem[Abdo \etal ~2009f]{AbdoSpectralPaper} Abdo, A.A. et al., 2009f,  \apj ~submitted
\bibitem[Acciari \etal  ~2009a]{Acciari09a} Acciari, V.A. \ et al. \ 2009, \apj ~693, 104L
\bibitem[Acciari \etal ~2009b]{Acciari09b} Acciari, V.A., 2009, \apj in press, ArXiv:0808.0889
\bibitem[Aharonian \etal ~2009]{Aharonian09} Aharonian, F. \ et al. \ 2009, \apj 696L, 150
\bibitem[Albert \etal ~2006]{Albert06} Albert, J. \ et al. \ 2006, \apj  639, 761
\bibitem[Albert \etal ~2007a]{Albert07a} Albert, J. \ et al. \ 2007, \apj 669, 862
\bibitem[Albert \etal ~2007b]{Albert07b}  Albert, J. \ et al. \ 2007, \apj 666, 17L
\bibitem[Ajello \etal ~2008]{ajello08} Ajello, M., \ et al.\ 2008, \apj, 673, 96
\bibitem[Ajello \etal ~2009]{ajello09} Ajello, M., \ et al.\ 2009, arXiv:0905.0472
\bibitem[Angelakis \etal ~2008]{Angelakis08} Angelakis, E., Fuhrmann, L., Marchili, N., Krichbaum, T.~P.,\& Zensus, J.~A. 2008, arXiv:0809.3912
\bibitem[Atwood \etal ~2009]{Atwood2009} Atwood, W.~B., et al. 2009, \apj, 697, 1071
\bibitem[Baars et al. 1977]{Baars77} Baars, J.~W.~M., Genzel,R., Pauliny-Toth, I.~I.~K., \& Witzel, A. 1977, \aap, 61, 99
\bibitem[Barthelmy \etal ~2005]{Barthelmy05} Barthelmy, S., Barbier, L.~M., Cummings, J., {et~al.} 2005, SSRv., 120, 95
\bibitem[Bessel ~1998]{bessel98} Bessel, \aap, 333, 231
\bibitem[Blandford \& Rees 1978]{bla78} Blandford, R.D. \& Rees, M.J. 1978, in Pittsburg
Conference on BL Lac Objects, Ed. A.M. Wolfe, Pittsburgh, University of Pittsburgh press, p. 328
\bibitem[Burrows \etal ~2005]{Burrows05} Burrows, D., Hill, J.~E., Nousek, J.~A., {et~al.} 2005, SSRv., 120, 165
\bibitem[Cardelli \etal ~1989]{cardelli89} Cardelli, J.A., Clayton, G.C. and Mathis, J.S., 1989, \apj, 345, 245
\bibitem[Cassandjian \& Grenier 2008]{EGR} Cassandjian J-M \& Grenier I A 2008, \aap, 489, 849
\bibitem[Colafrancesco \& Giommi 2006]{colaf06} Colafrancesco, S. \& Giommi, P., 2006, ChJAS, 6, 47
\bibitem[Costamante \& Ghisellini 2002]{costaghis02} Costamante, L. \&  Ghisellini, G., 2002, \aap,  384, 56
\bibitem[Costamante \etal 2007]{Costamante07} Costamante, L., Aharonian, F. \& Khangulyan D., 2007, AIPC, 921,157
\bibitem[Dermer \etal ~2002]{Dermer02} Dermer, C. D., \& Schlickeiser, R. 2002, \apj, 575, 667
\bibitem[Donnarumma \etal ~2009]{donnarumma09} Donnarumma, I., \ et al. \   2009, \apj,  691, 13L
\bibitem[Errando 2008]{Errando08} Errando, M. 2008, AIP Conf. Proc., 1085, 423
\bibitem[Fuhrmann \etal ~2007]{Fuhrmann07} Fuhrmann, L., Zensus, J.~A., Krichbaum, T.~P., Angelakis, E.,
\& Readhead, A.~C.~S.\  ~2007, The First GLAST Symposium, 921, 249
\bibitem[D\'Agostini ~1995]{dagostini95}G. D\'Agostini, Nuclear Instruments \& Methods A362 (1995), 487
\bibitem[Fitzpatrick ~1999]{Fitzpatrick1999} Fitzpatrick, N. 1999 \pasp, 111, 63
\bibitem[Fossati \etal ~1998]{fossati98}Fossati, G., Maraschi, L., Celotti, A., Comastri, A., Ghisellini, G., 1998 \mnras, 299, 433
\bibitem[Fuhrmann \etal  ~2008]{Fuhrmann08} Fuhrmann, L.\ et al.\ 2008, \aap, 490, 1019
\bibitem[Gehrels \etal ~2004]{Gehrels04} Gehrels, N. et al. 2004, \apj, 611, 1005
\bibitem[Ghisellini  \& Maraschi 1989]{GhiselliniMaraschi89} Ghisellini, G. \& Maraschi, L. 1989, \apj, 340, 181
\bibitem[Ghisellini ~1999]{ghisellini99}Ghisellini, G. 1999, ApL\&C, 39, 17
\bibitem[Gioia  \etal  ~1990]{emss} Gioia, I. M.,  Maccacaro, T.,  Schild, R. E., Wolter, A., Stocke, J. T.,  Morris, S. L.,  Henry, J. P., 1990 \apjs, 72, 576
\bibitem[Giommi \etal ~1995]{gioansmic} Giommi, P., Ansari, S. G. \& Micol, A., 1995, \aaps, 109, 267
\bibitem[Giommi  \etal  ~2001]{giommi01}Giommi, P., Ghisellini, G., Padovani, P., \& Tagliaferri, G.,  2001, AIPC, 599, 441
\bibitem[Giommi \etal ~2002]{giommisax} Giommi, P. \etal ~2002, babs.conf,  63.
\bibitem[Giommi \& Colafrancesco (2004)]{giocol04} Giommi, P. \& Colafrancesco, S., 2004 \aap, 414, 7
\bibitem[Giommi \etal  ~2005]{giommi05} Giommi, P., Piranomonte, S.,  Perri, M. \& Padovani, P., 2005, \aap, 434, 385
\bibitem[Giommi \etal  ~2006a]{giommi06} Giommi, P., Colafrancesco, S., Cavazzuti, E., Perri, M. and Pittori, C. 2006 \aap, 445, 843
\bibitem[Giommi \etal  ~2006b]{giommi454.3} Giommi, P.,  Blustin, A. J.,  Capalbi, M. \etal ~2006, \aap, 456, 911
\bibitem[Giommi \etal  ~2007]{GiommiWMAP07} Giommi, P., Capalbi, M.,  Cavazzuti, E.  \etal ~2007, \aap, 468, 571
\bibitem[Giommi \etal  ~2008]{giommi08} Giommi, P., \ et al.\ 2008, \aap 487, L49
\bibitem[Giommi \etal  ~2009]{giommiWMAP3} Giommi, P., \ et al.\ 2009 \aap, in press (arXiv:0908.0652)
\bibitem[Grandi \etal ~2004]{grandi04} Grandi, P. \& Palumbo, G.G.C., 2004, Science,  306, 998
\bibitem[Hartman \etal ~1999]{hartman} Hartman, R.C. et al., 1999, \apjs, 123, 79
\bibitem[Healey \etal ~2007]{crates}Healey, S.\ E.\ et al.\ 2007, \apjs, 171, 61
\bibitem[Healey \etal ~2008]{Healey08}Healey, S.\ E.\ et al.\ 2008, \apj, 175, 97
\bibitem[Hill  \etal ~2004]{Hill04}Hill, J. E., Burrows, D. N., Nousek, J. A., \ et al. \ 2004, Proc. SPIE, 5165, 217
\bibitem[Hunter \etal  ~1997]{Hunter97} Hunter S D \etal  1997 \apj ,  481, 205
\bibitem[Kalberla \etal ~2005]{Kalberla05}  Kalberla, P.M.W., Burton, W.B., Hartmann, Dap, et al. 2005, \aap, 440, 775
\bibitem[Komatsu \etal ~2009]{Komatsu}Komatsu, E.\ et al.\  2009,  \apjs, 180, 330
\bibitem[Korolkov \& Parijskij 1979]{RATANreview} Korolkov, D.~V., \& Parijskij, Yu.~N.\ 1979, Sky~Telesc., 57, 324
\bibitem[Kovalev \etal 1999]{Kov99} Kovalev, Y.~Y., Nizhelsky, N.~A., Kovalev, Yu.~A., Berlin, A.~B.,
Zhekanis, G.~V., Mingaliev, M.~G., \& Bogdantsov, A.~V.\ 1999, \aap Suppl., 139, 545
\bibitem[Kovalev \etal ~2002]{Kov02} Kovalev, Y. Y., Kovalev, Yu. A., Nizhelsky, N. A., Bogdantsov, A. B.\ 2002, PASA, 19, 83
\bibitem[Kovalev \etal  ~2007]{VCS5} Kovalev, Y.~Y., Petrov, L., Fomalont, E.~B., \& Gordon, D.\  2007, \aj, 133, 1236
\bibitem[Kovalev \etal  ~2009]{Kov09}  Kovalev, Y.~Y. \etal. 2009, \apj, 696L, 17
\bibitem[Kovalev ~2009]{Kov09a}  Kovalev, Y.~Y. 2009, \apj, 707, L56
\bibitem[Kubo \etal 1998]{kubo} Kubo, H., Takahashi, T., Madejski, G., Tashiro, M., Makino, F., Inoue, S., \& Takahara, F.\ 1998, \apj, 504, 693
\bibitem[March\~a \etal ~1996]{marcha96} March\~a, M. J. M., Browne, I.W.A., Impey, C.D., \& Smith, P.S., 1996, \mnras, 281, 425
\bibitem[Jones \etal ~1974]{Jones74} Jones, T. W., OÍdell, S. L., \& Stein, W. A. 1974, \apj, 188, 353
\bibitem[Lagage \etal ~2004]{2004Lagage} Lagage, P.O. \etal, 2004., Msngr, 117, 12
\bibitem[Lister \etal ~2009]{Lister09} Lister, M. L. \etal, 2009, \apj, 696L, 22
\bibitem[Massaro et al.~2004]{Massaro04} Massaro, E., Perri, M., Giommi, P., \& Nesci, R. 2004, \aap, 413, 489
\bibitem[Massaro et al.~2006]{Massaro06} Massaro, E., Tramacere, A., Perri, M., Giommi, P., \& Tosti, G. 2006, \aap, 448, 861
\bibitem[Massaro et al.~2009]{Massaro09} Massaro, E.,  Giommi, P.,  Leto, C.,  Marchegiani, P.,  Maselli, A.,  Perri, M.,  Piranomonte, S.,  Sclavi, S., 2009 \aap, 495, 691
\bibitem[Maselli \etal ~2009]{Maselli09} Maselli, A., Massaro, E. , Nesci, R., Rossi, C., Giommi, P., 2009, submitted to \aap
\bibitem[Mazziotta 2009]{Unfolding} M.N. Mazziotta, ICRC proceedings 2009
\bibitem[Monet  \etal  ~2003]{usno} Monet, D. G. et al. 2003, \aj,125, 984
\bibitem[Moretti \etal ~2005]{Moretti05} Moretti, A., Campana, S., Mineo, T., et al. 2005, Proceedings of SPIE, Vol. 5898, 360
\bibitem[Nieppola \etal ~2006]{nieppola}Nieppola, E., Tornikoski, M., Valtaoja, E., 2006, \aap, 445, 441
\bibitem[Nilsson \etal ~2007]{Nilsson07} Nilsson 2007, \aap, 475, 199
\bibitem[Ott \etal ~1994]{Ott94} Ott, M., Witzel, A., Quirrenbach, A., Krichbaum, T.~P., Standke, K.~J., Schalinski, C.~J., \& Hummel, C.~A.\ 1994, \aap, 284, 331
\bibitem[Padovani \& Giommi 1995]{padgio95} Padovani, P. \& Giommi, P., 1995, \apj, 444, 567
\bibitem[Padovani \& Giommi 1996]{padgio96}  Padovani, P. \& Giommi, P., 1996 \mnras, 279, 526
\bibitem[Padovani \etal ~2003]{pad03} Padovani, P., Perlman, E.S.,  Landt, E., Giommi, P., Perri, M., 2003, \apj 588, 128
\bibitem[Padovani \etal ~2006]{pad06} Padovani, P. \etal ~2006, \aap, 456, 131
\bibitem[Pittori \etal ~2009]{Pittori09} Pittori, C.\ et al.\  2009, \aap, in press
\bibitem[Pushkarev \etal ~2009]{Pushkarev09} Pushkarev, A.B.,  Kovalev, Y.Y., Lister, M.L., \& Savolainen, T.,  2009, \aap, 507, L33
\bibitem[Poole \etal ~2008]{poole08} Poole ? \ et al.\  2008 \mnras, 383, 627
\bibitem[Raiteri \etal ~2008a]{raiteri08a} Raiteri, C.\ et al.\ 2008, \aap 485, L17
\bibitem[Raiteri \etal ~2008b]{raiteri08b} Raiteri, C.\ et al.\ 2008, \aap 480, 339
\bibitem[Raiteri \etal ~2005]{raiteri05} Raiteri, C.\ et al.\ 2005 \aap 438, 39
\bibitem[Readhead \etal ~1989]{Readhead89} Readhead \ et al.\ 1989, \apj, 346, 566
\bibitem[Roming \etal ~2005]{Roming05} Roming, P. W.~A., Kennedy, T.~E., Mason, K.~O., {et~al.} 2005, SSRv, 120, 143
\bibitem[Sambruna \etal ~1996]{sambruna96}Sambruna, R.M., Maraschi, L., \& Urry, C.M., 1996, \apj, 463, 444
\bibitem[Savolainen \etal ~2009]{Savolainen09}Savolainen, T.,  \ et al.\  2009, submitted to \aap, arXiv:0911.4924
\bibitem[Schlegel \etal 1998]{schlegel1998} Schlegel, D.J., Finkbeiner D.P., and Davis, M., 1998, \apj, 500, 525
\bibitem[Sikora \etal ~1994]{Sikora94} Sikora, M., Begelman, M. C., \& Rees, M. 1994, \apj, 421, 153
\bibitem[Sikora \etal  ~2002]{Sikora02} Sikora, M., Blazejowski, M., Moderski, R., \& Madejski, G. M. 2002, \apj 577, 78
\bibitem[Stickel \etal ~1991]{stickel91} Stickel, M., Padovani, P., Urry, C. M., Fried, J. W., \& K\"uhr, H., 1991, \apj, 374, 431
\bibitem[Tavani \etal 2008]{tavani08} Tavani, M., Barbiellini, G., Argan, A. et al. 2008, Nucl. Instr. and Meth. in Phys. Res. A, 588, 52
\bibitem[Toffolatti \etal 1998]{toffolatti98} Toffolatti, L.,  Argueso Gomez, F.,  de Zotti, G.,  Mazzei, P.,  Franceschini, A.,  Danese, L., Burigana, C., 1998 \mnras, 297, 117
\bibitem[Tramacere \& Tosti 2003]{Tramacere03} Tramacere, A., \& Tosti, G. 2003, NewA Rev., 47, 697
\bibitem[Tramacere 2007a]{Tramacere07a} Tramacere, A. 2007a, PhD thesis, Spectral Variability in blazars High Energy Emission  La Sapienza University, Rome (2007)
\bibitem[Tramacere 2007b]{Tramacere07b} Tramacere, A.,Massaro, F., \& Cavaliere, A. 2007b, \aap, 466, 521
\bibitem[Tramacere 2009]{Tramacere09} Tramacere, A. et al., 2009, \aap, 501, 879
\bibitem[Teshima 2008]{Teshima08} Teshima, M. 2008, ATel 1500
\bibitem[Thompson \etal ~1993]{Thompson93}  Thompson, D. J. \etal ~1993 \apjs, 86, 629
\bibitem[Urry \& Padovani 1995]{Urry95} Urry, M. \& Padovani, P. 1995, PASP, 107, 803
\bibitem[Vercellone \etal ~2009]{vercellone09} Vercellone, S. et al. 2009, \apj, 690, 1018
\bibitem[Villata \etal ~2007]{villata07} Villata M.\ et al.\  2007,  \aap 464, L5
\bibitem[Villata \etal ~2008]{villata08} Villata M.\ et al.\  2008 \aap  481, L79
\bibitem[von Montigny \etal ~1995]{vonmontigny95}von Montigny, C \ et al.\  1995 \apj, 440, 525
\bibitem[Weekes 2008]{Weekes08}Weekes T. C. 2008, in AIP conf. series, ed. F. A. Aharonian, W. Hofmann, \& F. Rieger, 1085, 3

\end{thebibliography}
\end{document}